\documentclass[usenatbib]{aa}
\usepackage{graphicx}
\usepackage[varg]{txfonts}
\usepackage[english,english]{babel}
\usepackage{amsmath}
\usepackage{amssymb,amsfonts,textcomp}
\usepackage{array}
\usepackage{supertabular}
\usepackage{hhline}
\usepackage[usenames]{color}
\usepackage{hyperref}
\hypersetup{colorlinks=true, linkcolor=black, citecolor=blue, filecolor=blue, urlcolor=blue}
\usepackage{xspace}
\bibpunct{(}{)}{;}{a}{}{,}

\def\gtrsim{\lower.5ex\hbox{$\; \buildrel > \frac \sim \;$}}

\newcommand{\hagn}{\mbox{{\sc \small Horizon-AGN}}}

\newcommand{\nh}{\mbox{{\sc \small NewHorizon}}}
\newcommand{\nhs}{\mbox{{\sc \small NewHorizon\,}}}

\newcommand{\ramses}{\mbox{{\sc \small RAMSES}}}

\newcommand{\PSIS}{$\Psi_{\mathrm{star,BH}}$\,}

\newcommand{\PSIGAS}{$\Psi_{\mathrm{gas,BH}}$}

\newcommand{\crg}{gas-versus-stars counter-rotating galaxies}
\newcommand{\crgg}{gas-versus-stars counter-rotating galaxies\,}

\newcommand{\CR}{gas-versus-stellar kinematic misalignment}

\newcommand{\crd}{gas-versus-stars counter-rotating disks}
\newcommand{\crdd}{gas-versus-stars counter-rotating disks\,}

\newcommand{\Vsig}{$V_*/\sigma$}
\newcommand{\Vsigg}{$V_*/\sigma$\,}

\newcommand{\diagagee}{$\epsilon_*$-$r$-$age$\,}
\newcommand{\diagmetall}{$\epsilon_*$-$r$-$metal$\,}

\newcommand{\diagmetalgass}{$\epsilon_g$-$r$-$metal$\,}

\newcommand{\degrees}{$^\circ$}

\newcommand{\tuniv}{$T_{\mathrm{Universe}}$}

\newcommand{\dsbh}{$d_{\mathrm{star,BH}}$}

\newcommand{\epoch}[1]{{\bf{\textcircled{\tiny{#1}}}}}

\definecolor{grey}{rgb}{0.75,0.75,0.75}
\definecolor{Orange}{rgb}{1.0,0.5,0.15}
\definecolor{brown}{rgb}{0.7,0.25,0.0}
\definecolor{pink}{rgb}{1.0,0.5,0.5}
\definecolor{darkerred}{rgb}{0,0.5,0.5}
\definecolor{darkerblue}{rgb}{0,0,0.8}
\definecolor{lightblue}{rgb}{0.12, 0.56, 1.0}
\definecolor{Blue}{rgb}{0,0.08,0.65}
\definecolor{Red}{rgb}{0.65,0.08,0.05}
\definecolor{Green}{rgb}{0.15,0.45,0.25}

\begin{document} 

%\title{Dissecting the formation of counter-rotating gas-stellar disks \\ from the NewHorizon simulation}

\title{Dissecting the formation of gas-versus-star counter-rotating
galaxies from the NewHorizon simulation}

\titlerunning{Counter-rotating disks}
\authorrunning{S. Peirani et al.}

\author{
S\'ebastien Peirani\inst{1,2,3,4}
\and Yasushi Suto\inst{5,6,7}
\and Seongbong Han\inst{8}
\and Sukyoung K. Yi\inst{8}
\and Yohan Dubois\inst{4}
\and Katarina Kraljic\inst{9}
\and Minjung Park\inst{10}
\and Christophe Pichon\inst{4,11,12}
} 
\institute{
% 1 %
ILANCE, CNRS – University of Tokyo International Research Laboratory, Kashiwa, Chiba 277-8582, Japan, France\\\email{sebastien.peirani@cnrs.fr}
% 2 %
\and Kavli IPMU (WPI), UTIAS, The University of Tokyo, Kashiwa, Chiba 277-8583, Japan
% 3 %
\and Universit\'e C\^ote d'Azur, Observatoire de la C\^ote d'Azur, CNRS, Laboratoire Lagrange, Bd de l'Observatoire,\\
 CS 34229, 06304 Nice Cedex 4
% 4 %
\and Institut d'Astrophysique de Paris, CNRS and Sorbonne Universit\'e, UMR 7095, 98 bis Boulevard Arago, F-75014 Paris, France
% 5 %
\and
Research Institute, Kochi University of Technology, Tosa Yamada, Kochi
782-8502, Japan
% 6 %
\and Department of Physics, School of Science, The University of Tokyo, 7-3-1 Hongo, Bunkyo-ku, Tokyo 113-0033, Japan
% 7 %
\and Research Center for the Early Universe, School of Science, The University of Tokyo, 7-3-1 Hongo, Bunkyo-ku, Tokyo 113-0033, Japan
% 8 %
\and Department of Astronomy and Yonsei University Observatory, Yonsei University, Seoul 03722, Republic of Korea
% 9 %
\and Observatoire Astronomique de Strasbourg, Universit\'e de Strasbourg, CNRS, UMR 7550, F-67000 Strasbourg, France
% 10 %
\and Center for Astrophysics | Harvard \& Smithsonian, Cambridge, MA, USA
% 11 %
\and IPhT, DRF-INP, UMR 3680, CEA, L'Orme des Merisiers, B\^at 774, 91191 Gif-sur-Yvette, France
% 12 %
\and Kyung Hee University, Dept. of Astronomy \& Space Science, Yongin-shi, Gyeonggi-do 17104, Republic of Korea
% 13 %
%\and
%Astrophysics, University of Oxford, Denys Wilkinson Building, Keble Road, Oxford, OX1 3RH, UK
}

   \date{accepted (21/02/2025)}

% \abstract{}{}{}{}{} 
% 5 {} token are mandatory
 
  \abstract
  {Gas-versus-star counter-rotating galaxies
   are characterized by the presence of a disk of stars and a disk of gas 
   that are co-spatial but rotating in opposite directions.
   Using the \nhs simulation, we identified and studied ten such galaxies in field environments with a stellar mass of M$_*\sim$[1-5]$\times$10$^{10}$ M$_{\odot}$.
  For all of them, 
  the retrograde accretion of gas either from gas stripping
  from a nearby companion or from the circumgalactic medium is the 
  starting point of the formation process. This is followed by 
  the coexistence of two distinct disks of gas (or components) rotating in opposite directions, 
  with the pre-existing disk in the inner parts of the galaxy and the accreted gas in the
  outer parts. The latter
  progressively replaces the former, leading to the final gas-star kinetic misalignment configuration. During the process, star formation is first enhanced and then progressively decreases. We roughly estimate that a higher fraction of the pre-existing gas is converted into stars rather than being expelled. We also
  found that the black hole (BH) activity tends to be enhanced during
  the removal of the pre-existing gas.
  Furthermore, our analysis suggests that the formation of a counter-rotating
  gas component is always accompanied with the formation of counter-rotating stellar disks.
  These stellar disks can have diverse properties, but in general, they host a younger and more metal rich 
  population of stars  with respect to the main disk,
  depending on the star formation history and BH activity.
  The central part of counter-rotating disks also tend to be characterized by a younger population, an enhanced
  star formation rate, and a higher metallicity than their outer parts. 
  The high metallicity comes from the progressive metal enrichment of the accreted gas through mixing with the pre-existing gas and by supernovae activity
  as the accreted gas sinks toward the center of the galaxy.
  In case of major mergers, a large amount of
  accreted stars from the companion would be distributed at large distances from the remnant center due to conservation of the initial orbital angular momentum. This process might favor the observation of two distinct counter-rotating stellar disks, particularly in observed projected velocity fields 
  from integral field spectroscopy surveys, as well as
  stellar streams characterized by specific age-metallicity properties.
  }


   \maketitle
%
%-------------------------------------------------------------------
\section{Introduction}

In the traditional picture of galaxy formation, galaxies form
when baryonic gas falls into the gravitational potential well of
their dark matter halos. The gas initially heats up by shocks, cools radiatively, forming dense clouds which sink to the center
of the halo and in which stars are formed \citep{fall80,mo98,cole00}.
Thus, the stellar component of galaxies, especially the disk,
is expected to inherit most of the dynamical properties of the
gas component out of which it forms. In particular, the
average angular momentum (AM) vectors of both components are expected
to be well aligned. 

However, this traditional picture is seriously challenged by the presence  of kinematic misalignments between gas and stellar components, commonly observed in  nearby galaxies, mainly in $\sim$30-40\% of elliptical and lenticular galaxies. Indeed, the discovery of such galaxies has increased due to 
the advent of galaxy surveys using integral field spectroscopy 
 (IFS) such as SAURON \citep{sauron}, ATLAS$^{\mathrm{3D}}$ \citep{atlas3d}, CALIFA \citep{califa}, SAMI \citep{sami} or
 MaNGA \citep{bundy+15}.
In particular, these surveys permit to derive the projected two-dimensional maps
of both stellar and gas velocity from which 
one can estimate the kinematic position angle of the
stars ($\mathrm{PA}_{*}$) and the gas ($\mathrm{PA}_{\mathrm{gas}}$), defined as the orientation of the mean
stellar and gas motions on a map of velocities. A galaxy
is commonly defined as kinematically misaligned if
the difference of kinematic position, 
$\Delta \mathrm{PA}=|\mathrm{PA}_{*} - \mathrm{PA}_{\mathrm{gas}}|$ is greater than 30\degrees.
In some extreme cases, $\Delta \mathrm{PA}$ can be even close to 180\degrees
meaning that the stellar and gas components are counter-rotating.

Due to their atypical properties, \crgg have been the center of particular attention
in the last two decades. They indeed represent ideal laboratories to 
test theoretical models of galaxy formation. 
Most of the time, they do not only own a counter-rotating gaseous component but also
one or several counter-rotating stellar components. Thus, most observational studies have tried to detect and analyze
these specific galaxies to constrain the origin 
of the misaligned velocity fields between gas and stellar components,
by characterizing both their stellar populations, their gas-phase metallicity, their morphology, their kinematic properties and their cosmological environment \citep[e.g.,][]{sarzi+06,davis+11,coccato+11,johnston+13,garcia-lorenzo+15,jin+16,bryant+19,raimundo21,omori+21,li_manga+21,bevacqua+22,xu+22,zhou+22,bao+22,zinchenko+23,zhou+23,katkov+24}.

From the different analysis, counter-rotating components tend to have
younger stellar populations compared to the main stellar
discs, and their ionized gas rotates in the same direction as
the secondary stellar components, i.e. it also counter-rotates
with respect to the main disc. Moreover,
observations indicate that counter-rotating stars could have either a lower, similar or
higher metallicity with respect to stars belonging to the main disk
\citep{coccato+11,johnston+13,katkov+13,johnston+13,katkov+16}.
In view of such observational trends, a scenario in which counter-rotating components are assembled from gas accreted on retrograde orbits from the nearby environment
is privileged.
The sources of cold gas accretion may be dwarf
satellite merging \citep[e.g.,][]{kaviraj+11} or cosmological gas filaments \citep{keres+05,dekel+06,ocvirk+08} 
though it is often difficult to disentangle them.
The presences of different stellar populations between co- and counter-rotating stars seems therefore to rule out any internal origin of counter-rotating stars with for instance stars moving on retrograde orbits during a bar dissolution process \citep{evan&collett94}.
%, since only counter-rotating stellar discs with identical stellar population properties can be produced in the framework of this scenario.

Observational studies also suggest that the central part of the 
counter-rotating stellar disks has generally a younger population, an enhanced
star formation rate and a higher metallicity than their outer parts \citep{jin+16,chen+16}. The gas-stellar kinetic misalignment 
seems also more frequent in spheroidal dominated galaxies
than in late type galaxies \citep{kannappan+01,pizzella+04,davis+11,bryant+19,duckworth+20} and
also found in non-interacting systems \citep[e.g.,][]{califa,katkov+14}.

In parallel to observational analyzes, numerical simulations 
have allowed us to give new insights 
to explain the origin of the gas-star kinetic misalignment observed in 
some galaxies.
They first do support the idea that an external origin such as gas accretion, galaxy interaction and galaxy mergers, is generally needed.
They also confirm that the accreted gas (from nearby galaxies of filaments) exchanges AM with the pre-existing co-rotating gas causing it to fall toward the center of the galaxy \citep{thakar+96,roskar+10,vandevoort+15,taylor+18,kimmetal17,starkenburg+19,khim+20_paperI,khoperskov+21,lu+21,cenci+24}. This process may also trigger low-level AGN feedback \citep{raimundo+23}
and  boost the BH luminosity/growth and significantly inject more energy into the surrounding gas \citep{duckworth+19}.
Finally, according to theoretical studies, it may be possible to distinguish between different formation paths
such as galaxy merger, galaxy fly-by, 
interaction with dense environments or interaction with 
central galaxies in a galaxy cluster environment \citep{khim+21_paperII}
or several accretions from different filaments \citep{algorry+14}.

One interesting aspect which is still in debate concerns the co-evolution of the pre-exising gas disk and the accreted one. For instance, from IllustrisTNG  \citep{pillepich+18} and idealized simulations, \cite{khoperskov+21} suggest 
that the pre-existing gas might be either mixed with the infalling component or expelled. 
\cite{cenci+24} have studied Milky Way mass galaxies from the FIREbox cosmological simulation
\citep{feldmann+23,fire}. They presented the most dramatic case of kinetic misalignment in
which almost all of the pre-existing gas (95\%) is converted into stars.
Using the Illustris simulation \citep{illustris}, \cite{starkenburg+19} also
identified two main channels responsible for the gas loss: a strong feedback burst and gas stripping during a fly-by passage through a more massive group environment.

The aim of the present paper is to contribute to the theoretical effort
by proposing a detail analysis of the complex process of the formation of this class of galaxies. 
We therein 
use the \nhs simulation \citep{nhz} which allows a detailed analysis
of the stellar/gas dynamics thanks to  its high spatial resolution ($\sim$34 pc) that accurately resolves a typical scale height of
galactic disks \citep{park+21,sukyoung+24}. The high spatial resolution also
allows a reliable estimation of their stellar angular momentum while
the high mass resolution permit a better characterization of the stellar population. 
Moreover, \nhs permits to characterize the stellar populations in each simulated galaxy (through their age and global metallicity), which is essential to constrain and interpret the observations.

The paper is organized as follows. Section~\ref{sec:simu} briefly
introduces the \nh\, simulation and the numerical modeling used in
this work (simulations and post-process).
Section~\ref{sec:results} presents a detailed analysis of
the individual evolution of four galaxies that present differences in
their formation path. Then, we discuss
the main results and trends in section~\ref{sec:discussion}.
We summarize and conclude in
section~\ref{sec:conclusions}.

%%%%%%%%%%%%%%%%%%%%%%%%%%%%%%%%%%%%%%%%%%%%%%%%%%%%%%%%%%%%%%%%
\section{Methodology}
\label{sec:simu}

%\subsection{The NewHorizon simulation - general}

Throughout this paper, we analyze the results of the 
\nh\footnote{https://new.horizon-simulation.org/} simulation.
The details of the simulation has been described in many previous papers \cite[e.g,][]{nhz,peirani+24}, so 
we only summarize here its main features.

\subsection{The NewHorizon simulation}

\nh\, is a high-resolution zoom-in simulation from the \hagn\, simulation \citep{hagn}, which extracts a spherical sub-volume with a radius of 10 comoving Mpc.
A standard $\Lambda$CDM cosmology was adopted with the total matter density
$\Omega_{\rm m}$ = 0.272, the dark energy density $\Omega_\Lambda$ = 0.728,
the baryon density $\Omega_{\rm b}$ = 0.045, the Hubble
constant $H_0$=70.4 km s$^{-1}$ Mpc$^{-1}$, 
the amplitude of the matter power spectrum $\sigma_8$ = 0.81
and the power-law index of the primordial power spectrum $n_{\rm s}$ = 0.967,
according to the WMAP-7 data \citep{komatsu+11}. 
The initial conditions have been generated with 
{\mbox{{\sc \small MPgrafic}}} \citep{mpgrafic} at the resolution of 4096$^3$ for \nh\, in contrast to 1024$^3$ for \hagn.
The dark matter mass resolution reaches 1.2$\times$10$^6$ M$_\odot$ compared to
8$\times$10$^7$ M$_\odot$ in \hagn. As far as star particles are concerned, their typical mass resolution is $\sim$10$^4$ M$_\odot$ for  \nh.

Both simulations were run with the \ramses\, code \citep{ramses} in
which the gas component is evolved using a second-order Godunov scheme
and the approximate Harten-Lax-Van Leer-Contact~\citep[HLLC,][]{toro}
Riemann solver with linear interpolation of the cell-centered
quantities at cell interfaces using a minmod total variation
diminishing scheme.  In \nh, refinement is performed according to a
quasi-Lagrangian scheme with approximately constant proper highest resolution of 34 pc.
The refinement is triggered in a quasi-Lagrangian manner,
if the number of DM particles becomes greater than 8, or the total
baryonic mass reaches 8 times the initial DM mass resolution in a
cell. Extra levels of refinement are successively added at $z=$ 9, 4,
1.5 and 0.25 (i.e. for expansion factor $a=$ 0.1, 0.2, 0.4 and 0.8 respectively).  The
simulation is currently completed until the redshift $z=0.18$.

\subsection{Galaxy catalog}
\label{subsec:catalog}

%%%%%%%%%%%%%%%%%%%%%%%%%%%%%%%%%%%%%%%%%%%%%%%%%%%%%%%%%%%%%%
%     FIG 1
%%%%%%%%%%%%%%%%%%%%%%%%%%%%%%%%%%%%%%%%%%%%%%%%%%%%%%%%%%%%%%
\begin{figure}
\begin{center}
%\rotatebox{0}{\includegraphics[width=\columnwidth]{FIG_test.jpg}}
\rotatebox{0}{\includegraphics[width=\columnwidth]{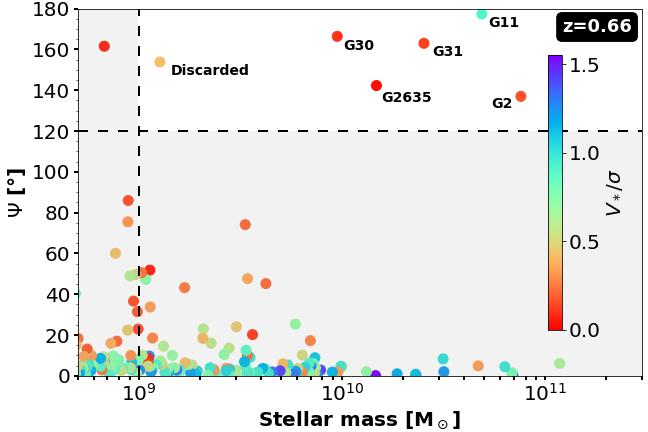}}
\rotatebox{0}{\includegraphics[width=\columnwidth]{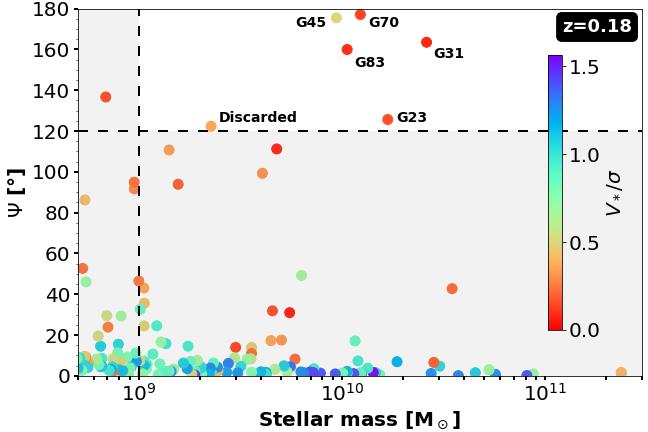}}
\caption{The 3-d angles $\Psi$ subtended between the stellar and gas angular momentum vectors, both estimated within one effective radius R$_e$, at $z=0.66$ (upper panel) and $z=0.18$ (lower
panel). Our final sample of \crgg consists in
 galaxies with non-disturbed morphology (or in an ongoing merging phase) 
 with a mass greater than 10$^9$M$_\odot$ and $\Psi$>120$^\circ$.
We also used a color coding according to the
quantity $V_*/\sigma$ (also estimated within one effective radius).
Gas-star kinetic misaligned galaxies are more likely to 
form in spheroidal dominated galaxies (associated with low $V_*/\sigma$ values). 
}
\label{fig1}
\end{center}
 \end{figure}
%%%%%%%%%%%%%%%%%%%%%%%%%%%%%%%%%%%%%%%%%%%%%%%%%

\begin{table}
\caption{\label{t7}The identity and mass of each galaxy identified
at $z=0.18$ from
the \nhs simulation and studied in this work. }
\centering
\begin{tabular}{lccc}
\hline\hline
Gal. id & Stellar mass & Comments\\
\hline
   G11  & 5.3$\times$10$^{10}$M$_\odot$   & 2 successive cosmic gas accretions; \\
   && Associated with BH1049\\
\hline   
   G23  & 1.7$\times$10$^{10}$M$_\odot$   & Minor Merger\\
   G45 & 9.4$\times$10$^{9}$M$_\odot$   & Minor Merger\\
   G70 & 1.2$\times$10$^{10}$M$_\odot$   & Minor Merger\\
\hline
   G31 & 2.6$\times$10$^{10}$M$_\odot$   & Major Merger; Ass. with BH549\\
   G83 & 1.1$\times$10$^{10}$M$_\odot$  & Major Merger \\
\hline
G2635 & 1.4$\times$10$^{10}$M$_\odot$   & Multiple Mergers + high SFR\\
   G2 & 2.4$\times$11$^{10}$M$_\odot$ &  Multiple Mergers; Ass. with BH796\\
  G30 & 2.8$\times$10$^{10}$M$_\odot$   & Multiple Mergers\\
G173 & 3.0$\times$10$^{9}$M$_\odot$ &   Multiple Mergers\\
\hline
\end{tabular}
\tablefoot{
Each of these galaxies display
one or several phases of gas-stars counter-rotation during
their lifetime. The last column add some information about
their formation path and also whether if they host
a massive central black hole. For instance, the galaxies G11
and G31 host respectively the black hole BH1049 and BH549, 
studied in detail in \cite{peirani+24}.
Note that G2 is merging with a more massive galaxy at $z\sim0.27$. Thus its identity and 
mass are taken just before the event.
}
\end{table}

%%%%%%%%%%%%%%%%%%%%%%%%%%%%%%%%%%%%%%%%%%%%%%%%%%%%%%%%%%%%%%
%     FIG 2
%%%%%%%%%%%%%%%%%%%%%%%%%%%%%%%%%%%%%%%%%%%%%%%%%%%%%%%%%%%%%%
\begin{figure}
\begin{center}
\rotatebox{0}{\includegraphics[width=\columnwidth]{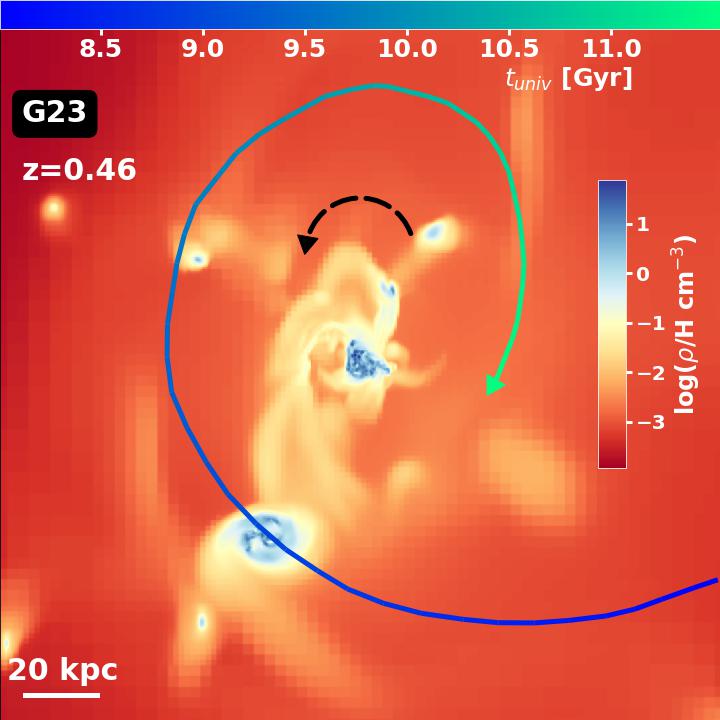}}
\rotatebox{0}{\includegraphics[width=\columnwidth]{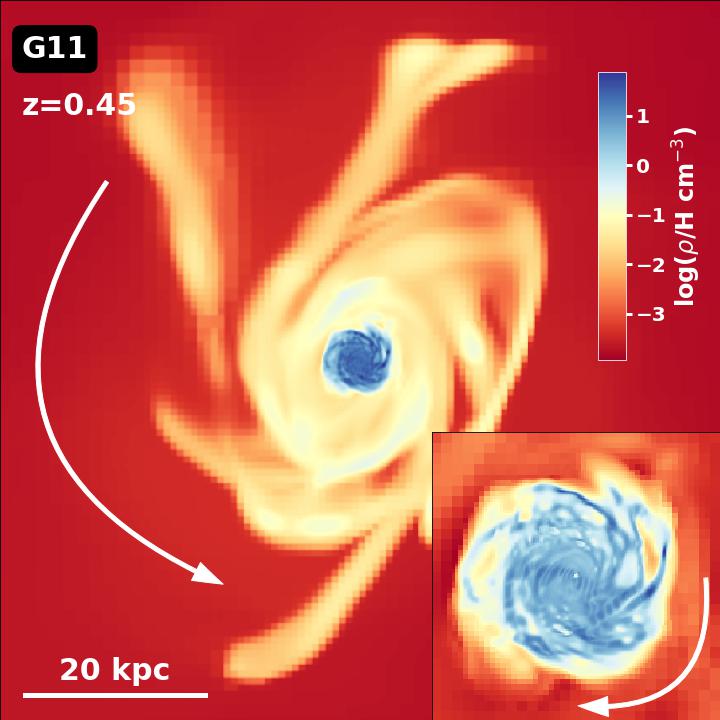}}
%\rotatebox{0}{\includegraphics[width=\columnwidth]{FIG_orbit3.jpg}}
%\rotatebox{0}{\includegraphics[width=9.1cm]{FIG_orbit.jpg}}
%\rotatebox{0}{\includegraphics[width=9.1cm]{FIG_orbit2.jpg}}
\caption{Projected gas density ($\rho$) around two galaxies from 
our sample, G23 (upper panel) and G11 (lower panel). These two panels
illustrate each of the two successive phases leading to the formation of \crd.  
In the upper panel, the central galaxy (G23) has a anti-clockwise rotation, as specified by the dashed circular arrow, and 
is accreting some gas from a companion galaxy through tidal stripping.
In this specific case, the orbit of the satellite galaxy is  
indirect (or retrograde) as illustrated by the orientation of the two arrows.
Then, the retrograde accretion of gas may lead afterward to the co-existence of two disks of gas that rotate in opposite sense, as shown in the lower panel by the spectacular case
of G11.
The accreted gas lies in the outer parts of the galaxy while the pre-existing one 
is concentrated in the inner parts. The process ends when the latter is totally replaced by the former. 
}
\label{fig2}
\end{center}
 \end{figure}
%%%%%%%%%%%%%%%%%%%%%%%%%%%%%%%%%%%%%%%%%%%%%%%%%

Galaxies are identified using the
structure finder {\mbox{{\sc \small Adaptahop}}} \citep{aubertetal04,tweed+09}
 an improved version of the {\mbox{{\sc \small HOP}}} algorithm \citep{hop}.
 This method identifies structures from the particle positions only.
 It determines first the local density of each stellar particle
from the 20 nearest neighbors using the standard SPH (smoothed particle hydrodynamics) spline kernel \citep{SPH}.
Then the algorithm jumps from one particle to its highest density neighbor until it reaches a local
maximum. Once all the local maxima of the field are found, a peak patch around each maximum
is defined as the set of particles above a specific density threshold.
We have used a density threshold equal to 178 times the average matter density at that redshift, a value commonly used in the literature to identify collapsed  and virialised objects.
Moreover, since \nh\, is a zoom simulation, low
mass-resolution dark matter particles might "pollute" some halos,
especially when they are located close to the boundary of the high
resolution area.  We remove those DM halos and their embedded
galaxies when the low-mass resolution particles budget is higher than
4\% of the total halo mass.
In the following, we define the mass of each galaxy by the value
returned by {\mbox{{\sc \small Adaptahop}}}.

\subsection{Identification of gas-versus-stars counter rotating galaxies}
\label{subsec:identification}

In the present work, we are mainly interested in studying the formation 
of galaxies in which the gas component is counter-rotating with respect
to a co-spatial stellar component. Following \cite{khim+21_paperII}, by "gas component" we refer here to the "galactic cold gas" separated from
the non-cold surrounding. To do so, we use the linear cut in the
logarithmic density-temperature plane using Equation (1) from \cite{torrey+12}:

\begin{equation}
    \mathrm{log}(T/[{\rm K}]) = 6 + 0.25\,\mathrm{log}(\rho/10^{10}[{\rm M_\odot}\,h^{2}\,{\rm kpc}^{-3}]).
\end{equation}

Depending on the density, the "galactic cold gas" is mainly characterized by a temperature of about 10\,000 - 30\,000 K. It is used for the star formation
model in the simulation and has similar properties to the gas component
observed in H$\alpha$ emission lines ($\simeq$10\,000 K) in IFS such as SAMI.

To identify such objects in the simulation, we have
computed the three-dimensional angle ($\Psi$) subtended between the stellar and gas AM vectors, at different redshifts.
These two vectors are calculated by taking all stars (cold gas cells) within
one effective radius ($R_e$), estimated by taking the geometric mean of the
half-mass radius of the projected stellar densities along each of the
simulation's Cartesian axes.
In our selection process, we first identified galaxies satisfying both a stellar mass greater than 10$^9$M$_\odot$ (corresponding to $\sim$10\,000 stellar particles) and $\Psi$>120$^\circ$. 
Then, we visually inspected both the edge-on projected velocity fields and the distribution of the circularity parameter (see the definition below) relative to the star and gas components.
Galaxies displaying complex velocity fields without any clear distinct peaks close to 1 and -1 in the distribution of the circularity parameter of stars and gas respectively were discarded.
%In general, these objects are in an ongoing merger process or in interacting systems resulting in disturbed galaxy shapes. Although this may lead to high $\Psi$ values, those galaxies are
%irrelevant for our study.

For instance, Fig.~\ref{fig1} shows the values of the 3-d angle $\Psi$ 
for galaxies extracted at $z=0.66$ and $z=0.18$. Among the six galaxies that satisfy
the above criteria at $z=0.18$,  five have been finally selected  (G23, G31, G45, G70 and G83). 
 Interestingly, the same number (and a similar fraction of galaxies) has been derived at $z=0.66$. The galaxy G31 is selected in the two samples suggesting either a long or several gas-star kinetic misalignment phases
Indeed, the galaxy identity is unique across all redshifts, so that the same
identity means the same galaxy at different redshifts.
We have repeated this procedure for snapshots regularly spaced in time (every $\sim$0.5 Gyr) from $z=3$ to $z=0.18$.
At the end, ten low-mass galaxies
in the simulation present or have presented clear gaseous counter-rotating component. We study in the next section this sample of galaxies whose
relevant information is summarized in Table.~\ref{tab1}.

Another way to identify counter-rotating galaxies is to use the so-called circularity
parameter, introduced by \cite{abadi+03}. 
The circularity parameter for each 
stellar particle ($\epsilon_*$) or gas cell ($\epsilon_{\mathrm{g}}$) is calculated based on the specific angular momentum ($J_z$) of each particle or cell along the galaxy stellar spin axis, divided
by the maximum specific angular momentum allowed on a circular orbit with the same specific
binding energy $E$, namely $\epsilon \equiv J_z/J_{\mathrm{max}}(E)$. In this case, $\epsilon$ ranges
between -1 and 1. Another definition can be found in the literature in which the specific angular momentum of each stellar particle 
is divided by the angular momentum corresponding to a circular orbit $J_{circ}$:
\begin{equation}
\epsilon \equiv \frac{J_z}{J_{\mathrm{circ}}},   \,\,\,\, \mathrm{with}  \,\,\,\, 
J_{\mathrm{circ}} = rV_{\mathrm{cir}} = r\sqrt{\frac{GM_{\mathrm{tot}}(<r)}{r}}
\end{equation}

\noindent
where $G$ is the gravitational constant and $M_{\mathrm{tot}}$ is the total mass (DM+gas+stars+BH)
within a sphere of radius "r" centered on the galaxy center. 
In each definition, stellar particles or gas cells with $\epsilon>0$ ($\epsilon<0$) are on co-rotating (counter-rotating) orbits with respects to the average rotation 
axis of the stellar component.
The two definitions lead to consistent trends and results, as already pointed out
by \cite{cenci+24}.
In the following, we use the latter definition as it
requires less computational effort. 
Then, to identify gas-star counter-rotating galaxies, 
one can select objects in which a large fraction of cold gas cells are in retrograde motion
within one effective radius by considering, for instance, the average satisfying $\langle\epsilon_{\mathrm{g}}\rangle<-0.5$.
We have also checked that this approach leads to the selection of the same sample of galaxies.

Note that there is potentially a third (though indirect) way to identify galaxies with the presence of stars-gas counter-rotating components.
As we will see in the next sections, these galaxies
also display clear stars-stars counter-rotating components
which can be revealed by the presence of a double-peaked in the stellar velocity dispersion maps \citep[e.g.][]{krajnovic+11,rubino+21,bevacqua+22,bao+24,katkov+24}.
We will discuss this aspect in section~\ref{subsec:stellarpop}.

In Fig.~\ref{fig1}, we also used a color coding according to the
quantity $V_*/\sigma$, where $V_*$ is the rotation stellar velocity
and $\sigma$, the stellar velocity dispersion. 
The ratio $V_*/\sigma$ can be used as a proxy of galaxy morphology 
\citep[see][for more
information]{peirani+24}.
The two panels of the figure clearly indicate that gas-vs.-stellar counter-rotating galaxies are 
more likely to be found in more spheroid dominated galaxies ($V_*/\sigma<0.5$) which
is consistent with observational analysis \citep{xu+22}.

%%%%%%%%%%%%%%%%%%%%%%%%%%%%%%%%%%%%%%%%%%%%%%%%%%%%%%%%%%%%%
%     FIG 3
%%%%%%%%%%%%%%%%%%%%%%%%%%%%%%%%%%%%%%%%%%%%%%%%%%%%%%%%%%%%%%
\begin{figure*}
\begin{center}
\rotatebox{0}{\includegraphics[width=3.6cm]{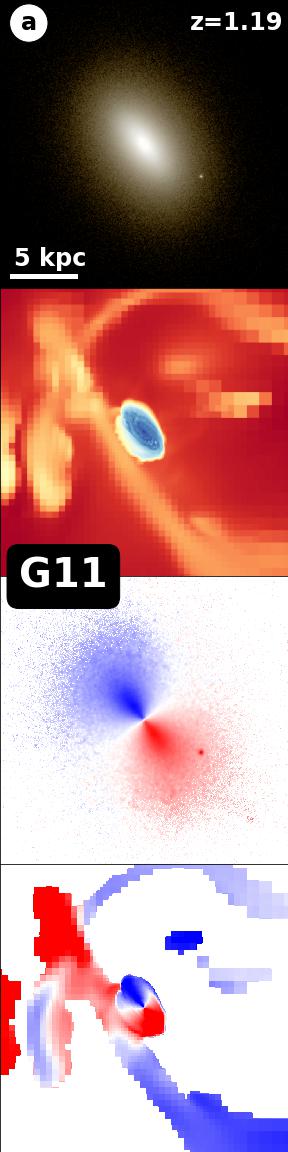}}
\rotatebox{0}{\includegraphics[width=3.6cm]{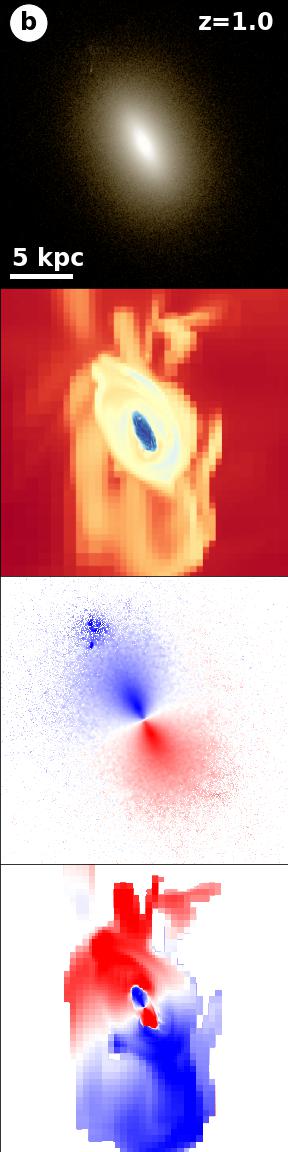}}
\rotatebox{0}{\includegraphics[width=3.6cm]{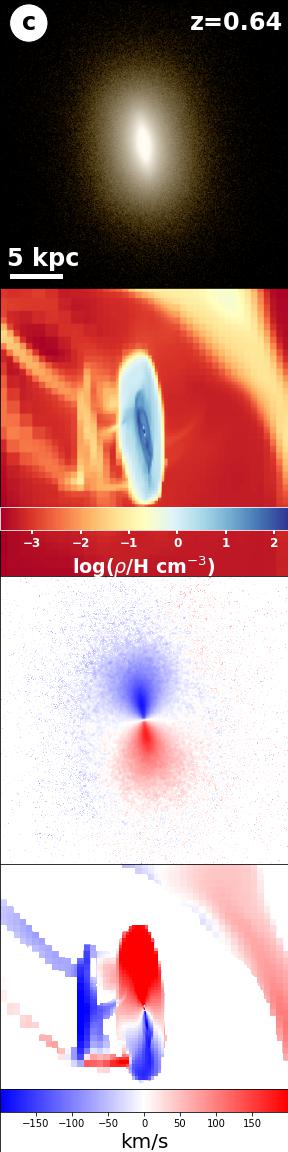}}
\rotatebox{0}{\includegraphics[width=3.6cm]{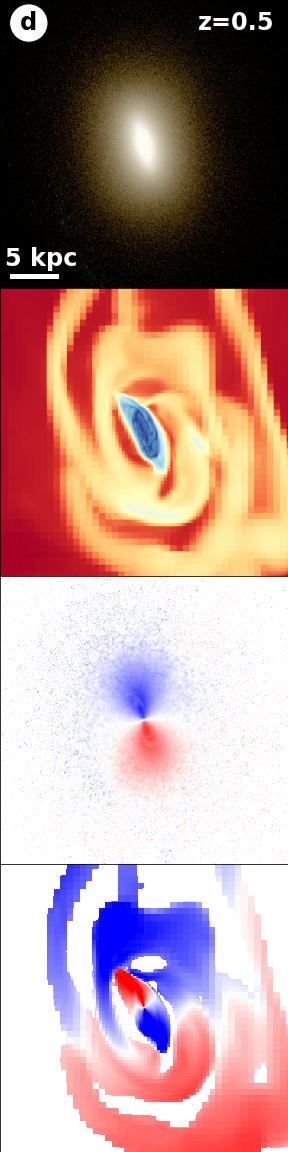}}
\rotatebox{0}{\includegraphics[width=3.6cm]{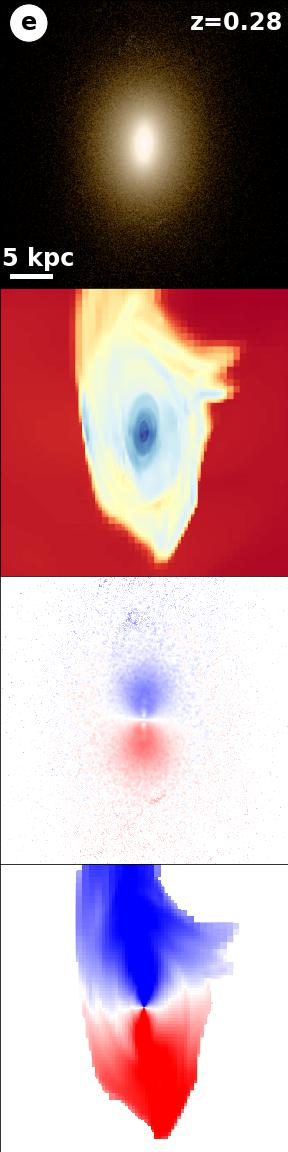}}
\caption{Galaxy G11. Projected u-g-r band images of the stellar component (first row), projected gas density ($\rho$, second row) and the associated projected velocity fields (third and fourth rows) at five different epochs (symbolized
by items \epoch{a}-\epoch{e}).
The values of rotation velocities are indicated by the color bar,
the red parts are moving away from us while the blue ones 
are approaching us.
The galaxy undergoes two successive episodes of retrograde gas accretion.
At epoch \epoch{a}, the stars and central gas component are co-rotating. Then the first gas accretion lead to the coexistence of
two disks of gas which rotate in opposite direction: the accreted material in
the outer parts and the pre-existing one on the inner region (\epoch{b}). At epoch \epoch{c}, the new disk of gas has totally replaced the 
pre-existing one leading to the stellar-gas counter-rotating configuration. However, a new
episode of retrograde gas accretion with respect to the inner disk (hence this new gas component is co-rotating with the stars) is already starting.
 Then epochs \epoch{d} and \epoch{e} repeat the same steps namely
 co-existence of two counter-rotating gas disks ending with a complete replacement. 
}
\label{fig3}
\end{center}
 \end{figure*}
%%%%%%%%%%%%%%%%%%%%%%%%%%%%%%%%%%%%%%%%%%%%%%%%%

%%%%%%%%%%%%%%%%%%%%%%%%%%%%%%%%%%%%%%%%%%%%%%%%%%%%%%%%%%%%%%
%     FIG 4   G11 COSMIC
%%%%%%%%%%%%%%%%%%%%%%%%%%%%%%%%%%%%%%%%%%%%%%%%%%%%%%%%%%%%%%
\begin{figure*}
\begin{center}
\rotatebox{0}{\includegraphics[width=17cm]{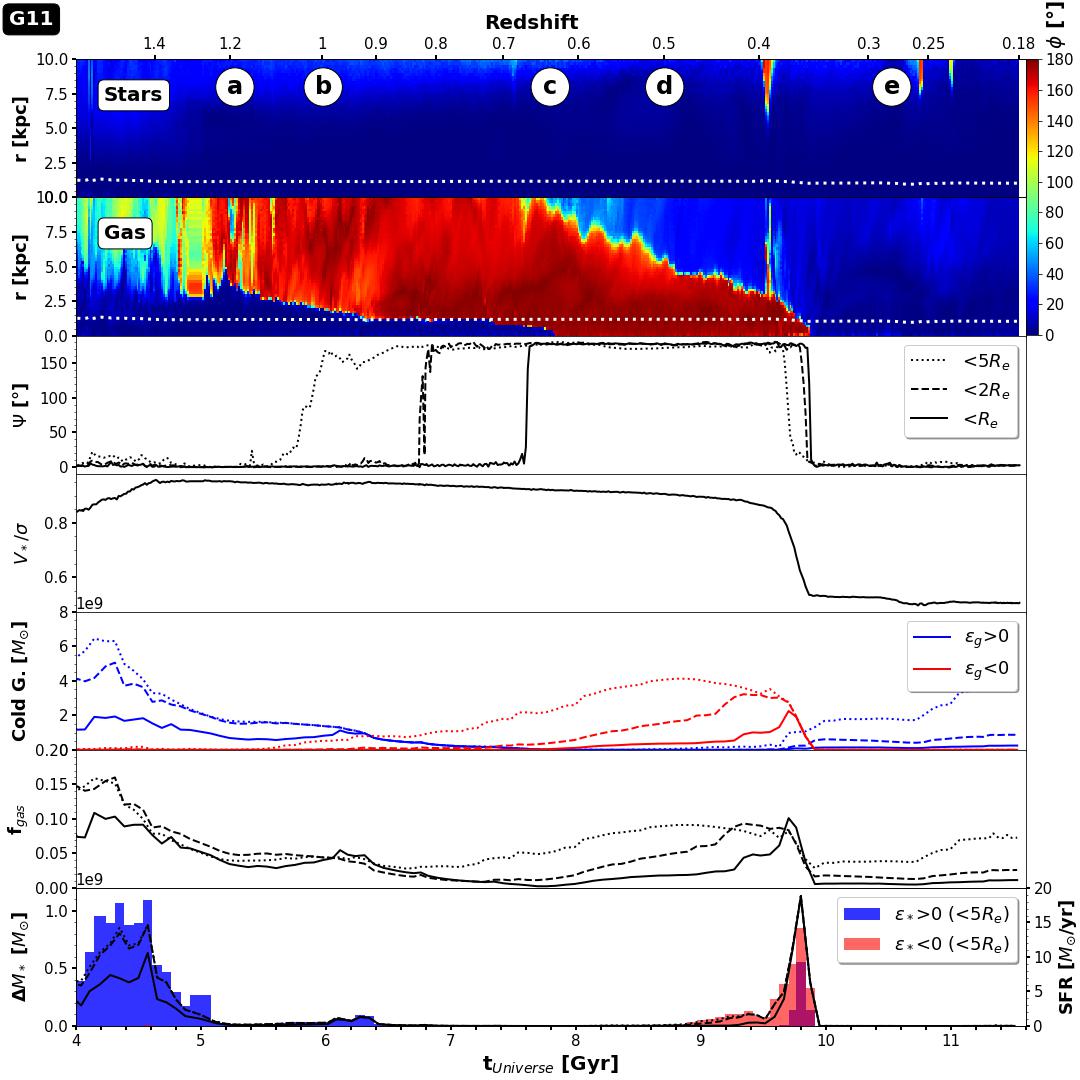}}
\caption{Cosmic evolution of relevant properties relative to G11.
From top to bottom: 
{\bf 1)} the mass-weighted angle $\Phi(r)$ between the
angular momentum vector of the stellar component (computed by taking all stars within one effective radius),
and the angular momentum vector of all stars located within two concentric spheres of
radius $r-dr$ and $r+dr$. $\Phi(r)$ is estimated within 0<$r$<10 kpc using
50 bins. The dotted line indicates the time evolution
of $R_e$; 
{\bf 2)} $\Phi(r)$ computed this time with all "cold" gas cells located within two concentric spheres of radius $r$ and $r+dr$;
{\bf 3)} the evolution of the  3-d angle $\Psi$ subtended between the stellar and gas angular momentum vectors estimated within one (solid line),
two (dashed line) or five (dotted line effective radius);
{\bf 4)} The ratio of the stellar rotational velocity to the 
stellar velocity dispersion $V_*/\sigma$, estimated within one effective
radius;
{\bf 5)} The masses of cold gas that co- or counter-rotate with the stellar component, by selecting all (cold) gas cells with a positive (blue lines) or negative (red lines) circularity parameters. We also estimate these two
quantities withing one, two or five effective radius;
{\bf 6)}
The evolution of the (cold) gas fraction, $f_{\rm gas}=M_{\rm gas}/(M_{\rm gas}+M_{\rm stars})$
also estimated withing one, two or five effective radius;
{\bf 7)} The mass of new stars formed in situ within 5 effective radii
in successive intervals, and color coded according to 
a positive (blue lines) and negative (red lines) circularity parameters.
We also plot in the same panel,
the cosmic star formation rate estimated by identifying
all stars born in situ within the last 100 Myr, within one, two or five
effective radii. 
Note that the z-axis of the cylindrical coordinate at any time
corresponds to the total angular momentum vector direction of the stellar component estimated
within one effective radius.
}
\label{fig_g11_cosmic}
\end{center}
\end{figure*}
%%%%%%%%%%%%%%%%%%%%%%%%%%%%%%%%%%%%%%%%%%%%%%%%%

%%%%%%%%%%%%%%%%%%%%%%%%%%%%%%%%%%%%%%%%%%%%%%%%%%%%%%%%%%%%%%%%
\section{Results}
\label{sec:results}

%\subsection{General counter-rotating galaxy Formation path}
%\label{subsec:path}

The ten galaxies identified in the last section develop during their lifetime 
at least one phase of \crd. As we will see in more detail 
in the next sections, the physical mechanism responsible for such configurations
is common for all them. It consists of two successive phases:

\noindent
i) the first phase starts with the cosmic accretion of gas either 
from a nearby companion, by tidal stripping, or from the circumgalactic medium.
This accretion is always retrograde.
%fly in by or in a retrograde pre-merger configuration. 
An example is displayed in the upper panel of
Fig.~\ref{fig2}. It shows the projected distribution of gas of an interacting galaxy system at $z=0.46$. The 
central galaxy is G23, with a mass of 1.7$\times$10$^{10}$ M$_\odot$, and has a anti-clockwise rotation, as
specified by the dashed circular arrow. The companion galaxy has a mass of
8.8$\times$10$^{7}$ M$_\odot$ located at a distance of $\sim$40 kpc and is moving on an indirect orbit, as suggested by its projected trajectory in the plane. Due to the relative short distance between the two objects and due to
the gravitational potential of the central and more massive galaxy, a bridge of 
gas is created: by tidal stripping,  G23 is accreting a fraction of gas of its companion. 

\noindent
ii) co-existence of two cold gas disks (or two gas components) that 
rotate in an opposite sense: in the inner part, the pre-existing gas which belongs to the central galaxy, and in the outer part, the accreted gas from the companion.
A clear and spectacular example of such a configuration is shown in the lower panel of Fig.~\ref{fig2}. In this galaxy, G11, the pre-existing disk is compressed by the surrounding gas and tends to have a higher density. It is displayed in
blue color while the accreted material, less dense, appears in orange color. 
A zoom-in of the central part is displayed is the lower right corner. 
The orientations of spiral arms clearly indicate that the two disks 
have opposite AM vectors. We will study in detail this galaxy in the next
section and show that the pre-existing disk is progressively  replaced
by the accreted gas leading to the existence of only one disk of gas that finally
counter-rotates with the main stellar component.

In the following, we study in details the formation process
of \CR, by considering four individual cases,  G11, G70, G31 and G2635,
which present some noticeable differences in their respective
formation paths.

\subsection{G11: two successive clear episodes of gas vs. gas counter-rotating disks}
\label{subsec:g11}

G11 is a very interesting galaxy, as it undergoes  two successive retrograde gas accretions during its lifetime.
This galaxy has been partially studied in \cite{peirani+24} and more specifically in
the evolution of its central black hole, BH1049 (see their Figs.~7 and 8). The following analysis
presents therefore some additional information about the formation path of this system.
First, in Fig.~\ref{fig3}, we show both the projected u-g-r band images of the stellar component, the projected gas density and the associated 2-d velocity fields,
at five different epochs, nicknamed 
\epoch{a}, \epoch{b}, \epoch{c}, \epoch{d} and \epoch{e}.
It is worth mentioning that the projected velocity fields presented throughout the paper 
are for qualitative purposes only. They are derived from the intrinsic mass distribution of stars (or gas), but more realistic 
modeling can be considered by calculating the light distribution produced by
the different stellar populations \citep[see, for instance, ][]{nanni+22,sarmiento+23}.
At epoch \epoch{a} ($z=1.19$),
the galaxy is still evolving through a complex formation phase, undergoing minor merger episodes and cosmic gas accretions. The galaxy has a clear stellar disk along with a rather small co-rotating disk of gas (due to a recent minor merger),  as suggested by the respective 2-d projected 
velocity fields. Then, at epoch \epoch{b} ($z=1.0$), 
a retrograde accretion of cosmic gas (most probably from gas stripping
from nearby galaxies) has taken place at larger radii and  
the co-existence of two gas components can be distinguished. The central one is the pre-existing gas disk which is still
co-rotating with the stellar component, contrary to the outer gas disk.
At epoch \epoch{c} ($z=0.64$), the former
has been totally replaced by the latter.
At this time, a clear \crdd can be clearly seen from the projected velocity fields.
However, a new episode of (indirect) gas accretion (mainly from fly-by galaxies) is already in place, and the scenario is repeating: coexistence of two noticeable gas component rotating
in opposite direction in \epoch{d}, ending with the replacement
by the new disk at epoch \epoch{e}.
Note that since this second gas accretion episode starts
from a \crdd configuration, 
the final product is a galaxy in which the main gas and stellar components are now co-rotating.

%%%%%%%%%%%%%%%%%%%%%%%%%%%%%%%%%%%%%%%%%%%%%%%%%%%%%%%%%%%%%
%     FIG 5
%%%%%%%%%%%%%%%%%%%%%%%%%%%%%%%%%%%%%%%%%%%%%%%%%%%%%%%%%%%%%%
\begin{figure*}
\begin{center}
\rotatebox{0}{\includegraphics[width=9cm]{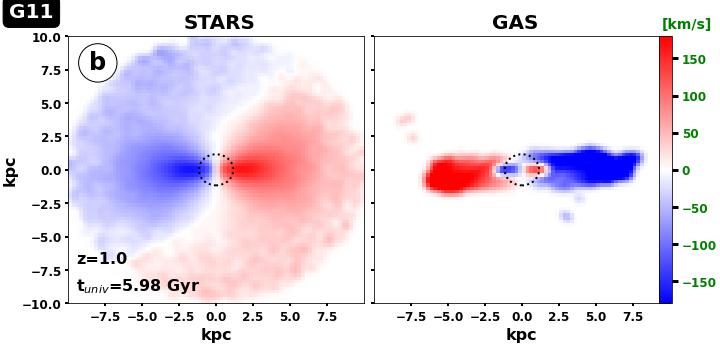}}
\rotatebox{0}{\includegraphics[width=9cm]{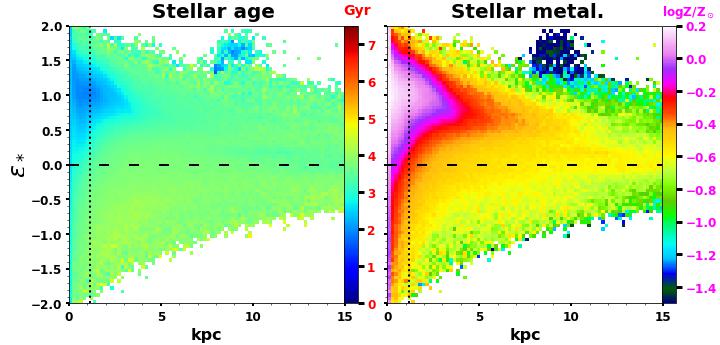}}
\rotatebox{0}{\includegraphics[width=9cm]{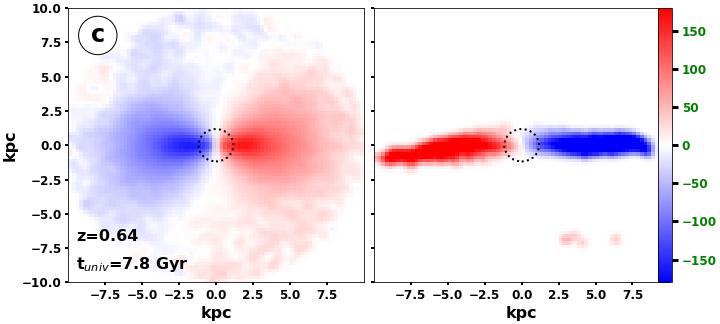}}
\rotatebox{0}{\includegraphics[width=9cm]{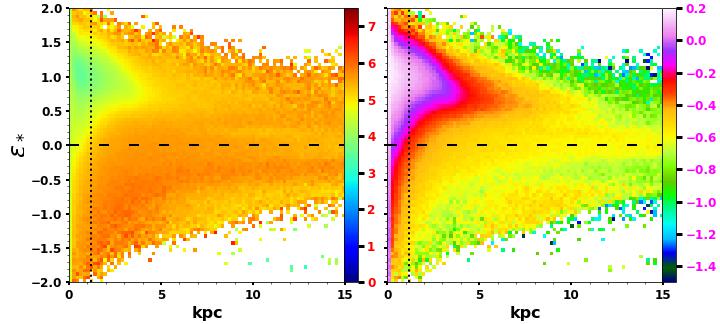}}
\rotatebox{0}{\includegraphics[width=9cm]{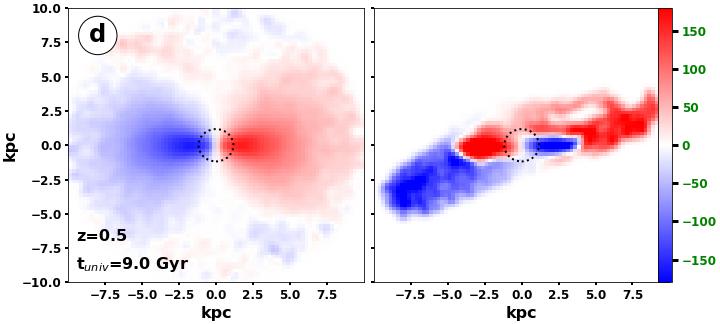}}
\rotatebox{0}{\includegraphics[width=9cm]{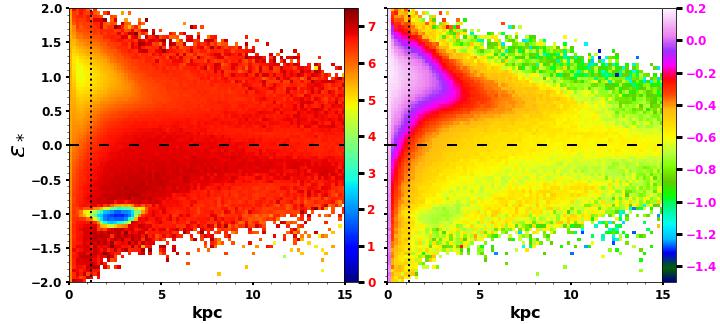}}
\rotatebox{0}{\includegraphics[width=18cm]{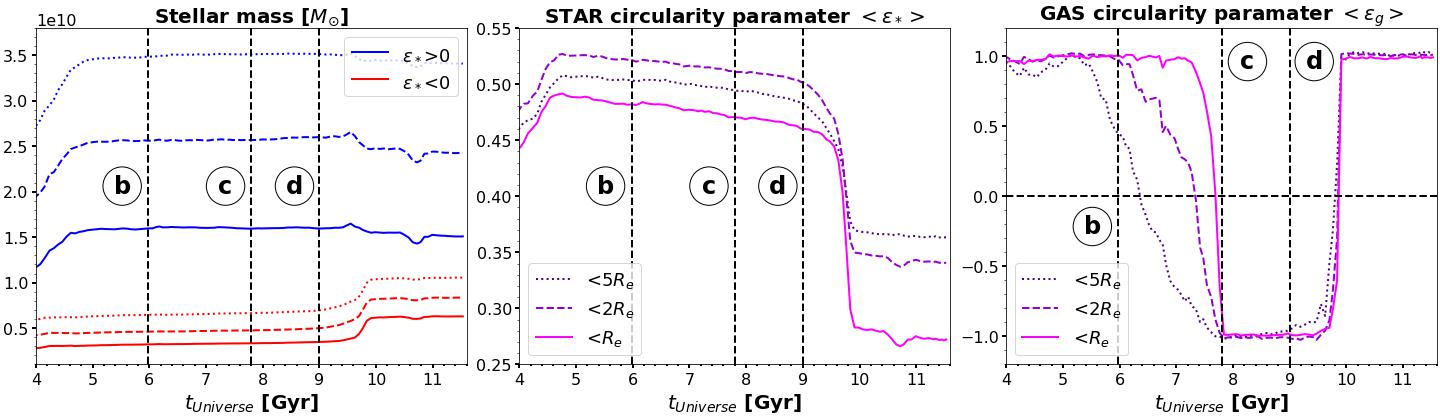}}
\caption{Projected edge-on mass-weighted velocity fields (first and second columns) and associated mass-weighted stellar circularity parameter ($\epsilon_*$)-radial distance-stellar age diagram (third column)
or $\epsilon_*$-radial distance-stellar metallicity diagram (fourth column), for galaxy G11
at epochs \epoch{b}, \epoch{c} and \epoch{d}. 
Here again, the values of rotation velocities are indicated by the color bar, the red parts are moving away from us while the blue ones 
are approaching us. The dotted circles and dotted lines indicate
values of the effective radius. The dashed lines is $\epsilon_*$=0. For each figure, we sample using
a regular grid of 80x80 pixels. Only pixels with at least 5 stellar
particles are taken into account.
This galaxy displays three different gas-stars misalignment configurations in its lifetime:
At epoch \epoch{b}, the outer parts of the gas disk are counter-rotating
with the stellar component, whereas at epoch \epoch{d}, it is 
the inner part of the gas disk which is in counter-rotation.
Meanwhile at epoch \epoch{c}, a clear \crdd is
obtained (within 10 kpc i.e. $\sim$8 R$_e$). 
Note also that in \epoch{d}, the two disks of stars are not entirely
co-planar but have misaligned angle of a $\Delta\mathrm{PA}\sim$198$^\circ$.
To facilitate 
the comparison with the trends presented in the previous figure, we plot 
in the last row  the time evolution
of the total stellar mass for stars satisfying either $\epsilon_*>0$ or $\epsilon_*<0$   (left panel) along with the evolution of the mean (mass-weighted) circularity parameters for stars (middle) and gas  (right), all quantities estimated within 
one (solid line), two (dashed line) or five (dotted line) effective radii.
}
\label{fig_g11_diag}
\end{center}
 \end{figure*}
%%%%%%%%%%%%%%%%%%%%%%%%%%%%%%%%%%%%%%%%%%%%%%%%%

To look deeper in the formation process of this galaxy, we show in Fig.~\ref{fig_g11_cosmic}
the cosmic evolution of some of its relevant properties. 
In the top panel, we compute $\Phi(r)$, the angle between the
angular momentum vector of the stellar component (computed by taking all stars within one effective radius),
and the angular momentum of all stars located within two concentric spheres of
radius $r-dr$ and $r+dr$. $\Phi(r)$ is estimated within 0<$r$<10 kpc using 50 bins (that is, $dr$=0.1 kpc).
In the second row, we also compute $\Phi(r)$ but considering this time
all "cold" gas cells located within two concentric spheres of
radius $r-dr$ and $r+dr$.
$\Phi(r)$ is simply defined by:
\begin{equation}
    \mathrm{cos}\, \Phi_i(r\pm dr) = \hat{J}_{\mathrm{stars}}  (\leq\mathrm{R_{eff}})\cdot \hat{J}_i(r\pm dr)
\end{equation}

\noindent where "$i$" refers either to the stars or the cold gas component.
These two plots give a precise idea about the 
relative orientation of the AM of stars (or gas) at a specific radial distance with respect
to the total AM of the galaxy.
It is worth mentioning that the z-axis of the cylindrical coordinate at any time corresponds to the total angular momentum vector direction
of the stellar component estimated within one effective radius.
%And according to the color coding used, blue colors indicate
%regions where $\Phi(r)$ has low values (i.e. close to 0$^\circ$) namely the angular momentum
%of stars (or gas) at distance $r$ tends to be aligned with the angular momentum of the main stellar component. On the contrary, red colors suggest
%regions where $\Phi(r)$ has high values (i.e. close to 180$^\circ$)
%i.e. anti-alignment.
In the present case, the time evolution of 
the distribution of $\Phi(r)$ for stars is almost always "blue". This simply means that 
all layers of the stellar component, up to 10 kpc, are 
generally co-rotating all together.
In the second row, the gas component displays a different behavior. 
Indeed, in the outer parts, the red color indicates here the existence of
accreted gas in counter-rotation with a dominant contribution to the AM
budget. 
The two co-existing disks of gas
can be identified until the epoch \epoch{c} where the "red" has
totally replaced the "blue" regions. 

The second episode of (indirect) gas accretion, as discussed above, incidentally starts roughly at the same period (if we limit the analysis to 10 kpc). The "blue" component (i.e., co-rotating stellar component) started to  dominate in the outer parts and progressively replaced the "red" one
(i.e., counter-rotating stellar component)
until \tuniv$\sim$10 Gyr. From this time, the first and second rows are characterized by $\Phi(r)$ with low values,
suggesting that the stellar and gas component are now globally
co-rotating at every layers up to a radial distance of 10 kpc.

%%%%%%%%%%%%%%%%%%%%%%%%%%%%%%%%%%%%%%%%%%%%%%%%%%%%%%%%%%%%%
%     FIG 6
%%%%%%%%%%%%%%%%%%%%%%%%%%%%%%%%%%%%%%%%%%%%%%%%%%%%%%%%%%%%%%
\begin{figure}
\begin{center}
\rotatebox{0}{\includegraphics[width=\columnwidth]{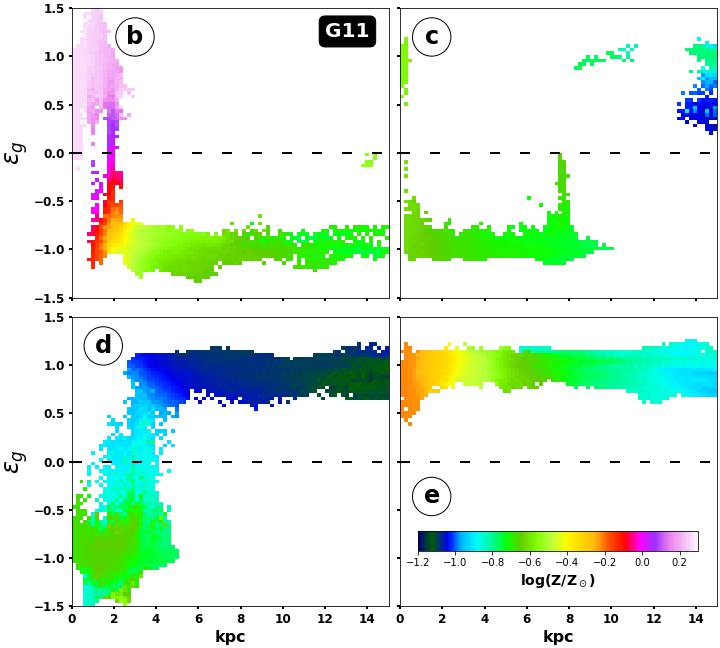}}
\caption{Gas circularity parameter-radial distance-stellar metallicity
diagrams at four different times. 
At epoch \epoch{b}, the pre-existing disk ($\epsilon_*>0$) has a high metallicity 
(log(Z/Z$_\odot$)>0)
compared to the accreted gas (log(Z/Z$_\odot$)$\sim$-0.7). Note that
at $r\sim$1 kpc, there is some gas mixing at the interface of the two disks.
At epoch \epoch{c}, the accreted gas has totally replaced the pre-existing one.
Due to the lack of star formation and therefore to a subsequent low supernovae activity,
the new disk tend to keep its original metallicity. 
In epoch \epoch{d}, the same phenomenon repeats: co-existence of two disks. The accreted gas is characterized here by a very low metalliciy 
(log(Z/Z$_\odot$)<-1). 
Then in epoch \epoch{e}, the new disk has totally replaced the older one, but due
to a star formation activity slightly more pronounced during the gas accretion, the new gas disk 
has a metallicity that gradually increases from smaller to larger radii.
}
\label{fig_g11_gas}
\end{center}
 \end{figure}
%%%%%%%%%%%%%%%%%%%%%%%%%%%%%%%%%%%%%%%%%%%%%%%%%

To date the precise time of the formation of the \crd, we plot in the third row, the  cosmic evolution of 3-d angle $\Psi$ subtended between the stellar and gas angular momentum vectors, introduced in section~\ref{subsec:identification}.  This angle is estimated by taking all stars and cold gas
cells within one, two and five effective radii. 
The transition between co-rotating ($\Psi$ close to 0$^\circ$)  and counter-rotating disks
($\Psi$ close to 180$^\circ$) generally depends on the 
definition of the radial distance within which the stellar and gas AM are computed.
In the present case, if the angular momenta are estimated within 5 effective
radii, the transition happens earlier since the retrograde accreted gas 
tends to have in first place a higher contribution to the AM budget at large radii.

To complete the analysis, in the next rows, we show 
the evolution of $V_*/\sigma$ which gives information on the morphology evolution of the
galaxy. Next, we present
the evolution of the mass of the cold gas 
estimated within one, two and five effective radii. We also separate the mass of
gas that co- or counter-rotates with the stellar component, by selecting all (cold) gas cells with positive (blue lines) or negative (red lines) circularity parameters. 
The evolution of the (cold) gas fraction, f$_{\mathrm{gas}}=M_{\mathrm{gas}}/(M_{\mathrm{gas}}+M_{\mathrm{stars}})$
is also shown.
The last row shows the cosmic star formation history 
by selecting
all stars born in situ within the last 100 Myr, within one, two or five
effective radii. We also plot in the same panel, the mass of new stars, $\Delta \mathrm{M}_*$
formed within successive intervals, and color coded according to 
 positive (blue lines) and negative (red lines) circularity parameters.

%%%%%%%%%%%%%%%%%%%%%%%%%%%%%%%%%%%%%%%%%%%%%%%%%%%%%%%%%%%%%
%     FIG 7   G11  MASS-AGE-Z
%%%%%%%%%%%%%%%%%%%%%%%%%%%%%%%%%%%%%%%%%%%%%%%%%%%%%%%%%%%%%%
\begin{figure}
\begin{center}
\rotatebox{0}{\includegraphics[width=\columnwidth]{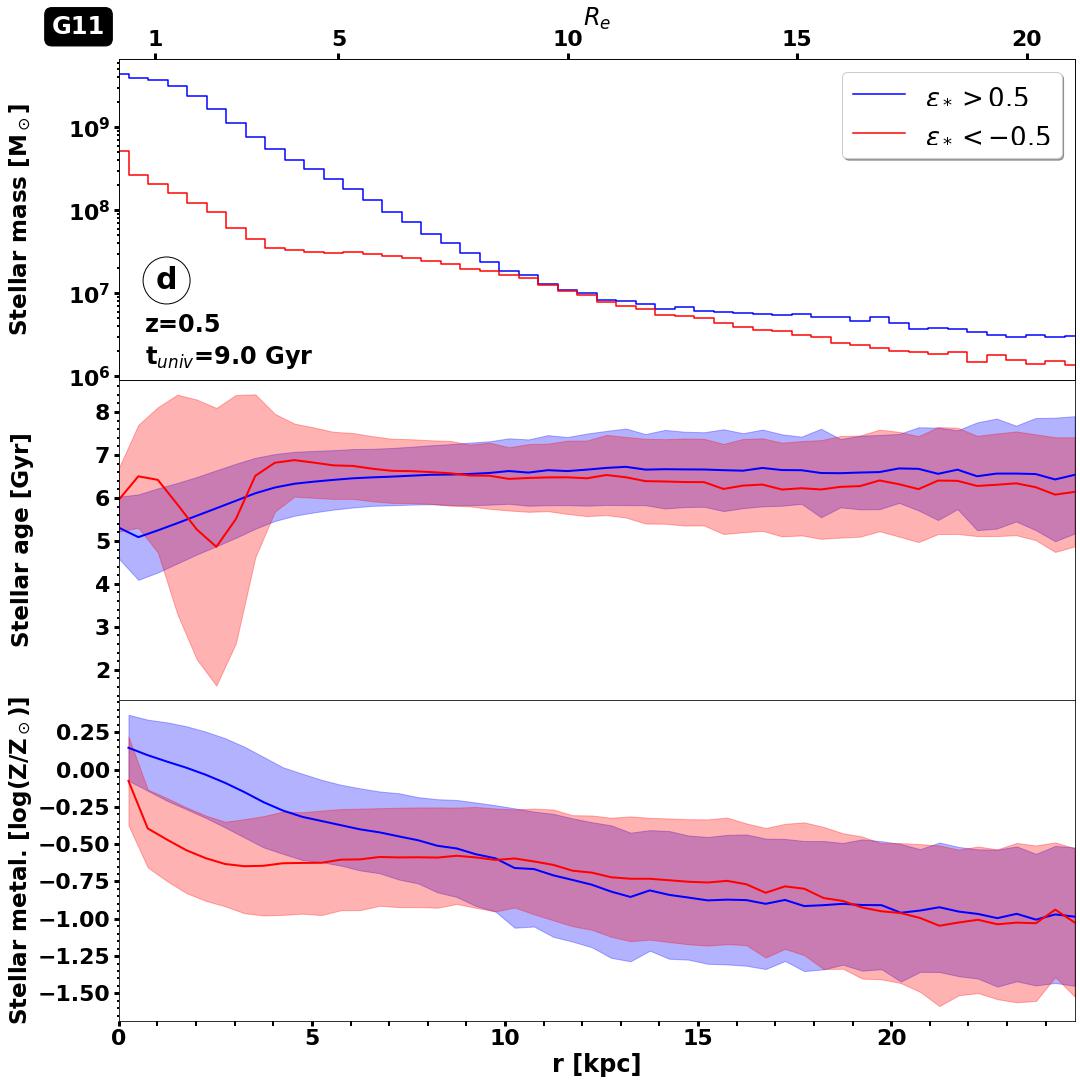}}
\caption{The 3-d radial distributions of the mass (upper panel),
age (middle panel) and metallicity (lower panel) of stars with
circularity parameters satisfying $\epsilon_*$>0.5 (blue lines) $\epsilon_*$<-0.5 (red lines), for G11 at z=0.5 (epoch \epoch{d}). 
We also plot on the top the corresponding scale with
respect to the effective radius. Shade areas indicate the dispersion.
Epoch \epoch{d} happens 1.2 Gyr after the total removal of the
pre-existing disk at \epoch{c}. During this time, the star formation is very low and
the new stars mainly formed at the interface of the inner disk and the new accreted one. This explains the "V-shape" seen in the distribution 
of the stellar age of counter-rotating stars within 5 kpc.
}
\label{fig_g11_maz}
\end{center}
 \end{figure}
%%%%%%%%%%%%%%%%%%%%%%%%%%%%%%%%%%%%%%%%%%%%%%%%%

From the last three rows, we note that during the two phases in which the new disk is progressively replacing the pre-exising
one, the star formation is quite high especially between \tuniv$\sim$4 and 5 Gyrs.
The mass of cold gas of the inner disk is also decreasing along with the gas fraction.
The same phenomenon is observed during the second episode of gas infall. 
A question that naturally arises concerns the fate of the inner disks of
gas. Are they mainly removed by the accreted gas component as suggested by
\cite{khoperskov+21}?  Or have they been mainly converted into stars, as advocated by \cite{cenci+24}?
Unfortunately, it is rather challenging to give a clear answer here since 
in the \ramses\, code, the motion properties of gas are treated using an Eulerian approach. Therefore, it is rather difficult to follow precisely each portion of gas in time\footnote{The possibility to use accurate tracer particles of baryon dynamics in \ramses\, simulation \citep{cadiou+19}
has not been considered in the completion of \nh.}. However, we can
try to provide some clues.
Indeed, from the evolution of the mass of cold gas (fifth row), we can 
estimate the mass of cold gas which has "disappeared" during the process, by estimating the
difference between the maximum mass value and the minimum one within five effective radii. The choice of five
effective radii is motivated to delimit reasonably the inner disk of gas.
Regarding the first accretion phase, in the time interval [4.3 - 7.8] Gyr, the
mass of cold gas decreases from $\sim$6.3$\times$10$^9$ M$_\odot$ to 0 M$_\odot$
while the mass of new stars that formed  with $\epsilon_*>0$ (within 5 effective radii) during
the same interval is $\sim$6.21$\times$10$^9$ M$_\odot$.
Regarding the second and late retrograde gas accretion phase, we estimate the same amounts
in the time interval \tuniv=[9 - 10] Gyr.  The
mass of cold gas decreases from $\sim$4.02$\times$10$^9$ to 0 M$_\odot$  
while the mass of new stars that formed  with $\epsilon_*>0$ during
the same interval is $\sim$3.03$\times$10$^9$ M$_\odot$.
These estimations suggest that the cold gas has been almost fully converted into
stars during the first accretion phase
while the percentage drops to 75.4\% during the second phase.

Also, to facilitate the comparison with observational analysis, we show in Fig.~\ref{fig_g11_diag}
the edge-on projected stellar and gas velocity fields within 10 kpc,
at epoch \epoch{b}, \epoch{c} and \epoch{d}.  
G11 is particularly interesting, as during its lifetime, it displays three different 
gas-star misalignment configurations.
Indeed, the stellar component is counter-rotating with the outer gas disk
in \epoch{b}, while it is counter-rotating with the inner gas disk in \epoch{d}. Between these two epochs, it reaches a \crdd configuration, if we limit our analysis to r$\leq$10 kpc.
Note that the use of term counter-rotating gas disks might not be 
totally suitable here. Indeed, 
the two disks are not necessarily rigorously coplanar. 
For instance, from Fig~\ref{fig_g11_diag} it can be seen that 
the  two gas components have a clear misalignment angle   
of $\Delta$PA$\sim$198$^\circ$ at epochs \epoch{d}.

On the contrary, the edge-on projected stellar velocity fields look very similar at every epoch, and are dominated by the co-rotating component. Indeed, as shown in the left panel of the last row of the 
figure, the total mass of stars owing positive 
circularity parameters  are 
significantly higher than the total mass of stars with negative $\epsilon_*$ values
(and irrespectively to the choice of the
radius within which the calculation is done).
Note that the two noticeable increases of the stellar
mass before \tuniv$\sim$4.7 Gyrs (blue lines) and before \tuniv$\sim$9.8 Gyrs (red lines) are the consequence of
the enhanced star formation during the compaction and removal of the inner gas component.
In the middle panel, the time evolutions of the mean stellar
circularity parameter, estimated within one, two and five $R_e$, are shown. They follow the same trends as the 
evolution of \Vsigg seen in Fig.~\ref{fig_g11_cosmic}. 
An interesting aspect is that the first phase of retrograde gas accretion lasts over several Gyrs. Consequently, the evolution
of $\langle\epsilon_*\rangle$ first increases and then slowly decreases. However, the second phase 
of retrograde accretion that leads to the removal of the inner
gas (here with $\epsilon_{\rm g}<0$) at \tuniv$\sim$9.8 Gyrs, is much more rapid leading to a sudden decrease of 
$\langle\epsilon_*\rangle$. Although this case is quite atypical, the results are 
in good agreement with the two phases of the kinematic misalignment
described in \cite{cenci+24}: 1) a first phase of gas compaction with an increase 
of both the star formation rate and $\langle\epsilon_*\rangle$; 2) a second phase in which both $\langle\epsilon_*\rangle$ and $\langle\epsilon_{\mathrm{g}}\rangle$ decrease (\tuniv$\sim$>5 Gyrs).
We will see that these trends are also found for the other galaxies e.g., G70 and G31.
Finally, note that the evolution of the mean circularity parameters for the gas
is strongly correlated with  the evolution of $\Psi$.

Moreover, for each epoch displayed in Fig.~\ref{fig_g11_diag}, we also show at the respective redshift, the 
distribution of the circularity parameter of stars ($\epsilon_*$) with respect to the radial distance, and color coded according to the stellar age (third column) or the global metallicity (fourth column).
In the following, we refer to these two plots as 
\diagagee and \diagmetall diagrams respectively.

%%%%%%%%%%%%%%%%%%%%%%%%%%%%%%%%%%%%%%%%%%%%%%%%%%%%%%%%%%%%%%
%          FIG 8      G-70  COSMIC
%%%%%%%%%%%%%%%%%%%%%%%%%%%%%%%%%%%%%%%%%%%%%%%%%%%%%%%%%%%%%%
\begin{figure*}
\begin{center}
\rotatebox{0}{\includegraphics[width=18cm]{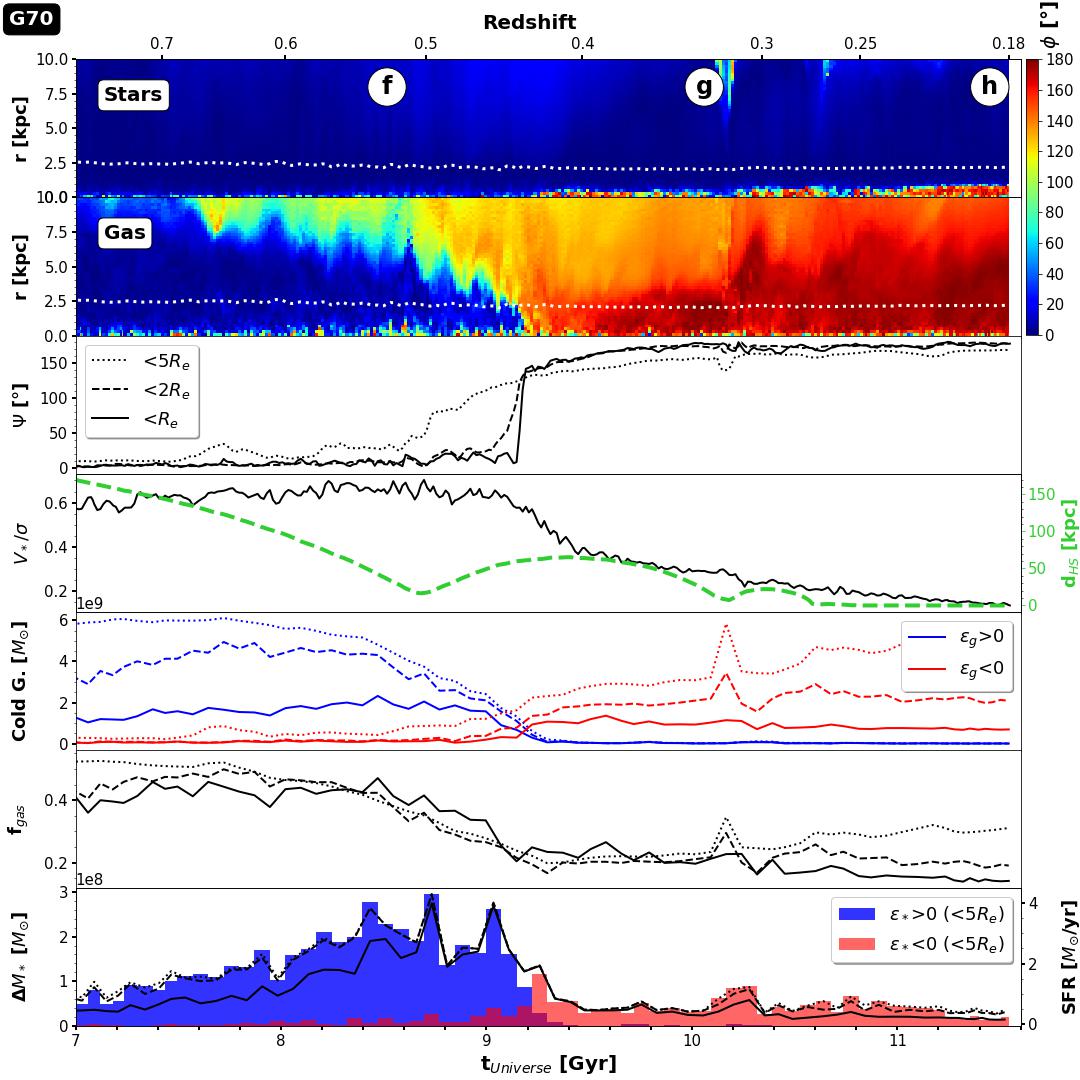}}
\caption{Same as Fig.~\ref{fig_g11_cosmic} but we have added here in the fourth row the time evolution of the distance
(d$_{\mathrm{HS}}$)between the host (G70)
and the satellite galaxy. This shows clearly the different passages at the pericentric
distance and the time of the final plunge (\tuniv$\sim$10.6 Gyr).
Here again, we choose the z-axis of the cylindrical coordinate as the total angular momentum vector direction of the stellar component estimated
within one effective radius.
}
\label{fig_g70_cosmic}
\end{center}
 \end{figure*}
%%%%%%%%%%%%%%%%%%%%%%%%%%%%%%%%%%%%%%%%%%%%%%%%%

At epoch \epoch{b},  the inner and pre-existing gas 
is progressively replaced by the accreted
one. The fact that the inner disk of gas is compacted during this process should normally induce some star formation activity  owing to positive circularity parameters. 
In the present case, some clear star formation does happen before \tuniv$\sim$5 Gyr and therefore
prior to epochs \epoch{a} and \epoch{b} while the star formation activity is quite low afterward probably due to AGN activity.
Some discussions about the effect of BH activity and AGN feedback in this system are presented in section~\ref{subsec:bhactivity}.
The low star formation at these epochs could also be attributed to the low content of the cold gas as suggested by rows 5 and 6 in Fig.~\ref{fig_g11_cosmic}.
The young stars are mainly
located at small radii with positive circularity parameters. On the contrary,
at epoch \epoch{d}, the central disk of gas is counter-rotating with respect to the 
stellar component. In this case, the new stars form with negative circular
parameters with a high contribution at radii 2 kpc<$r$<4 kpc i.e. close to the interface
of the two disks (see Fig.~\ref{fig_g11_diag}).

%%%%%%%%%%%%%%%%%%%%%%%%%%%%%%%%%%%%%%%%%%%%%%%%%%%%%%%%%%%%%
%     FIG 9    G70 diagrams
%%%%%%%%%%%%%%%%%%%%%%%%%%%%%%%%%%%%%%%%%%%%%%%%%%%%%%%%%%%%%%
\begin{figure*}
\begin{center}
\rotatebox{0}{\includegraphics[width=9cm]{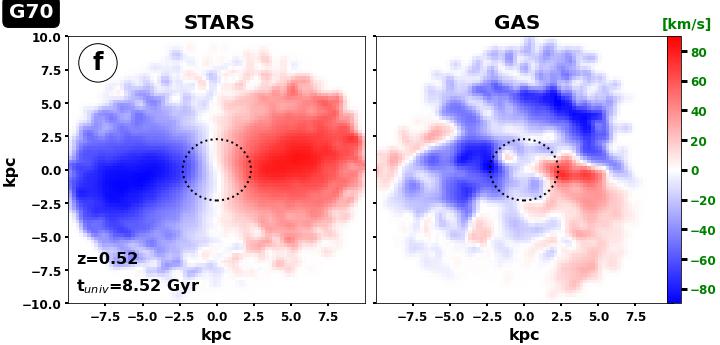}}
\rotatebox{0}{\includegraphics[width=9cm]{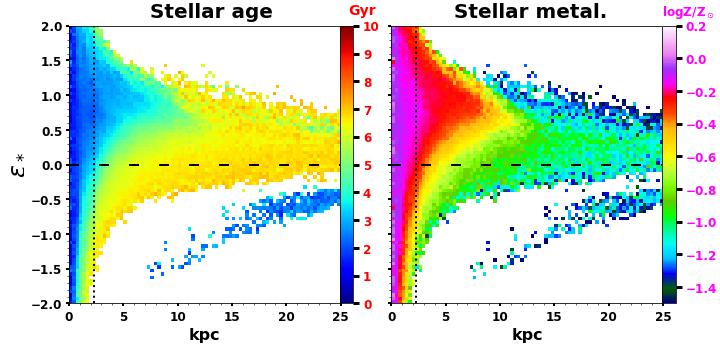}}
\rotatebox{0}{\includegraphics[width=9cm]{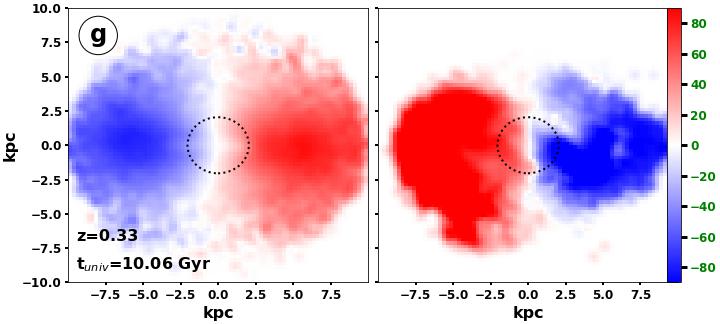}}
\rotatebox{0}{\includegraphics[width=9cm]{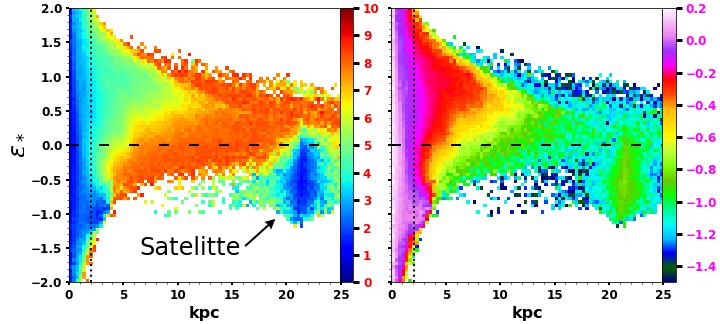}}
\rotatebox{0}{\includegraphics[width=9cm]{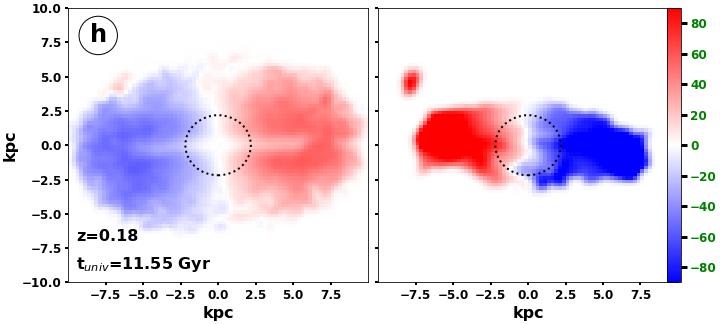}}
\rotatebox{0}{\includegraphics[width=9cm]{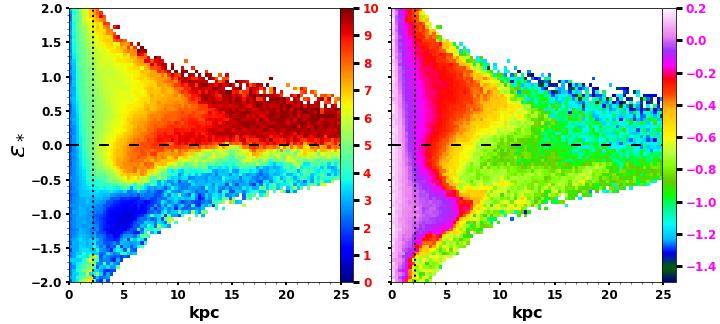}}
\rotatebox{0}{\includegraphics[width=18cm]{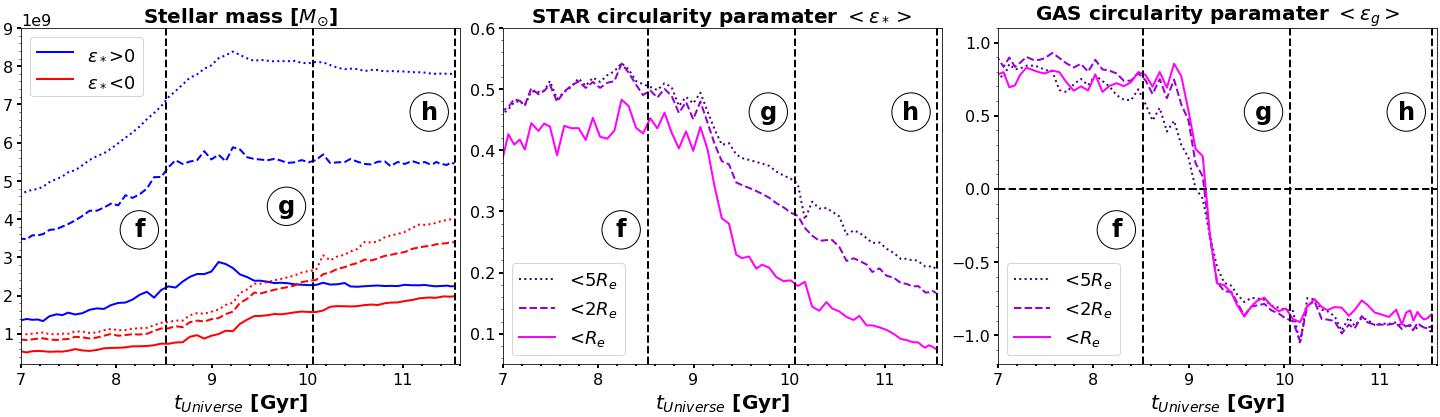}}
\caption{Same as Fig.~\ref{fig_g11_diag} but for the galaxy G70. This
latter undergoes a minor merger at $z\sim$0.27. We thus take an interest
in three different epochs: in the cosmic gas accretion phase prior the merger (\epoch{f}),
close to the final plunge (\epoch{g}) and at z=0.18 or 2.3 Gyr after the 
removal of the pre-existing disk (\epoch{g}). At epoch \epoch{g},
the satellite can be identified in both the \diagagee and \diagmetall diagrams. It is composed of a population of young stars with a global
metallicty of log(Z/Z$_\odot)\sim$-0.85 (green color). Note that some of the stars formed in the satellite are accreted by the main galaxy by tidal stripping at  \epoch{f}.
Then after the final plunge at epoch \epoch{h}, the counter-rotating stellar component ($\epsilon_*<0$) is composed of two main components: the in-situ formed stars at r<7.5 kpc, characterized by a high
metalicity, and the stars formed in the satellite displayed at larger radii owing a 
lower metalicty.
}
\label{fig_g70_diag}
\end{center}
 \end{figure*}
%%%%%%%%%%%%%%%%%%%%%%%%%%%%%%%%%%%%%%%%%%%%%%%%%

%%%%%%%%%%%%%%%%%%%%%%%%%%%%%%%%%%%%%%%%%%%%%%%%%%%%%%%%%%%%%
%     FIG 10  G70 GAS
%%%%%%%%%%%%%%%%%%%%%%%%%%%%%%%%%%%%%%%%%%%%%%%%%%%%%%%%%%%%%%
\begin{figure}
\begin{center}
%\rotatebox{0}{\includegraphics[width=\columnwidth]{FIG_GAS2.jpg}}
\rotatebox{0}{\includegraphics[width=\columnwidth]{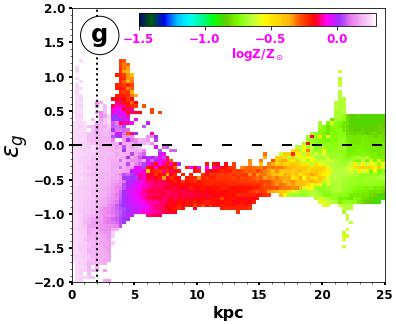}}

\caption{Variations of the gas circularity parameter $\epsilon_{\rm g}$ with respect to
the radial distance $r$ and color coded according to the global
gas metallicity at $z=0.33$ (\epoch{g}). The gas in the satellite
has a metallicy of log(Z/Z$_\odot)\sim$-0.85 (green) which is
consistent with the metalicity of young stars displayed in 
Fig.~\ref{fig_g70_diag} (second line - fourth column). Then some of the gas
of the satellite is accreted by G70 and is progressively enriched in
metal as the radial distance decreases due to supernovae activity dominated
by young stars in the central part. Consequently, the accreted gas reaches
the galactic plane with a much higher metallicy and then forms stars
that inherit the same high metallicity, consistent with
trends found in Fig.~\ref{fig_g70_diag} at the same epoch.
}
\label{fig9}
\end{center}
 \end{figure}
%%%%%%%%%%%%%%%%%%%%%%%%%%%%%%%%%%%%%%%%%%%%%%%%%

Note that the three respective \diagmetall diagrams look quite similar. At epoch \epoch{b}
the diagram is typical for a disk galaxy with a strong rotational
support.
Indeed, the younger stars form with
a high circularity parameter ($\epsilon_*$>0.5) and a rather high metallicity,
log(Z/Z$_\odot$)>0. Moreover, the stars that belong to the spheroidal
component of the galaxy, characterized by $\epsilon_*$ close to 0, tend 
to be older with a much lower metalicity.
At epoch \epoch{d}, the new stars form with a negative circularity parameter
($\epsilon_*$<0.5) and are characterized by a lower metallicity
log(Z/Z$_\odot$)$\sim$-0.7.

To understand the differences in the distribution of the stellar metallicity,
especially for the younger stars in epochs \epoch{b} and \epoch{d},
we plot  in Fig.~\ref{fig_g11_gas}
the \diagmetalgass diagrams relative to the (cold) gas component, at epochs
 \epoch{b}, \epoch{c}, \epoch{d}, \epoch{e}.
First, at epoch \epoch{b}, the pre-existing disk of gas is located at small distance $r<3$ kpc and
is characterized by a positive circularity parameter ($\epsilon_{\rm g}>0$) and a high metallicity.
On the contrary, the cosmic accreted gas is characterized by
a lower metallicity (i.e., log(Z/Z$_\odot$)$\sim$-0.7) and $\epsilon_{\rm g}<0$. We can also notice that
at $r\sim$1-2 kpc, there is some gas mixing at the interface of the two disks.
At epoch \epoch{c}, the accreted gas has almost completely replaced the pre-existing one.
Due to a low star formation activity and therefore to a subsequent low supernovae activity,
the new disk of gas tends to keeps its original metallicity. 
At epoch \epoch{d}, the same phenomenon is repeating. At this time, the
cosmic accreted gas is also characterized by a rather low metallicity (log(Z/Z$_\odot$)<-1). We also observe here some gas mixing between the two gas
components.
Then at \epoch{e}, the new gas disk has totally replaced the older one, but since
the star formation is slightly more pronounced in the
interval 9 Gyr<\tuniv<10 Gyr, 
the new gas disk has a metallicity which tends to increase from smaller 
to larger radii.

Additional information can also be extracted from Fig.~\ref{fig_g11_maz}
where we plot for $z=0.64$ (epoch \epoch{d}), the
3-d radial distributions of the mass, age and metallicity 
of stars with
circularity parameters satisfying either $\epsilon_*$>0.5 (blue lines) or $\epsilon_*$<-0.5 (red lines)
to identify co- and counter-rotating disks.
In this galaxy, co and counter-rotating stars tend to have similar ages with respect
to the radial distance. The "V-shape" seen in the distribution
of the stellar age of counter-rotating stars within 5 kpc is due to recent
star formation induced at the interface of the two disks. The 
counter-rotating stars are also less metal rich compared to co-rotating stars
within 10 kpc. This is most
probably a 
consequence of the low star formation activity prior epoch \epoch{c}.
Also, counter-rotating stars  tend to be slightly older but 
slightly more metal rich in the inner parts than in the outer parts.
Since the star formation rate is very low after the replacement
of the pre-existing gas, the mass of co-rotating stars ($\epsilon_*>0.5$) is 16 times larger than counter-rotating stars ($\epsilon_*<-0.5$)
within 5 R$_e$.

Finally, it is interesting to make the link with observations and analysis 
of the presence of two disks of gas in two galaxies presented
in \cite{cao+22} from the MaNGA survey. Such observations support the scenario
of multiple gas acquisition events which is also the case for G11. 
However, it is even more remarkable in their case
as each disk of gas presents a misalignment angle of $\sim$40$^\circ$ with respect to 
the stellar component compared to a single one in G11 (e.g. at epoch \epoch{d)}).

%%%%%%%%%%%%%%%%%%%%%%%%%%%%%%%%%%%%%%%%%%%%%%%%%%%%%%%%%%%%%
%     FIG 11  G70  MASS-AGE-Z
%%%%%%%%%%%%%%%%%%%%%%%%%%%%%%%%%%%%%%%%%%%%%%%%%%%%%%%%%%%%%%
\begin{figure}
\begin{center}
\rotatebox{0}{\includegraphics[width=\columnwidth]{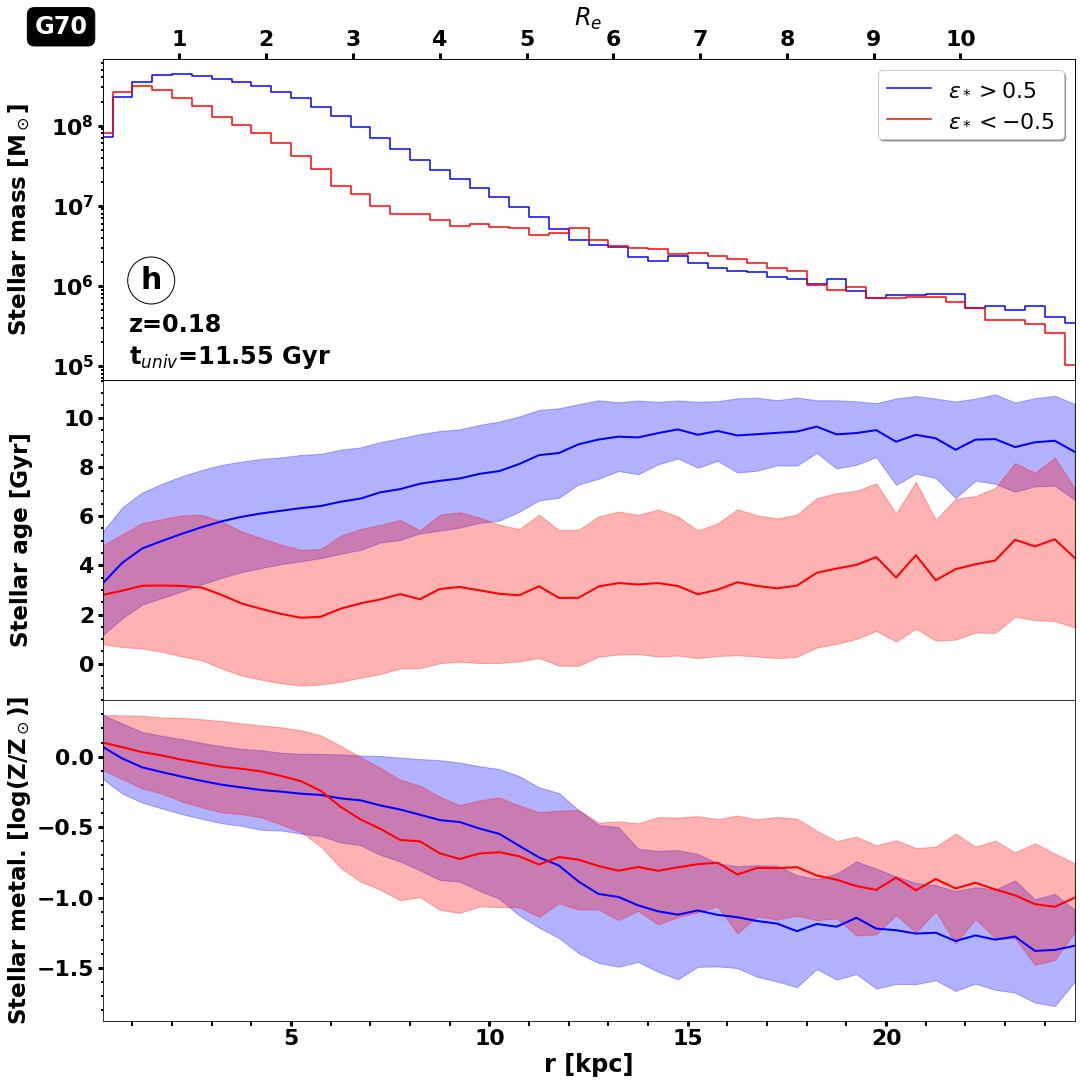}}
\caption{Same as Fig.~\ref{fig_g11_maz} but for G31. The trends are 
different than those from G11 since the star formation activity after
the total replacement of the pre-exising gas is here more pronounced.
And in this specific object, counter-rotating stars tend to be
younger and very slightly more metal rich than co-rotating stars
in the inner parts. Otherwise, counter-rotating stars tend to be 
younger and more metal rich in the inner parts than in the outer parts.}
\label{fig_g70_maz}
\end{center}
 \end{figure}
%%%%%%%%%%%%%%%%%%%%%%%%%%%%%%%%%%%%%%%%%%%%%%%%%

\subsection{G70: a typical gas vs. star counter-rotating galaxy formed from a minor merger event?}
\label{subsec:g70}

We present now the formation of G70 which has undergone a minor merger
with a mass ratio of 1:70 at late time.
Similarly to G11,
we show the cosmic evolution of some relevant properties in Fig.~\ref{fig_g70_cosmic}. 
However, in the fourth row we have added 
the evolution of the separation distance between the main galaxy, G70, 
and its satellite. This plot is quite typical in theoretical interacting galaxies studies
as it shows clearly the different passages at pericentric distances
as well as the accurate time of the final plunge, 
estimated in the present case at \tuniv$\sim$10.6 Gyr ($z\sim$0.27).

The evolutions seen in Fig.~\ref{fig_g70_cosmic} are quite similar to those
obtained from G11. In particular, in the second row, the accreted gas 
from tidal stripping is clearly seen, starting from the outer parts (\tuniv$\sim$7.5 Gyr) and progressively reaching the inner parts. 
Thus the presence of two misaligned gas components can be identified  in
the time interval 7.5<\tuniv<9.2 Gyr. Note that, in the present case,
the accretion of gas is done preferentially through an angle of $\sim$110$^\circ$.

The \crdd configuration occurs at \tuniv$\sim$ 9.1 Gyr.
One interesting point is that the replacement of the pre-existing disk by the new one is complete
while the satellite is still orbiting around the main galaxy.

Also, as seen for G11, the cold gas mass with a positive circularity parameter is 
progressively decreasing prior to the final gas-star counter-rotating disks configuration. The gas fraction follows the same trend while the replacement of the pre-existing gas is accompanied with an induced star formation (blue histograms). Here also, we can estimate the ratio between the mass of cold
gas (with positive $\epsilon_{\rm g}$) that disappeared between 7.7 and 9.4 Gyr and the mass of 
new stars (with positive $\epsilon_*$) formed in the same interval, both quantities
estimated within 5 effective radii. We find that 61.4\% of the cold gas has
been converted into stars.
Moreover, the evolution of the \Vsigg ratio starts to decrease when the accreted gas reaches the effective radius (at \tuniv$\sim$9 Gyr).
This is quite expected since the quantity \Vsigg is estimated within one effective radius, and the notable formation of counter-rotating stars starting at this time can compensate for the angular momentum of co-rotating stars, also reducing \Vsig.
At $z=0.18$, \Vsig$\sim$0.2, which means  that the galaxy has lost some angular momentum
during the process and is more likely to turn to an S0 galaxy. 
A visual inspection of the u-g-r bands images (not shown in this paper)
suggests indeed a spheroidal galaxy without any clear spiral arms.

In Fig.~\ref{fig_g70_diag} we show the 2-d projected stellar and gas velocity
fields at three specific times:
during the gas stripping phase (z=0.52, \epoch{f}),
close to the final plunge (z=0.33, \epoch{g})
and at z=0.18 or 2.3 Gyr after the \crdd formation (\epoch{h}).
At epoch \epoch{f}, the inner part of the gas disk is perturbed by 
the cosmic accretion of gas whereas the stellar velocity still displays a typical  distribution. 
Unlike G11, the coexistence of two clear gas disks is less obvious.
Also, as expected, the total mass of stars with $\epsilon_*>0$ is increasing before \tuniv$\sim$9.2 Gyrs
due to compression of the pre-existing gas with subsequent star formation enhancement.
The trends again agree well with two phases of kinematic misalignment proposed by
\cite{cenci+24}. We indeed first observe an increase of the star formation rate 
accompanied by an increase of $\langle\epsilon_*\rangle$ (while $\langle\epsilon_{\rm g}\rangle$ decreases).  The second phase is then characterized by a decrease of both $\langle\epsilon_*\rangle$ and $\langle\epsilon_{\rm g}\rangle$.

Here also, the \diagagee diagram is typical
of a galaxy with a strong rotational support ($V_*/\sigma$=0.65): the young stars have been produced at small radial 
distances with a high metallicity while stars that belong to the spheroidal
component (e.g. $\epsilon_*$ close to 0 with high radial distances) tend to be
older with a lower metallicity.

At epoch \epoch{g}, the satellite galaxy can be identified at a radial distance around
22 kpc from the center of G70 in the \diagagee diagram. Since the orbit of the satellite is indirect in the reference frame of G70, its stars
have negative circularity parameters.
The satellite is also characterized by a young population of stars with globally a lower metallicy, log(Z/Z$_\odot$)$\sim$-0.8, than its host.
This is consistent with \cite{nhz} who found that 
massive galaxies tends to be more metal 
rich than lower mass galaxies in \nhs (see their Figs. 13 and 14).
This is also in agreement  with observational trends \citep{gallazzi+14}.
Since the kinetic misaligned star-gas configuration is already in place at this time, the gas is by definition counter-rotating with respect to the stellar component. Therefore, 
the new stars inherit the dynamical properties of their parent gas,
and, in particular, they form with negative circularity parameters $\epsilon_*$. This is what we 
observe precisely in the \diagagee diagram at small radii (i.e. r<4 kpc).
Now if we look at the \diagmetall diagram, these young stars formed
with a high metalicity (purple color or log(Z/Z$_\odot$)>-0.1).
This result might be a bit non-intuitive since the gas in the satellite
has a lower metallicy (green color or  log(Z/Z$_\odot)\sim$-0.85) and this
gas is accreted by the main galaxy to form these new stars. 
One would naively expect that the new stars forming at  r<4 kpc 
should have a metallicty close to log(Z/Z$_\odot)\sim$-0.85.
To understand this, we have plotted  in the next figure, Fig.~\ref{fig9},
the same diagram but for the gas component i.e.
the distribution of the gas circularity parameter $\epsilon_{\rm g}$ with respect to
the radial distance $r$ and using a color coding according to the global
gas metallicity, at $z=0.33$ (\epoch{g}). 
First, we indeed find that the gas in the satellite
has a metallicy of log(Z/Z$_\odot$$\sim$-0.85) (green) which is
consistent with the metalicity of young stars displayed in 
Fig.~\ref{fig_g70_diag} at the same location in the diagram (second line, fourth column). Then, 
some of the gas of the satellite is accreted by G70 (tidal stripping) 
and is progressively enriched in
metal as the radial distance decreases. This can be explained by the supernovae activity which tends to be more pronounced at small 
radii (where most of the younger stars form). 
Consequently, the accreted gas reaches
the galactic plane with a much higher metallicy and then form stars
that inherit the same high metallicity, consistent with the
trend found in Fig.~\ref{fig_g70_diag} at the same epoch.

Then back to Fig.~\ref{fig_g70_diag}, the epoch \epoch{h} displays 
different interesting trends well after the formation of the gas-versus-stellar counter-rotating galaxy.
First, such a configuration can remain stable for several Gyrs. In this case, the in situ star formation induced by the counter-rotating gas
disk is still clearly visible and extends almost up to $r\sim$7.5 kpc. The metallicity of these young stars is again generally quite high (purple color or log(Z/Z$_\odot$)>-0.1).  Stars located at larger radii ($r$>7.5 kpc)  with
$\epsilon_*<0$, are also young. But they did not form in situ: they have been formed
in the satellite and are displayed in the outer parts due to the conservation of the orbital angular momentum. One efficient way to disentangle between these two populations of stars is
to look at their metallicity. Indeed, stars with $\epsilon_*<0$ in the outer parts are characterized by a lower metalicity (green color) that clearly reveals their origin.

%%%%%%%%%%%%%%%%%%%%%%%%%%%%%%%%%%%%%%%%%%%%%%%%%%%%%%%%%%%%%%
%          FIG 12  G-31  COSMIC
%%%%%%%%%%%%%%%%%%%%%%%%%%%%%%%%%%%%%%%%%%%%%%%%%%%%%%%%%%%%%%
\begin{figure*}
\begin{center}
\rotatebox{0}{\includegraphics[width=18cm]{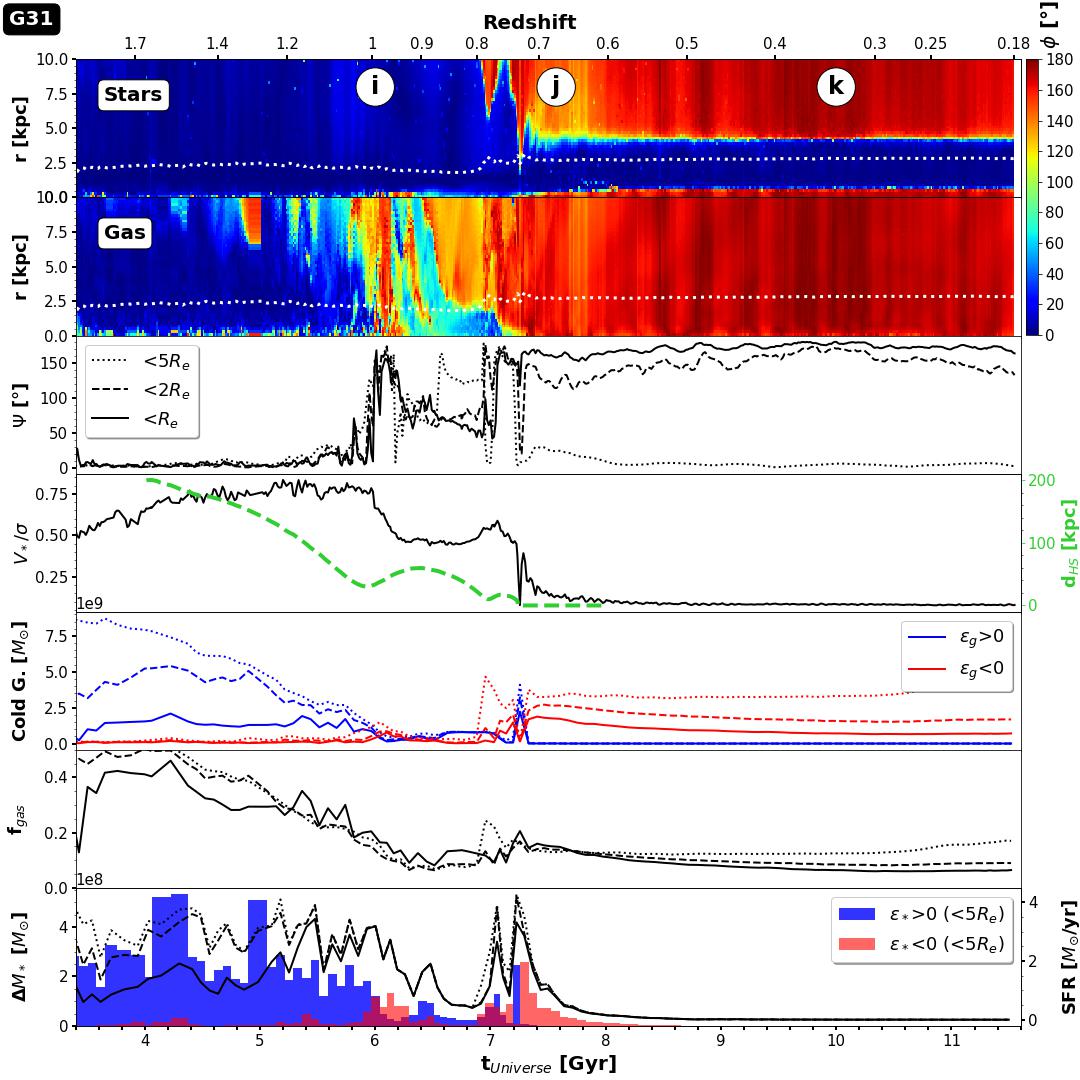}}
\caption{Same as Fig.~\ref{fig_g11_cosmic} but we have interposed here at the fourth row the time evolution of the distance separation between the galaxy satellite from the main galaxy, G31.
Note that the evolution of $\Psi$ estimated within five effective radii 
shows a completely opposite trend than the two other ones (i.e. <R$_e$ and
<2R$_e$) after the merger as encompassing distant counter-rotating stars flip the average stellar angular momentum.
}
\label{fig_g31_cosmic}
\end{center}
 \end{figure*}
%%%%%%%%%%%%%%%%%%%%%%%%%%%%%%%%%%%%%%%%%%%%%%%%%

%%%%%%%%%%%%%%%%%%%%%%%%%%%%%%%%%%%%%%%%%%%%%%%%%%%%%%%%%%%%%
%     FIG 13  G31  diagrams
%%%%%%%%%%%%%%%%%%%%%%%%%%%%%%%%%%%%%%%%%%%%%%%%%%%%%%%%%%%%%%
\begin{figure*}
\begin{center}
\rotatebox{0}{\includegraphics[width=9cm]{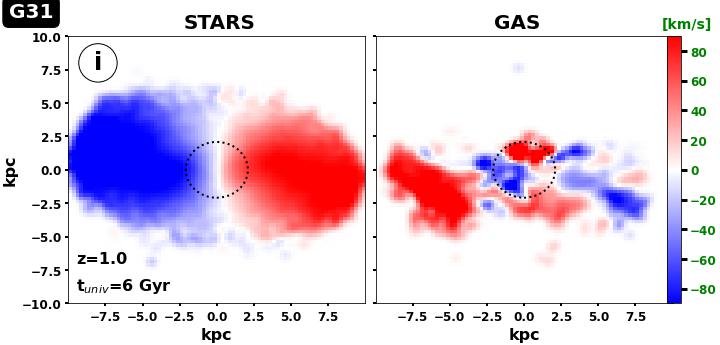}}
\rotatebox{0}{\includegraphics[width=9cm]{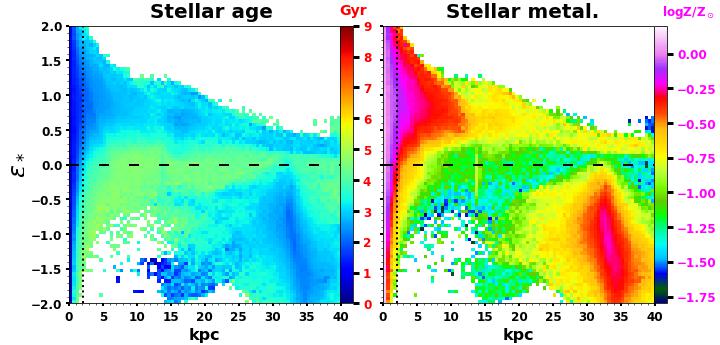}}
\rotatebox{0}{\includegraphics[width=9cm]{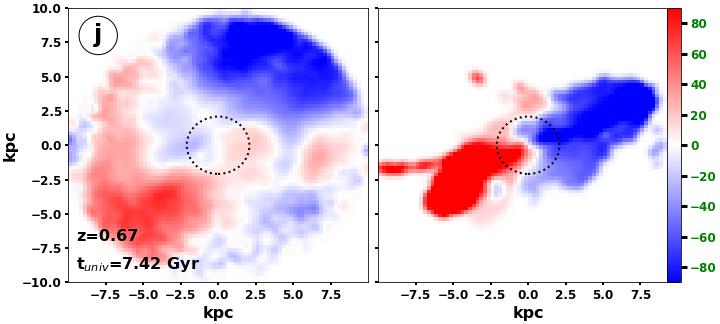}}
\rotatebox{0}{\includegraphics[width=9cm]{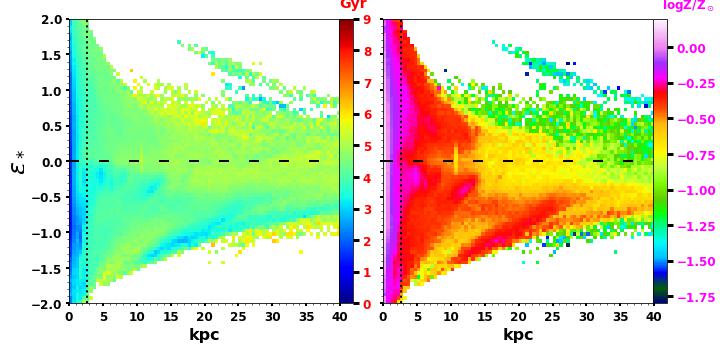}}
\rotatebox{0}{\includegraphics[width=9cm]{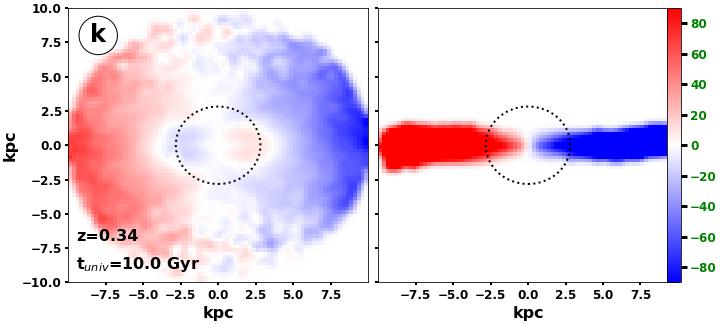}}
\rotatebox{0}{\includegraphics[width=9cm]{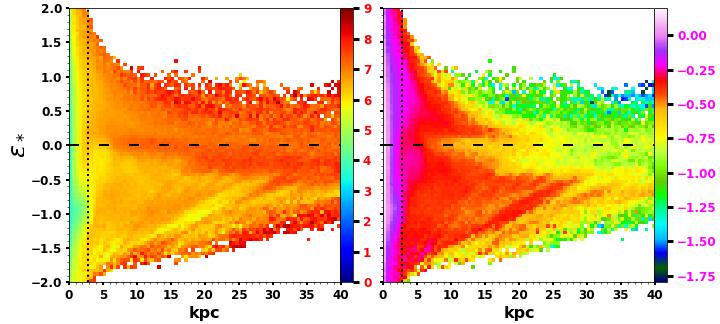}}
\rotatebox{0}{\includegraphics[width=18cm]{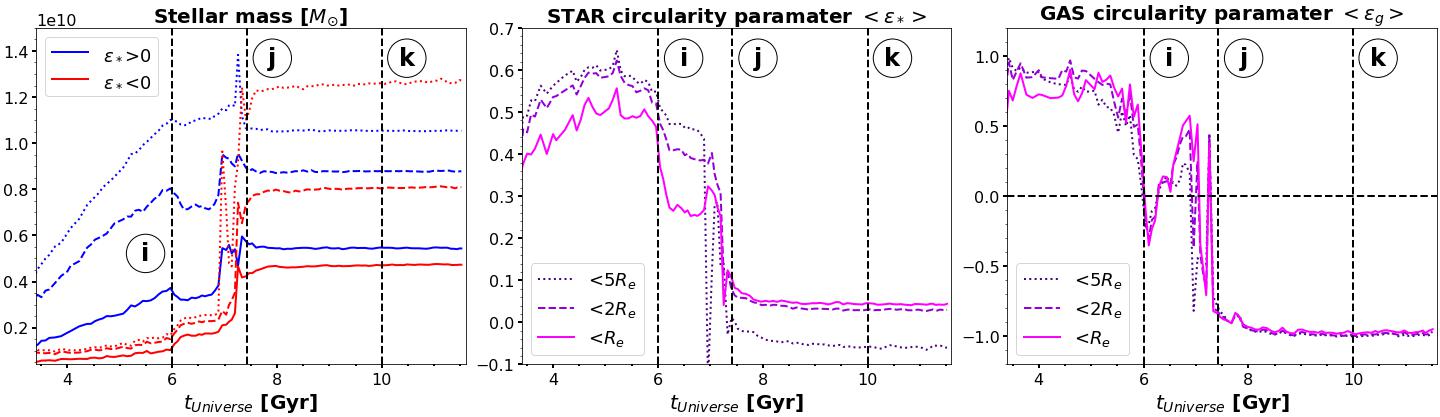}}
\caption{Same as Fig.~\ref{fig_g11_diag} but for G31. Before the final plunge
(\tuniv$\sim$7.1 Gyr), the gas velocity field is perturbed by the accretion of gas from tidal stripping. Right after the final plunge, epoch \epoch{j}, two stellar components
can be identified: in the inner part, the original stars from the host galaxy G31 and in the outer part, the star originated from the satellite galaxy, which
are displayed due to the orbital angular momentum. When the satellite is
passing by the pericentric distance, the tidal disruption of the process lead
to the formation of stellar streams clearly visible in 
the \diagagee and \diagmetall diagrams. Note that the misalignment between the two stellar components at epoch \epoch{j} is due
to the orbital configuration of the major merger. Interestingly, this misalignment
seems to weaken in the time, as seen in epoch \epoch{k}.
It is also remarkable that at this epoch, three distinct populations of stars can
be identified: the main stellar component in the middle and two counter-rotating stellar components,
in the very inner parts (stars formed in-situ) and in the outer parts (accreted stars).
}
\label{fig_g31_diag}
\end{center}
 \end{figure*}
%%%%%%%%%%%%%%%%%%%%%%%%%%%%%%%%%%%%%%%%%%%%%%%%%

%%%%%%%%%%%%%%%%%%%%%%%%%%%%%%%%%%%%%%%%%%%%%%%%%%%%%%%%%%%%%
%     FIG 14     G31     MASS - AGE - Z
%%%%%%%%%%%%%%%%%%%%%%%%%%%%%%%%%%%%%%%%%%%%%%%%%%%%%%%%%%%%%%
\begin{figure}
\begin{center}
\rotatebox{0}{\includegraphics[width=\columnwidth]{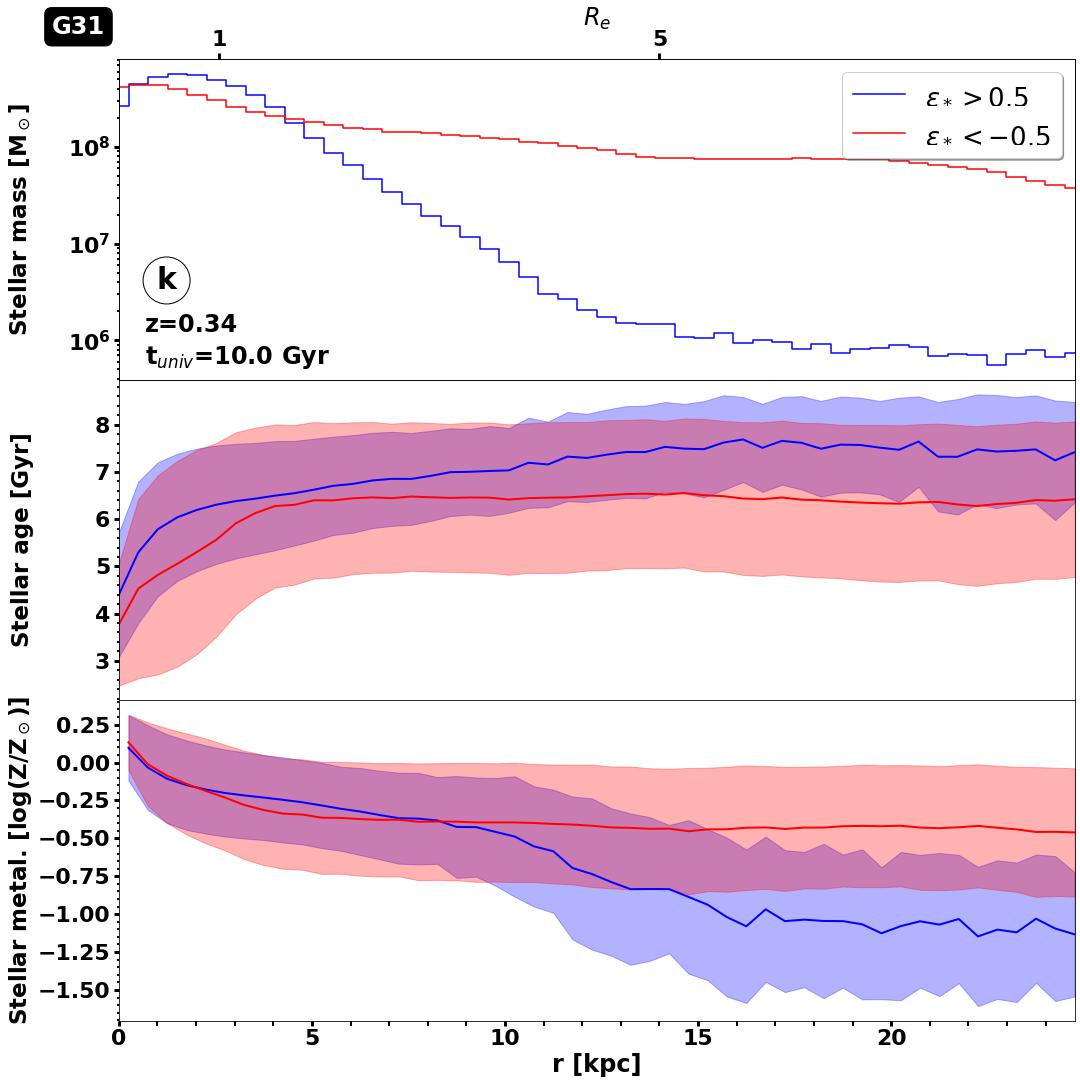}}
\caption{Same as Fig.~\ref{fig_g11_maz} but for G31.
Since G31 experienced a major merger, a large fraction of CR
stellar disk (i.e. $\epsilon_*$<-0.5) is displayed in the outer parts. A
clear signature of this
is the higher fraction in mass of counter-rotating stars in the outer parts with
respect to co-rotating stars (upper panel, red line). Other
potential signatures lie in the 
mean age and mean mellaticity of counter-rotating stars with constant values at larger radius (here r>4 kpc). We note that the mean metallicity
of log(Z/Z$_\odot)\sim$-0.25 is proved to match with
the mean metallicity of the satellite prior the merger event.}
\label{fig_g31_maz}
\end{center}
 \end{figure}
%%%%%%%%%%%%%%%%%%%%%%%%%%%%%%%%%%%%%%%%%%%%%%%%%

Finally, in Fig.~\ref{fig_g70_maz}, we plot for $z=0.18$ (or epoch \epoch{h}), the
3-d radial distributions of the mass, age and metallicity 
of stars with
circularity parameters satisfying either $\epsilon_*$>0.5 or $\epsilon_*$<-0.5.
The trends are different
from G11. Since there is  a pronounced star formation activity which takes place after the replacement of the pre-existing gas (\tuniv>9.2 Gyr),
the mass of co-rotating stars ($\epsilon_*>0.5$) is here "only" 2 times larger than counter-rotating stars ($\epsilon_*<-0.5$) within 5 R$_e$.
In this case, counter-rotating stars tend to be
younger and slightly more metal rich than co-rotating stars
in the inner parts. They also tend to be 
younger and more metal rich in the inner parts than in the outer parts.

Very similar results and trends are obtained for galaxies G23 and G45 (see Appendix~\ref{appendix1}). The only difference is that, in the case of 
G23 and G45, the final plunge of their respective satellite does not happen before the end of the simulation.

\subsection{G31: a typical gas vs. stars counter-rotating galaxy formed from a major merger event?}
\label{subsec:g31}

In our sample, at least two galaxies, G31 and G83, have experienced a major merger in their formation history (we define the limit
of the mass ratio as 1:4) leading to the formation of
a gas-star decoupled component.
In this section, we only focus on G31 since the two galaxies present clear similarities in their formation. But all the results and trends obtained
for G83 can be found in the Appendix. 
It is worth mentioning that G31 has also been partially studied in \cite{peirani+24} in
which the formation and evolution of its central black hole, BH549, has been studied in detail (see their Figs.~9 and A.1).

G31 is merging with another galaxy of similar mass. Before the final
plunge, estimated at \tuniv$\sim$7.2 Gyr, the mass ratio of the system  is 1:1.2 at \tuniv=4.5 Gyr.
In  Fig.~\ref{fig_g31_cosmic}, we plot the evolution of relevant physical 
properties relative to G31, similarly to Fig.~\ref{fig_g70_cosmic}.
Again, some similarities with the previously studied galaxies can
be seen. In particular, the accretion of gas by tidal stripping is here again
accompanied with  a decrease of both the mass of the cold gas (with positive
circularity parameters) and the fraction of the gas. In parallel, the star formation
activity is enhanced during this phase which ends at the final plunge.
However, some differences with trends obtained from galaxies that have undergone minor mergers can be also noticed.
First, the tidal stripping of the gas starts to be efficient only when the satellite galaxy is passing close to the pericentric distance ($\sim$30 kpc) at \tuniv$\sim$5.9 Gyr. Consequently, the formation of a counter-rotating gas disk seems to happen more closely to the final plunge (this is more visible in the case of G83).
Second, after the final plunge, from the first row of Fig.~\ref{fig_g31_cosmic}, the outer parts of the galaxy (r$\succsim$5 kpc) are dominated by stars that are  
counter-rotating. This is due to the conservation of orbital AM during the merger: a large amount of stars from the 
satellite is expelled and diffused in the outer parts after the merger process.
From this time, the top panel of Fig.~\ref{fig_g31_cosmic} indicates the existence
of three distinct stellar disks/components which dominate the AM budget
at three different regions:

\noindent
i) at very small radii (r<0.5 kpc): a population of young stars that have
formed in situ from the accreted gas and counter-rotate with respect to the main stellar component, 

\noindent
ii) 0.5<r<5 kpc: a population of stars originated from the host galaxy, G31, that co-rotate with respect to the main stellar component.

\noindent
iii) r>5 kpc: a population of stars that previously belonged to the satellite galaxy and are now located at larger radii after the merger episode, and counter-rotate with respect to the main stellar component.

The existence of these three populations of stars can be potentially distinguished from
the projected edge-on velocity fields presented in Fig.~\ref{fig_g31_diag}.
Another consequence of these lies in the \diagmetall diagram. Indeed, at epoch \epoch{i} 
before the final plunge, the satellite consists of young stars with a metallicity of log(Z/Z$_\odot)\sim$-0.25 (red/purple). Then, after each passage at the pericentric distance, the tidal disruption process produces
the formation of stellar streams that are clearly visible in the diagrams
at epochs \epoch{j} and  \epoch{k} (see also the trend for G83 in the Appendix). 
Since a larger amount of stars is more likely to be expelled in major merger
scenario, these stellar streams with specific age and metallicity might be more easily seen and detected in observational studies.

In fact, at all radii, two populations of stars most of the time coexist  with opposite rotation. This is 
illustrated in the upper panel of Fig.~\ref{fig_g31_maz}, 
where the 3-d radial distributions of the mass of stars 
with circularity parameters satisfying either $\epsilon_*$>0.5 or $\epsilon_*$<-0.5
are shown at z=0.34 (or epoch \epoch{k}).
For instance, counter-rotating stars have a higher mass fraction in the very
inner parts (r<1 kpc) and for r>4.8 kpc with respect to co-rotating stars.
Also, interestingly, as the merger process displays a large fraction of stars of the satellite in the
outer parts of the remain object, we find that the stellar age and metallicity tend to 
have constant mean values (for r$\succsim$5 kpc), 
6 Gyr and log(Z/Z$_\odot)\sim$-0.25 respectively for G11.
Similar trends are obtained for G83.

To finish, we can also take a look at the possible fate of the cold gas
that was present in G31 before the merger. If we consider the 
the interval [3.6 - 6.2] Gyr,
we estimate that  86.6\% of the mass of cold gas with positive $\epsilon_{\rm g}$ within 5 R$_e$ 
has been converted into news stars. A similar investigation 
for G83 leads to a fraction of 45.9\%.
Also, the time evolution of \Vsigg indicate that after the final plunge,
the system evolves toward a spheroidal dominate galaxy, consistent in
major merger scenario.

\subsection{G2835. An example of angular momentum orientation reversal}
\label{subsec:g2825}

The last galaxy studied in detail is G2835. 
This galaxy has already been introduced in \cite{park+19} (see, for instance, their
Figs~10 and 11). Before the formation of the \crg, G2835 has experienced
several minor mergers and one major merger (1:4 at z=1.56).

\begin{figure*}
\begin{center}
\rotatebox{0}{\includegraphics[width=18cm]{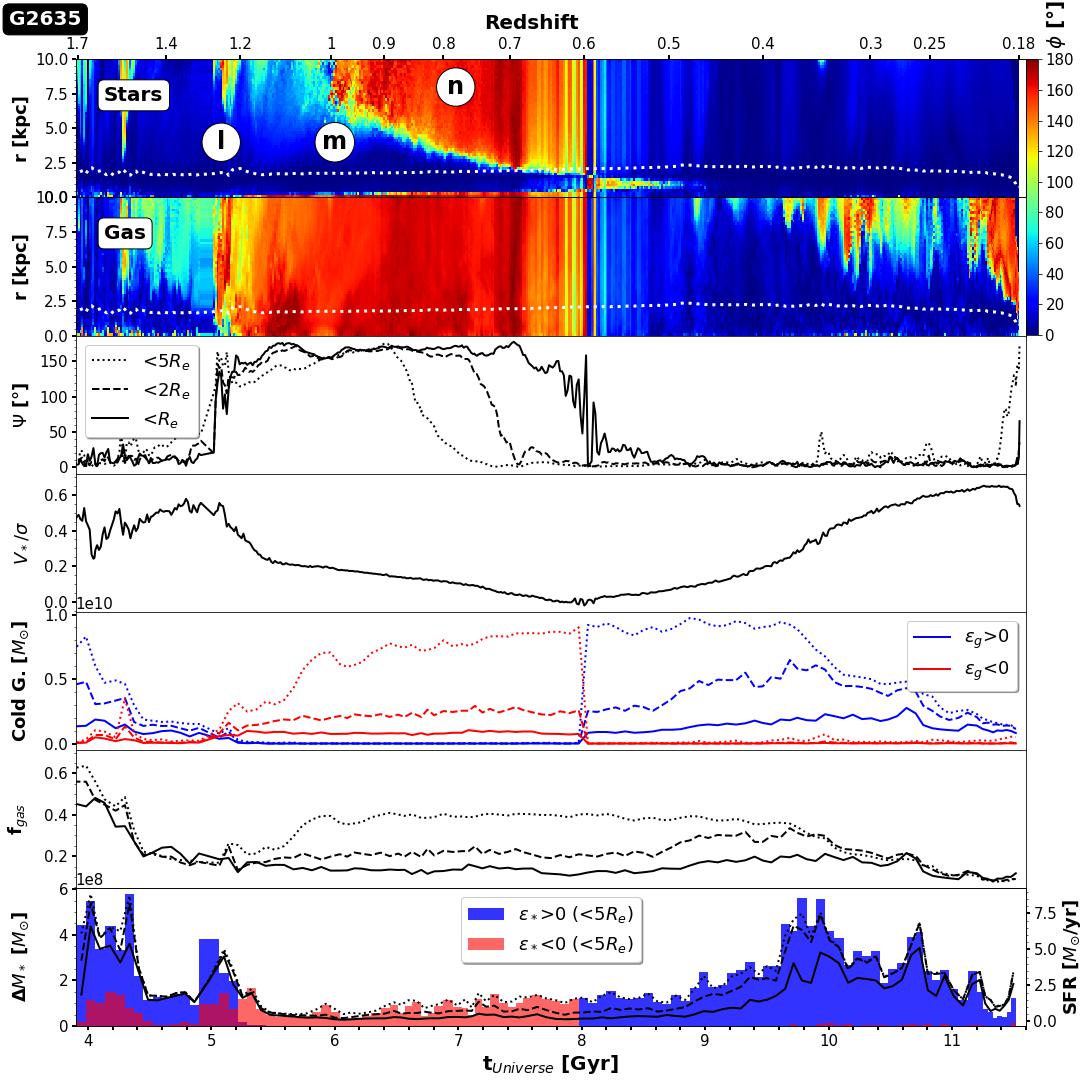}}
\caption{Same as Fig.~\ref{fig_g11_cosmic} but for G2835.
Here the interesting feature happens at \tuniv$\sim$8 Gyr 
where the orientation of the AM of the stellar component 
(estimated at the effective radius) flips. This is mainly due
to the contribution of the newly form stars in the disk.}
\label{fig_g2635_cosmic}
\end{center}
 \end{figure*}

Its formation pathway presents similar aspects to the previous cases studied in the last section.
 According to the evolution of $\Psi$ in the third row presented in Fig.~\ref{fig_g2635_cosmic}, the \crdd formation happens at
 \tuniv$\sim$5 Gyr (z$\sim$1.28), which is again preceded by decrease of the mass
 of cold gas (with $\epsilon_{\rm g}>0$) along with the gas fraction. The star formation
 rate is also more pronounced during this period.
 
Afterward, this galaxy presents an interesting feature in its evolution
just before   \tuniv$\sim$8 Gyr (z$\sim$0.6). Indeed,
the orientation of the angular momentum of the stellar component estimated within one effective radius flips.
As noted by \cite{park+19}, this is caused by to two main factors.
First, the AM of the system is small, suggested by the low values of \Vsig. Second, 
during the period of 5 Gyr < \tuniv<8 Gyr (1.27>z>0.6),
a large amount of stars form in situ with negative circularity parameters,
inherited from the cosmic gas accretion from several satellite galaxies. Since
they form in a disk, this population of young stars
progressively induces an angular momentum
whose contribution to the total stellar AM (within $R_e$) 
becomes at some point (\tuniv$\sim$8 Gyr) high enough to produce the AM flipping. 
The transition is seen, for instance, in the two first rows of Fig.~\ref{fig_g2635_cosmic}
where the blue color regions suddenly replaced the red ones.
As we will see in the next figure, the new stars form in the whole disk up to r$\sim$15 kpc.
The latter tends to dominate
first the AM budget at large radii  (high $\phi(r)$ values i.e. red colors in
the first row of Fig.~\ref{fig_g2635_cosmic})
and then progressively dominates the lower layers.

\begin{figure*}
\begin{center}
\rotatebox{0}{\includegraphics[width=9cm]{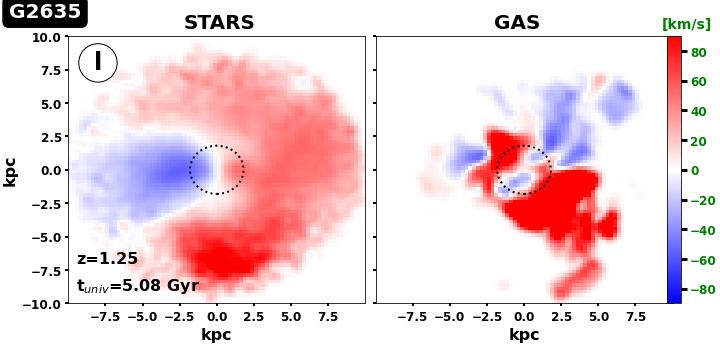}}
\rotatebox{0}{\includegraphics[width=9cm]{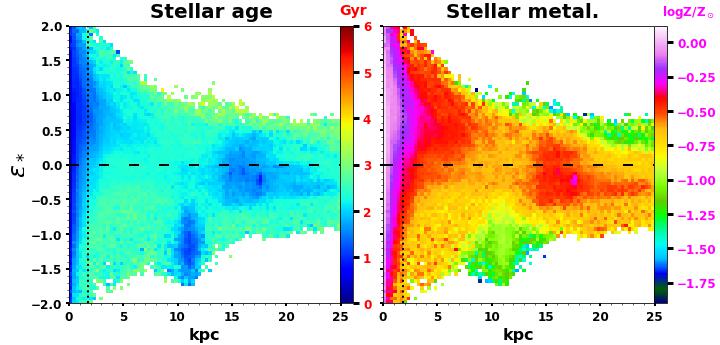}}
\rotatebox{0}{\includegraphics[width=9cm]{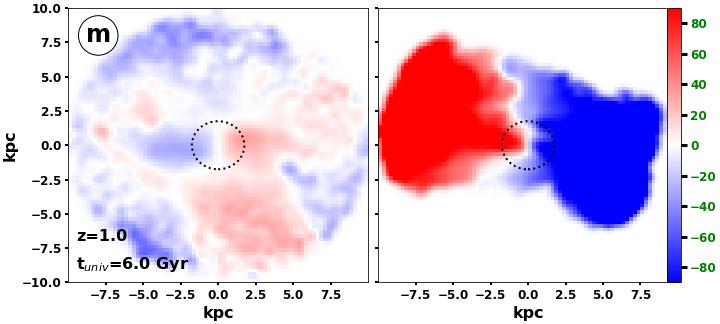}}
\rotatebox{0}{\includegraphics[width=9cm]{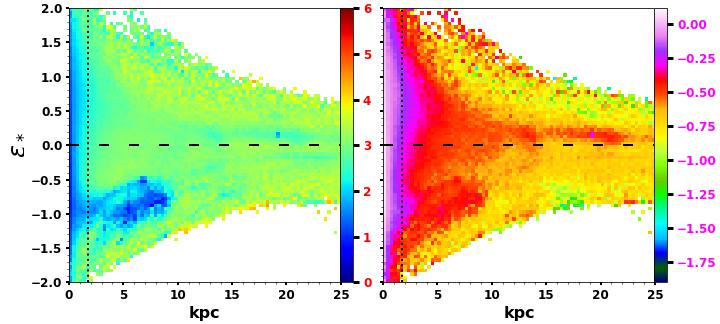}}
\rotatebox{0}{\includegraphics[width=9cm]{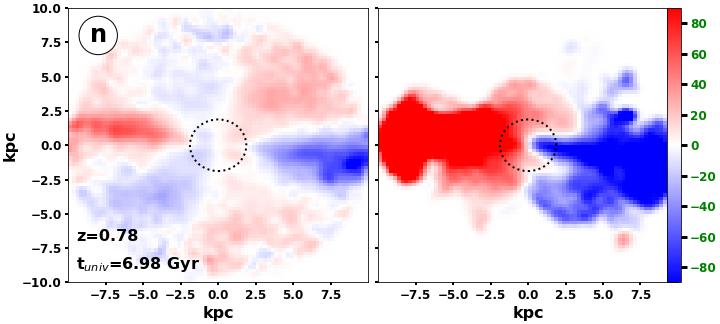}}
\rotatebox{0}{\includegraphics[width=9cm]{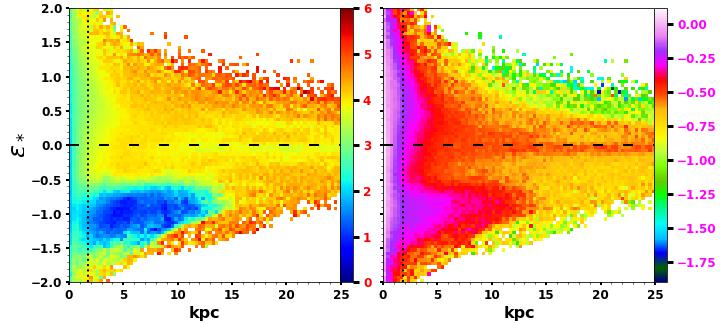}}
\rotatebox{0}{\includegraphics[width=18cm]{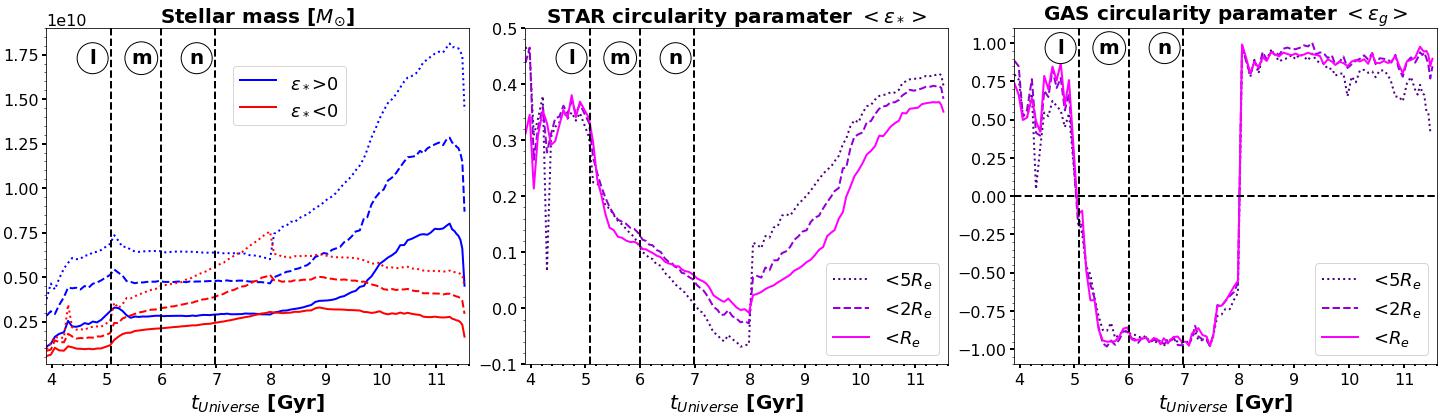}}
\caption{Same as Fig.\ref{fig_g11_diag} but for G2835.
At \epoch{l}, both edge-on stellar and gas projected velocity fields are quite disturbed due to the interaction 
of several galaxies. At this epoch, at least two galaxy satellites are present within 25 kpc. They can be identified
in the $\epsilon_*$-$r$-age and $\epsilon_*$-$r$-metal diagram at r$\sim$11 kpc and r$\sim$17 kpc, where their stars are 
generally young (age<2 Gyrs) with a global metalicity of $\sim$-1.2 (green) and $\sim$-0.35 (red) respectively.
A clear counter-rotating stellar-gas disks are obtained at \epoch{m} within 5 kpc. Some new stars are forming with $\epsilon_*\sim$-1 and r<10 kpc with
a high metalicity.
at epoch \epoch{n}, the new stars have now a high contribution within 15 kpc ($\epsilon_*\sim$-1) and tend to dominate at r>5.5 kpc
in the stellar projected velocity field.
}
\label{fig_g2635_diag}
\end{center}
 \end{figure*}

One consequence of this can be observed in the edge-on projected velocity fields presented in Fig.~\ref{fig_g2635_diag}.
At epoch \epoch{m}, the \crgg is clearly present within 5 kpc (i.e. two effective radius).
At epoch \epoch{n}, however, the newly formed stars, tend to have a higher contribution at larger radii
(r>2.5 kpc) which leads to the presence of two stellar component rotating
in opposite direction.
As mentioned above, the new stars are located up to 15 kpc with a high metallicity.
Note that the \diagagee diagram at epoch \epoch{n} is quite similar to the
ones generated and studied in \cite{khoperskov+21}.

\begin{figure}
\begin{center}
\rotatebox{0}{\includegraphics[width=\columnwidth]{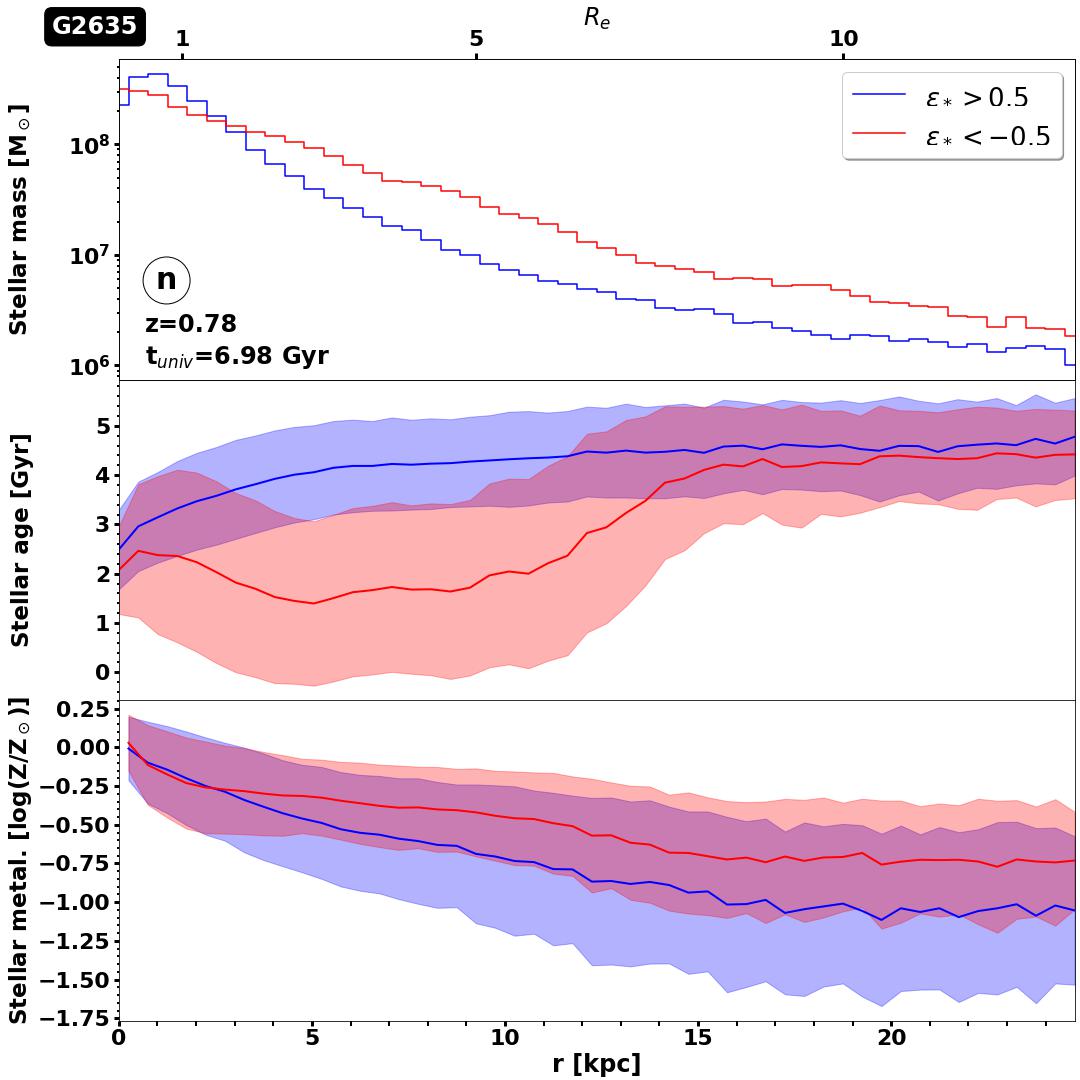}}
\caption{Same as Fig.~\ref{fig_g11_maz} but for G2635.}
\label{fig_g2635_maz}
\end{center}
 \end{figure}

Now we explore the 3-d radial distributions of the mass, age and metallicity  of stars with
circularity parameters satisfying either $\epsilon_*$>0.5 or $\epsilon_*$<-0.5,
in Fig.~\ref{fig_g2635_maz}, at z=0.78 (epoch \epoch{n}). 
Here the counter-rotating stars tend to be much younger and slightly more metal
rich than co-rotating stars, due to the recent and high star formation
activity. The counter-rotating stars also dominate the mass budget at large radius.
Contrary to G31, this is not due to a merger process but due 
to the long star formation activity.

Finally, it is worth mentioning that after the flip of the AM, the \Vsigg is
progressively increasing, suggesting an evolution toward a more
disk dominated galaxy.
Moreover, G2835 is also experiencing a new episode of 
\crgg formation at late time (\tuniv>9 Gyr).

%%%%%%%%%%%%%%%%%%%%%%%%%%%%%%%%%%%%%%%%%%%%%%%%%%%%%%%%%%%%%%
%     FIG 15      G2635    COSMIC  
%%%%%%%%%%%%%%%%%%%%%%%%%%%%%%%%%%%%%%%%%%%%%%%%%%%%%%%%%%%%%%
%%%%%%%%%%%%%%%%%%%%%%%%%%%%%%%%%%%%%%%%%%%%%%%%%

%%%%%%%%%%%%%%%%%%%%%%%%%%%%%%%%%%%%%%%%%%%%%%%%%%%%%%%%%%%%%
%     FIG 16     G2635  diagrams
%%%%%%%%%%%%%%%%%%%%%%%%%%%%%%%%%%%%%%%%%%%%%%%%%%%%%%%%%%%%%%
%%%%%%%%%%%%%%%%%%%%%%%%%%%%%%%%%%%%%%%%%%%%%%%%%

%%%%%%%%%%%%%%%%%%%%%%%%%%%%%%%%%%%%%%%%%%%%%%%%%%%%%%%%%%%%%
%     FIG 17     G2635   MASS-AGE-Z
%%%%%%%%%%%%%%%%%%%%%%%%%%%%%%%%%%%%%%%%%%%%%%%%%%%%%%%%%%%%%%
%%%%%%%%%%%%%%%%%%%%%%%%%%%%%%%%%%%%%%%%%%%%%%%%%

\section{Discussion}
\label{sec:discussion}

\subsection{Stellar populations}
\label{subsec:stellarpop}

The formation process of \crgg has significant impacts on
stellar populations.
First, during the compression of the pre-existing disk, stars usually form 
with positive $\epsilon_*$ in the main disk. Then, once it has been totally replaced by the accreted gas, some new stars may also form but, this
time, with negative $\epsilon_*$. According to this scenario, one should expect the presence of two co-spatial populations of young stars in the disk with opposite rotation. counter-rotating stars
are expected to be on average younger than co-rotating stars.
This is what we generally observe for our galaxy sample. However,
the opposite trend can also be obtained, for instance in G11 and G2.
In these two objects, the main disk tends to have a slightly younger stellar population (measured here within 5 kpc) than the counter-rotating one. In both cases,  the star formation is very low after
the removal of the pre-existing gas disk . This lack of star formation coincides 
with recent high BH activities induced by the compression of
the pre-exising disk causing it to fall in the centre of the galaxy (see section~\ref{subsec:bhactivity}). {The low star formation in these two objects could also be attributed to the low content of the cold gas along with
a low gas fraction.}

%%%%%%%%%%%%%%%%%%%%%%%%%%%%%%%%%%%%%%%%%%%%%%%%%%%%%%%%%%%%%
%     FIG 18     SIG Two peaks features
%%%%%%%%%%%%%%%%%%%%%%%%%%%%%%%%%%%%%%%%%%%%%%%%%%%%%%%%%%%%%%
\begin{figure}
\begin{center}
\rotatebox{0}{\includegraphics[width=\columnwidth]{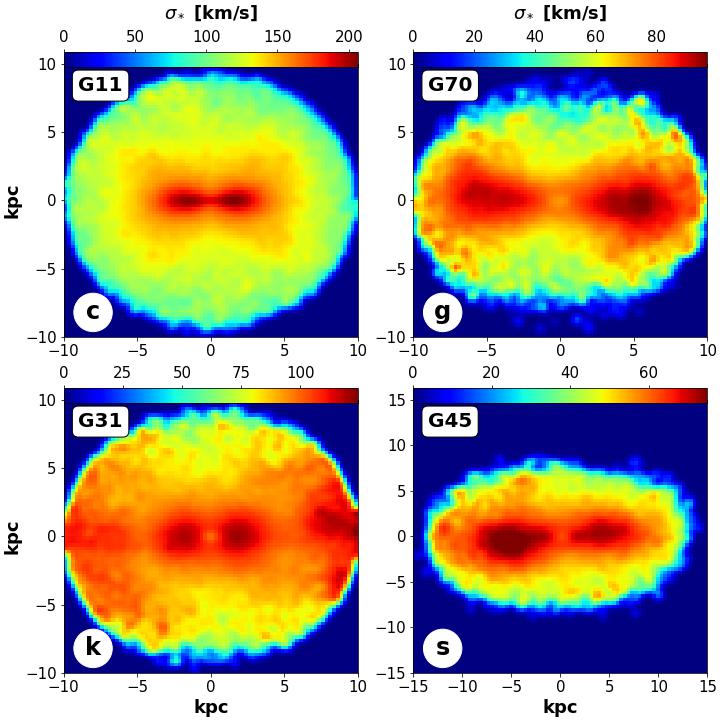}}
\caption{Projected edge-on mass-weighted stellar velocity dispersion fields ($\sigma_*$) for G11, G70, G31 and G45. Each
map is derived at a specific epoch in which the galaxy is displaying a clear stars-gas counter-rotating configuration. For G11, G70 and G45, the presence of the counter-rotating stellar component
cannot be seen in their projected stellar velocity fields but may be suggested by the
presence of the double peak feature exhibited in these maps.
}
\label{fig_Vsig}
\end{center}
 \end{figure}
%%%%%%%%%%%%%%%%%%%%%%%%%%%%%%%%%%%%%%%%%%%%%%%%%

The young and co-rotating stars generally have a high metallicity. Regarding the young and counter-rotating stars,
they form from the accreted gas and inherit its metallicity. Although the
parent gas can have a low metallicity before being accreted to the main galaxy, it can be progressively enriched 
by supernovae activity when getting closer to the center. In this case, it might be difficult to distinguish  co- and counter-rotating stars from
the analysis of their metallicity alone. This is what we get for most of the galaxies
in our sample, except again for G11 and G2. Here, one fundamental parameter
is the cold gas fraction $f_{\mathrm{gas}}$. In these two latter galaxies, $f_{\mathrm{gas}}$
is quite low, with values lower than 0.2 while the other galaxies of
the sample present values higher than 0.4 before the removal of the
pre-existing gas disk. Low values of $f_{\mathrm{gas}}$ may suggest that the reservoir
of gas is low to form new stars which will affect the supernovae
activity afterward.
We cannot totally exclude the role  of mixing between pre-existing gas and the newly-accreted one. Indeed, for galaxies with a high mass content of cold gas, metal-poor accreted gas would interact with the metal-rich
pre-existing gas and the dilution of metallicity leads to the higher metallicity in accreted gas without significantly reducing the metallicity of pre-existing gas due to its higher mass. On the contrary, for gas-poor galaxies (possibly lacking star formation), the mixing would not come into play,  simply because galaxies do not possess much gas.

Our results are in good agreement with observational analysis from
MaNGA galaxies that found younger stellar populations,  enhanced
star formation rates and higher metallicity in central region of
misaligned galaxies compared to their outer parts
\citep{jin+16, chen+16}.

Also, during a minor or major  merger, the stars of the satellite
can be expelled in the outer parts of the final object. This can 
produce some stellar streams with clear age-metallicity signatures.
This is more pronounced in major merger scenario where
each satellites's pericentric passage
induces successive
stellar streams.

From a dynamical point of view, the presence of two stellar components with opposite rotation
may lead to the so-called double-peaked feature seen in the observational stellar velocity dispersion maps 
\citep[e.g.][]{krajnovic+11,rubino+21,bevacqua+22,bao+24,katkov+24}.
We show examples of such maps  derived for G11, G70, G31 and G45 in Fig.~\ref{fig_Vsig}, confirming
the presence of a double peak at epochs of clear stars-gas counter-rotating
configurations. This feature is therefore a simple and efficient
diagnostic to detect the presence of the
counter-rotating stellar components, which are, on the contrary, not seen in the respective projected stellar velocity fields of e.g. G11, G70 and G45.

\subsection{Disappearance of the pre-existing disks: star formation or gas removal?}
\label{subsec:disappearance_gas}

%%%%%%%%%%%%%%%%%%%%%%%%%%%%%%%%%%%%%%%%%%%%%%%%%%%%%%%%%%%%%%
%              FIG 19
%%%%%%%%%%%%%%%%%%%%%%%%%%%%%%%%%%%%%%%%%%%%%%%%%%%%%%%%%%%%%%
\begin{figure}
\begin{center}
%\rotatebox{0}{\includegraphics[width=\columnwidth]{FIG_Cconversion.jpg}}
\rotatebox{0}{\includegraphics[width=\columnwidth]{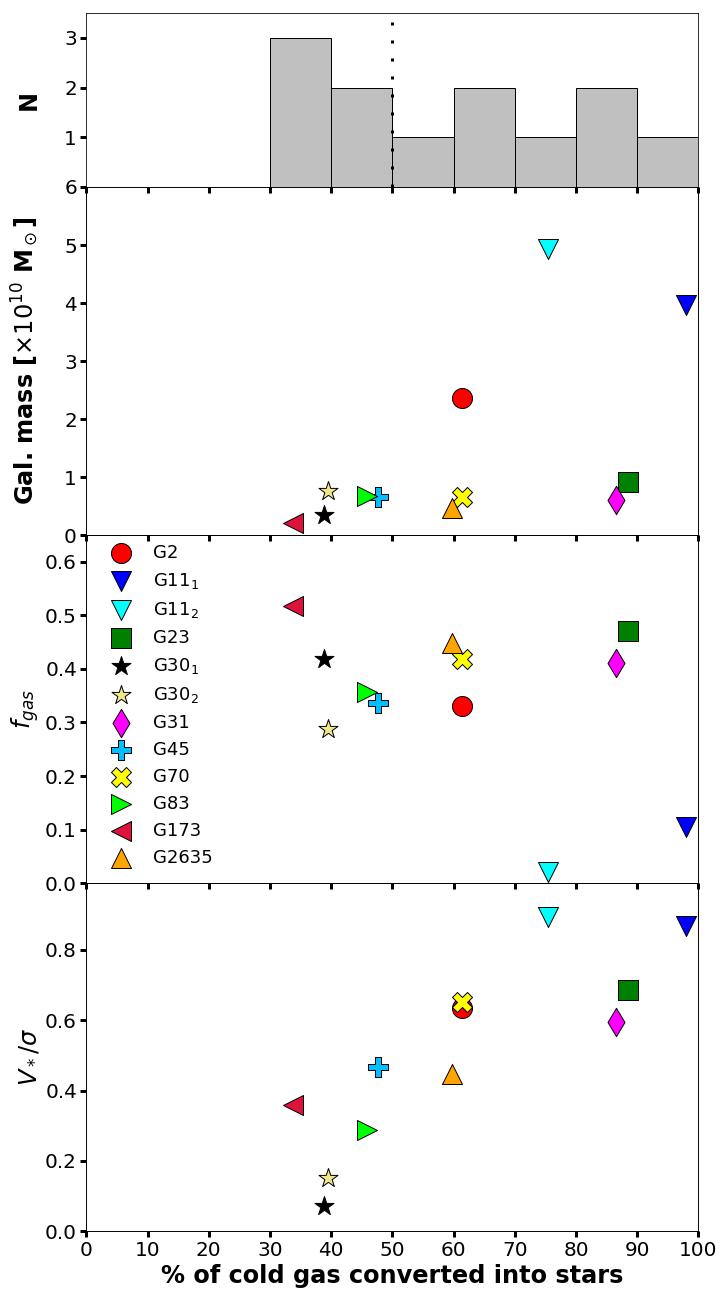}}
\caption{ Estimation of the  percentage of cold gas with
$\epsilon_{\rm g}>0$ converted into stars during the replacement of the inner gas disk
which preceded the formation of the \crd, for all the studied galaxies. 
  We also show the variations with respect to the stellar mass of the main galaxies
(upper panel), the gas fraction $f_{gas}$ (middle panel) and \Vsig,
all quantities estimated just before the retrograde gas accretion phase.
$f_{\rm gas}$ and \Vsigg have been estimated both within one effective radius but the
same trends are obtained when estimated within  5$R_e$.
These plots first suggest that, during
the removal of the pre-existing gas,
on average, a higher fraction of cold gas is converted into stars rather than
being expelled. Secondly, there is a clear correlation between the percentage
of gas converted into stars and \Vsig. The
original morphology of galaxies seems therefore to 
impact the efficiency of the star formation during
the compaction of the central gas component.}
\label{fig15}
\end{center}
 \end{figure}
%%%%%%%%%%%%%%%%%%%%%%%%%%%%%%%%%%%%%%%%%%%%%%%%%

%%%%%%%%%%%%%%%%%%%%%%%%%%%%%%%%%%%%%%%%%%%%%%%%%%%%%%%%%%%%%
%     FIG 19     DM densitity < 1 kpc
%%%%%%%%%%%%%%%%%%%%%%%%%%%%%%%%%%%%%%%%%%%%%%%%%%%%%%%%%%%%%%
%\begin{figure}
%\begin{center}
%\rotatebox{0}{\includegraphics[width=\columnwidth]{FIG_CDM.jpg}}
%\caption{Evolution of the mean dark matter density within one physical kpc
%for G11 (upper panel) and G70 (lower panel). The vertical red dashed lines %correspond to the time when the inner pre-existing gas disk has been totally
%replaced by the accreted gas. In case of G11, there are two episodes of gas %replacement at \tuniv$\sim$7.6 Gyr et  \tuniv$\sim$9.9 Gyr, respectively.
%In general, the replacement of the pre-existing gas is associated with a %strong induced
%star formation in the inner part due to the compression of the gas.
%Consequently, a compaction of the DM component is also observed 
%in the center of the galaxy.
%}
%\label{fig_dmdens}
%\end{center}
% \end{figure}
%%%%%%%%%%%%%%%%%%%%%%%%%%%%%%%%%%%%%%%%%%%%%%%%%

The fate of the pre-existing gas has been studied in past theoretical works.
\cite{khoperskov+21}, using toy models, suggest that the gas is mainly
removed while \cite{cenci+24} identified central starburst-driven depletion as the main reason for the removal of the pre-existing co-rotating gas component, with no need for feedback from, e.g., a central active black hole.
Our study does not allow to answer precisely to this question. 
The gas in the \ramses\, code relies on an Eularian approach which makes it difficult to follow the motion of the gas accurately,
contrary to a Lagrangian approach such as a smoothed-particle hydrodynamics model (SPH). However, we have made a rough 
estimation of the conversion of the cold gas into stars, by estimating
the amount of cold gas with positive circularity parameter that
disappears during the process, compared to the newly formed stars 
with positive $\epsilon_*$ during the same period. This quantity has been estimated
within five effective radii.
Fig.~\ref{fig15} shows the estimation of the percentage of cold gas
converted into stars during the retrograde accretion phase of all galaxies. Note that for G11, we estimate this quantity during
the first (G$11_1$) and second (G$11_2$) retrograde accretions, while
G30 undergoes two episodes of counter-rotating formation, noted here as G$30_1$ and G$30_2$. 
The plot suggests that during the process, a higher fraction of gas is generally converted into stars rather than being expelled. 
We also took an interest in any potential dependencies of
the estimated fraction of gas being converted into stars with e.g. the
stellar mass, the gas fraction, the galaxy morphology, $R_e$, etc...
In Fig.~\ref{fig15}, for instance, we additionally  
show the variations with respect to the mass of the main galaxies
(upper panel), the gas fraction $f_{\rm gas}$ (middle panel) and the quantity \Vsig.
These latter quantities have been estimated just before the retrograde gas accretion phase
where the mass of the cold gas (with $\epsilon_{\rm g}$>0) reaches a maximum value.
f$_{\rm gas}$ and \Vsigg have been estimated within one effective radius but the
same trends are obtained when estimated within five $R_e$.
Only one clear correlation appears between the percentage
of cold gas converted into stars and \Vsig. This suggests that the 
initial morphology of galaxies, especially disk-dominated galaxies with a strong
rotational support 
favor the conversion of gas into stars during the retrograde gas accretion
phase, rather than more spheroid dominated galaxies.

\subsection{Galaxy morphology evolution}
\label{subsec:morphology}

The mechanism responsible for the formation of \crgg implies 
the loss of angular momentum of the pre-existing gas
which reverberates on the angular momentum of the stellar component.
This can be clearly seen in the time evolution of \Vsigg for each galaxy.
\Vsigg tends indeed to decrease during the formation of counter-rotating
gas component and leads to the formation of spheroid-dominated
galaxies.
Once the \crgg has formed, as suggested by a visual inspection of the projected u-g-r band images of the stellar component of these galaxies,
they are more likely S0 galaxy.
In a companion paper, \cite{han+24} have also analyzed \nhs galaxies 
to study two potential formation pathways of lenticular galaxies (S0s) in field environments.
Among them, counter-rotating gas inflow is a promising channel as it causes
gas angular momentum cancellation through hydrodynamic collisions.
This is in agreement with \cite{zhou+23} who have studied MaNGA galaxies
and found misaligned gas accretion to be an important formation pathway for S0s.
\cite{beom+24} also found from MaNGA galaxies that the majority of 
counter-rotators are early-types.
Finally, \cite{katkov+14} studied a sample of completely 12 isolated S0 galaxies
by means of long-slit spectroscopy at the Russian 6-m telescope.
They found that 5 galaxies show a visible counter-rotation of the ionized-gas component with respect to the stellar component.

\subsection{Black hole activity}
\label{subsec:bhactivity}

During the first phase of the formation of a gaseous CR
component, the pre-existing gas is compressed by the
accreted gas. This may have some consequences on the BH feeding and activity.
Among the ten galaxies studied in our sample, 
three of them, G2, G11 and G31, have a massive black hole
well centered, while in the other ones, the BHs are off-centered 
\citep[as a result of intense stellar feedback creating shallow potential wells, e.g.,][]{dubois+15,bellovary+19,pfister+19}, and are not relevant for the present analysis.

First, BH1049 is associated with G11 and its properties are summarized in
Fig.~8 in \cite{peirani+24}. In particular, between \tuniv$\sim$4.5 Gyr and \tuniv$\sim$7.8 Gyr, namely
during the first retrograde gas accretion phase undergone by G11,
the radiative efficiency of the gas accretion disk, $\epsilon_r$, is very
high, which means that the BH accretes gas efficiently. The spin parameter
is also close to its maximum value (0.998) and the Eddington ratio $\chi$ indicates
that the black hole is most of the time in the quasar mode.
During the second retrograde gas accretion phase, between \tuniv$\sim$7.8 Gyr and
\tuniv$\sim$10 Gyr, the story is different. Here $\epsilon_r$ is low 
down to \tuniv$\sim$9.4 Gyr. The reason is that after the first phase of retrograde gas accretion, the orientation of the BH spin has flipped following the
orientation of the accreted gas \citep{peirani+24}. Consequently,
between  \tuniv$\sim$7.8 Gyr and
\tuniv$\sim$9.4 Gyr, the BH spin and the accreted gas are misaligned 
(see the evolution of \PSIGAS) which decreases $\epsilon_r$. But once the BH spin orientation flips
again to follow the orientation of the accreted gas, we can observe
the same behavior as in the first accretion phase:
$\epsilon_r$ and the BH spin parameter become high which suggests 
that the gas accretion onto the BH is now very efficient. Here again
the BH is mostly in the quasar mode.

As far as G31 is concerned, its central black hole, BH549 is studied
in Fig.~A.1 in \cite{peirani+24}. According to the second row of Fig.~\ref{fig_g31_cosmic}, the replacement of the
pre-existing gas takes place between \tuniv$\sim$5.8 Gyr and
\tuniv$\sim$7.2 Gyr. As shown in Fig.~A.1 in \cite{peirani+24},
this period corresponds precisely to a sudden increase of the radiative
efficiency ($\epsilon_r$) values
along with a high spin parameter. 

Regarding BH796, the massive central BH associated with G2, 
the evolution of its properties is shown 
in the Appendix~\ref{appendix2}. From Fig.~\ref{fig_app7},  
we found similar trends to those of BH1049 and BH549. 
More specifically,  $\epsilon_r$ and the spin parameter
are very high from \tuniv$\sim$4.4 Gyr to \tuniv$\sim$5.2 Gyr which
corresponds to the replacement of the pre-existing gas component. 

Overall, our results suggest an enhancement of the BH activity during the removal of the pre-existing gas disk.
Such trends are consistent with recent theoretical and observational analysis suggesting
a correlation between BH/AGN activity and the formation
of counter-rotating structures \citep{duckworth+19,raimundo+23}.

Furthermore, once the \crgg has formed, the orientation of the BH spin is more likely
to become anti-aligned with respect to AM of the stellar component
\citep{peirani+24}.

\subsection{Fate of counter-rotating galaxies}
\label{subsec:fate}

Once a gas-versus-stars counter-rotating galaxy has formed, what could be its fate?
Although we have considered only ten galaxies, we found different
possible evolutions.

First, without any strong perturbation from the nearby environment such as
galaxy interaction and mergers, the counter-rotating gas-stellar disks 
can have a stable configuration which can last for several giga years. For instance, after experienced a major merger, the counter-rotating disks in G31 is stable at least for more than 4 Gyrs. Also, G70 keeps this configuration for more than 2 Gyrs (i.e., until the end of the simulation).

Second, if a galaxy undergoes another retrograde gas accretion, which is the case for G11, the final object becomes a galaxy with co-rotating stars and gas. If the galaxy undergoes a major merger, these potential mechanisms may destroy the counter-rotating disks configuration.

Third, there is a more subtle evolution. In the case of G2635 (or G30 and G173), 
the new stars that form in the disk with $\epsilon_*<0$ can progressively
dominate the AM budget of the galaxy estimated at the effective radius (but same behavior is find when considering 2 and 5 effective radii).
In this case, the stellar angular momentum orientation will flip and $\Psi$ will be close to 0$^\circ$. According to our definition, the galaxy is not classified as a gas-versus-star counter-rotating galaxy anymore.

We also note that a galaxy can experience more than one episode
of \crgg formation, for instance, in G2635, G30, but also in G11 with two successive retrograde cosmic gas accretions.

\subsection{The frequency of counter-rotating galaxies}
\label{subsec:fraction}

At $z=0.18$, we found that $\sim$4.5\% of galaxies with  a mass greater than 10$^9$M$_\odot$ present distinct stars-gas counter-rotating components.
Although the volume covered by \nhs is relatively small,
it is still instructive to compare this trend with those
obtained from other numerical works or from observational
analyzes.

First, our derived fraction is similar to the one obtained by \cite{cenci+24} with about 4\% of FIREbox galaxies with stellar masses larger than 5$\times$10$^9$M$_\odot$ at $z=0$  exhibiting gas–star kinematic misalignment. 
Similarly, \cite{baker+24} 
analyzed the EAGLE simulation \citep{eagle} and found that gas counter-rotation
can be present in 5.7\% of the galaxy sample
with mass greater than  10$^{9.5}$M$_\odot$ at $z\approx 0.1$
\citep[see also,][]{casanueva+22}.
A lower value is however obtained
by \cite{starkenburg+19}
from the Illustris simulation \citep{illustris}. They
indeed studied the origin of such counter-rotation in low mass galaxies (2$\times$10$^9$ - 5$\times$10$^{10}$ M$_\odot$)
and found that 
only $\sim$1\% of the sample shows a counter-rotating gaseous disk at $z=0$. Moreover, although the corresponding fraction value is not explicitly
reported, \cite{khoperskov+21} selected only 25 
galaxies from illustrisTNG100
with a mass greater than 10$^{9}$M$_\odot$ at $z=0$.
This number may appear quite low given the fact that 
the simulation has a box side of 75 Mpc/h and then 
a volume much bigger than \nhs ($\sim$300 times).
Nevertheless, the discrepancies between these numerical works may be attributed to existing differences
in the galaxies selection strategy,
the spatial/mass resolutions and treatment of feedback processes (in particular the efficiency of AGN feedback in depleting the gas content in massive galaxies and
groups) as well as the different studied environments.
For instance, \nhs has been
dedicated to study the formation of galaxies in the field
environments. It does not consider 
cluster environments in which interaction of galaxies is
more frequent and may favor the appearance of 
episodes of counter-rotating components while
reducing their lifetime.

As far as observations are concerned, 
stars-gas counter-rotating galaxies are generally 
selected when the PA offset between the stars and gas components
ranges from 150$^\circ$ to 180$^\circ$.
For example, \cite{beom+24} found a fraction 
of $\sim$3.0\% by 
identifying 303 gaseous
counter-rotators out of 9992 galaxies in MaNGA (>10$^9$ M$_\odot$). This is consistent with \cite{li_manga+21} 
who have also considered the MaNGA sample and focused on
star forming and quiescent galaxies samples
to get a fraction of counter-rotators of $\sim$3.3\%.
Note that a lower value of  $\sim$1.9\% was previously obtained
by identifying 10 stars-gas counter-rotating components
among 523 edge-on disk galaxies in MaNGA (MPL-11) \citep{beom+22}.
These trends are consistent with 
\cite{bryant+19} using the  SAMI data with
a fraction of about 3.5\% (including late-type and
early-type galaxies).

%\subsection{Central DM density evolution}
%\label{subsec:profile}

%Although a detailed analysis of evolution of the inner DM density is beyond the scope of the present paper, it is interesting to investigate the potential behavior of this component.
%Indeed, during the replacement of the pre-existing gas, two mechanisms
%are of potential interest. First, the inner gas is more and more compacted
%by the accreted gas. Second, this compaction generally leads to an enhancement
%of the star formation in the center of galaxies. Therefore, one can expect a dynamical response from the dark matter component.
%In Fig.~\ref{fig_dmdens}, we show the evolution of the DM density,
%estimated within one physical kpc, for both G11 and G70.
%It appears that the DM density noticeably increases during the removal of
%the pre-existing gas, especially during periods of strong star formation.
%This may have some consequences in the evolution of the 
%total density profile of S0 galaxies \citep{peirani19} as well as
%on the indirect detection of dark matter with potentially boosted 
%signals

\section{Summary and conclusions}
\label{sec:conclusions}

Using the \nhs simulation, we have studied the formation
of \crgg with a stellar mass of $\succsim$10$^{10}$ M$_\odot$. 
The identification of these galaxies is based on the 
3 dimensional angle  $\Psi$ subtended between the stellar and gas angular momentum
vectors (estimated within one effective radius). Our final
sample consists of ten galaxies satisfying $\Psi$>120$^\circ$
which present similarities and
differences in their evolution. 
Our main conclusions can be summarized as follows:

%$\bullet$
{\bf 1.}
The physical mechanism responsible for the formation of \crdd is common
for all galaxies of our sample. It follows two successive phases:
i) a retrograde cosmic gas accretion from gas stripping in galaxy interaction or from the circumgalactic medium; 
ii) coexistence of two gas components rotating
in opposite direction, the pre-existing one being progressively replaced by the  accreted one.

{\bf 2.}
During the second phase, the pre-exiting gas is compressed by the accreted gas.
A large fraction of the gas is consumed and converted into stars. The 
remaining fraction is more likely to be expelled or diluted in the 
accreted disk. 
In very good agreement with \cite{cenci+24}, this phase can be decomposed into two stages:
1) a gas compaction associated with an enhancement of the star formation rate resulting in an increase and a decrease of the
mean circularity parameter of star and gas respectively;
2) the newly accreted gas gradually forms a counter-rotating disc with respect to the main stellar component and lead 
 both the average circularity of the stellar and gas components to decrease. 
Also, during this phase, if a central BH is present in the center of the galaxy,
the gas accretion onto the BH is very efficient and its activity enhanced.

{\bf 3.}
The formation of a counter-rotating gas component is always accompanied by the formation
of counter-rotating stellar components. The latter ones can have different properties
depending on the star formation histories and merger process.
However, our results suggest that in most cases,
once the accreted gas has totally replaced the pre-exising disk,
some in-situ star formation takes place to produce counter-rotating
stars in the central part of the galaxy. This may induce a positive age gradient.
Moreover, it should be possible
to accurately date the formation of the counter-rotating disk by identifying such stars in observational analyzes.

{\bf 4.}
Depending on the star formation/supernovae activity that is generally more pronounced
in the inner parts of the galaxies, the accreted gas might be gradually enriched in
metal as its sinks toward the galaxy center. Consequently, counter-rotating stars that form in-situ 
could be metal-rich even though the accreted gas originally had a lower metallicity.

{\bf 5.}
Counter-rotating stars are generally slightly younger and more metal-rich than co-rotating ones. This trend is more pronounced at larger radii.
Also, counter-rotating stars located in the inner parts of the galaxies tend to be younger and more metal rich than those displayed in the outer parts.

{\bf 6.}
In major merger scenario, a larger fraction of stars are more likely to
be expelled and diffuse in the outer part of the remnant object. In this case,
different populations of stars rotating in opposite directions are more
likely to be observed or detected. 
Also, major mergers redistribute the stars along all directions
which is expected to smooth any significant age/metal gradient.

Cosmological filaments and gas-rich satellites are considered as the main candidate sources of external cold gas. However, it is particularly difficult to separate them
in observational analyzes. 
Our analysis of a sample of ten galaxies is limited since all the galaxies
develop counter-rotating components due to galaxy satellite interaction. However, there are clear
signatures inherent to galaxy interaction and merger.
On the one hand, merger is expected to display a fraction of stars in the outer parts.
These stars may have different age and metal properties from the in situ formed stars in the center of the galaxy. In case of gas accretion from cosmological filaments, we do not expect such trends.
Moreover, as specified in Point 6, any age/metal gradient is expected to
be smoothed during a merger event. This is what we found with G31 and G83 where the mean age and mean metallicity tend to be constant at larger radii.
{This is also the case with G173 which underwent two successive minor mergers.}

Although our analysis gives detailed insights on the formation of \crg, several limitations remain. For instance, the \nhs simulation has been dedicated to study
the formation of galaxies in the field environments. Therefore, we have not studied
the formation process in cluster environment in which interaction of galaxies is more frequent.
Consequently, episodes of counter-rotating components might also be more frequent and their lifetime might be reduced.
Moreover, our sample only consists of only ten low-mass galaxies.
It would be then interesting to consider a lower mass galaxy range to 
investigate whether the same mechanisms operate.  
Other high resolution simulations are therefore needed 
 to lead more robust statistical analyzes and help interpreting
 forthcoming high spectral and spatial resolution data from, e.g., 
 MUSE \citep[Multi Unit Spectroscopic Explorer,][]{muse} 
 or JWST \citep[James Webb Space Telescope,][]{jwst}.

\begin{acknowledgements}
We warmly thank the referee for an insightful review that
considerably improved the quality of the original manuscript.
It is a pleasure to thank St\'ephane Colombi for interesting discussions. 
S.P. acknowledges the support from the JSPS (Japan Society for the
Promotion of Science) long-term invitation program, and is grateful for the
hospitality at Department of Physics, the University of Tokyo.
This research is partly supported by the JSPS KAKENHI grant Nos.
23K25908 (Y.S.).
This work was granted access to the HPC resources of CINES under the allocations c2016047637, A0020407637 and A0070402192 by Genci, KSC-2017-G2-0003, KSC-2020-CRE-0055 and KSC-2020-CRE-0280 by KISTI, and as a “Grand Challenge” project granted by GENCI on the AMD Rome extension of the Joliot Curie supercomputer at TGCC. The large data transfer was supported by KREONET, which is managed and operated by KISTI. S.K.Y. acknowledges support from the Korean National Research Foundation (2020R1A2C3003769,RS-2022-NR070872).
This work was carried within the framework of the
Horizon project (\href{http://www.projet-horizon.fr}{http://www.projet-horizon.fr}).
Most of the numerical modeling presented here was done on the Horizon cluster at Institut d'Astrophysique de Paris (IAP).
\end{acknowledgements}

\bibliographystyle{aa}
\bibliography{author}

%\vfill

%\appendix
\begin{appendix}

\section{Other individual evolutions}
\label{appendix1}

We present the results for the other galaxies of the sample. Although our
sample is limited to ten galaxies, the latter have diverse properties
at different stages of their evolution. This could be relevant and useful for
the interpretation of observational data.

%%%%%%%%%%%%%%%%%%%%%%%%%%%%%%%%%%%%%%%%%%%%%%%%%%%%%%%%%%%%%%
%          FIG APPENDIX 1
%%%%%%%%%%%%%%%%%%%%%%%%%%%%%%%%%%%%%%%%%%%%%%%%%%%%%%%%%%%%%%
\begin{figure*}
\begin{center}
\rotatebox{0}{\includegraphics[width=9cm]{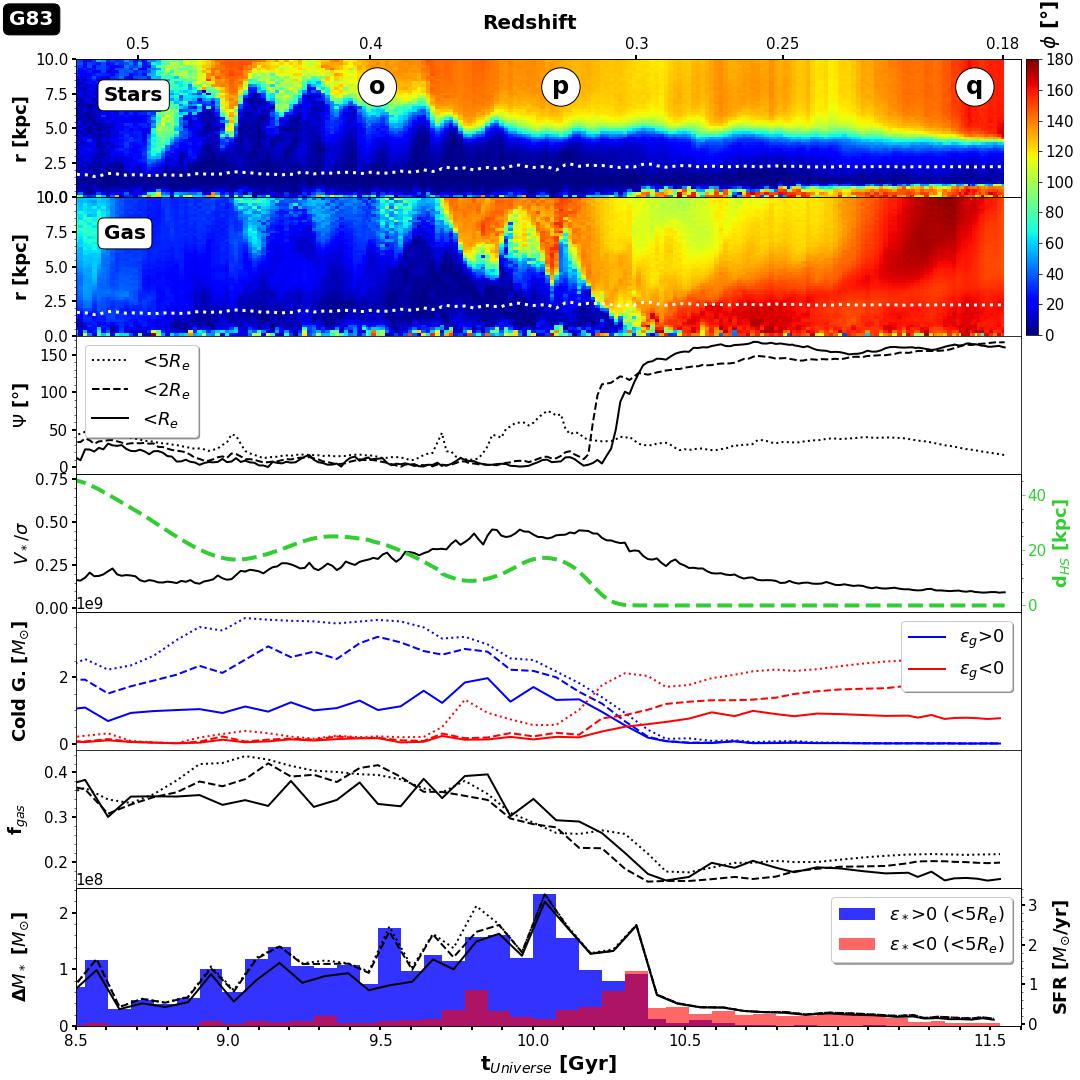}}
\rotatebox{0}{\includegraphics[width=9cm]{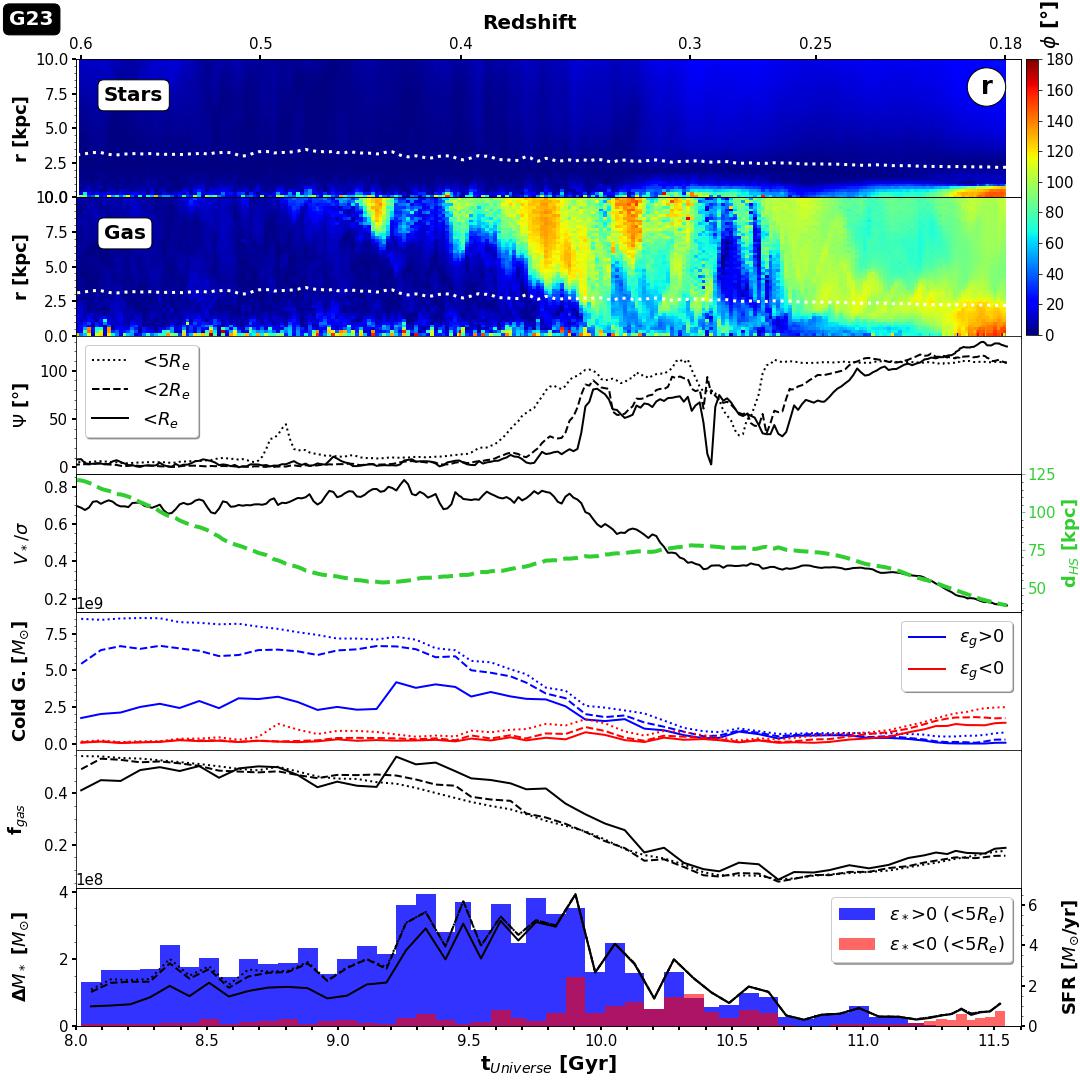}}
\rotatebox{0}{\includegraphics[width=9cm]{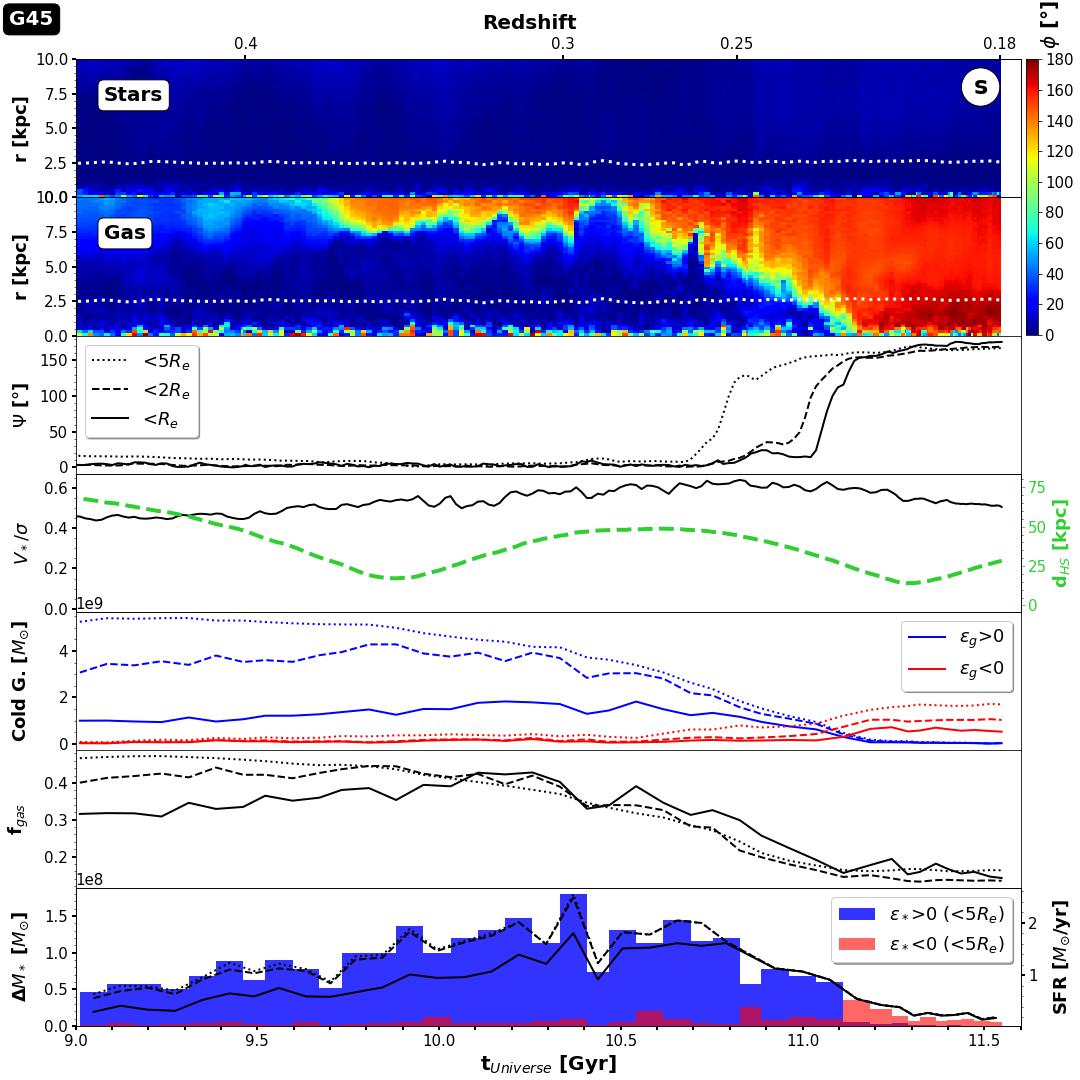}}
\rotatebox{0}{\includegraphics[width=9cm]{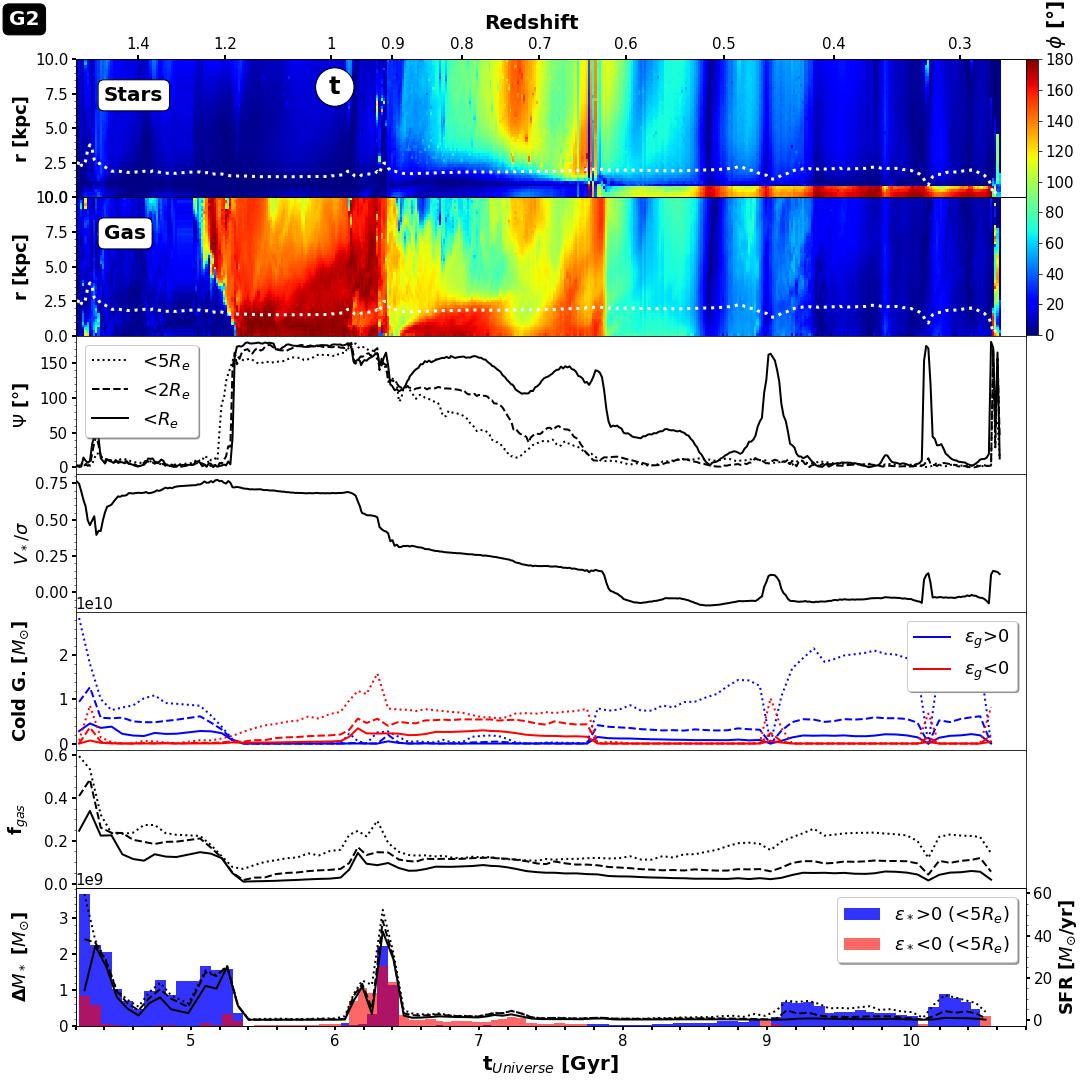}}
\caption{Same as Fig.~\ref{fig_g11_cosmic} but for G83, G23,
G45 and G2.
}
\label{fig_app1}
\end{center}
 \end{figure*}
%%%%%%%%%%%%%%%%%%%%%%%%%%%%%%%%%%%%%%%%%%%%%%%%%

%%%%%%%%%%%%%%%%%%%%%%%%%%%%%%%%%%%%%%%%%%%%%%%%%%%%%%%%%%%%%%
%          FIG APPENDIX 2
%%%%%%%%%%%%%%%%%%%%%%%%%%%%%%%%%%%%%%%%%%%%%%%%%%%%%%%%%%%%%%
\begin{figure*}
\begin{center}
\rotatebox{0}{\includegraphics[width=9cm]{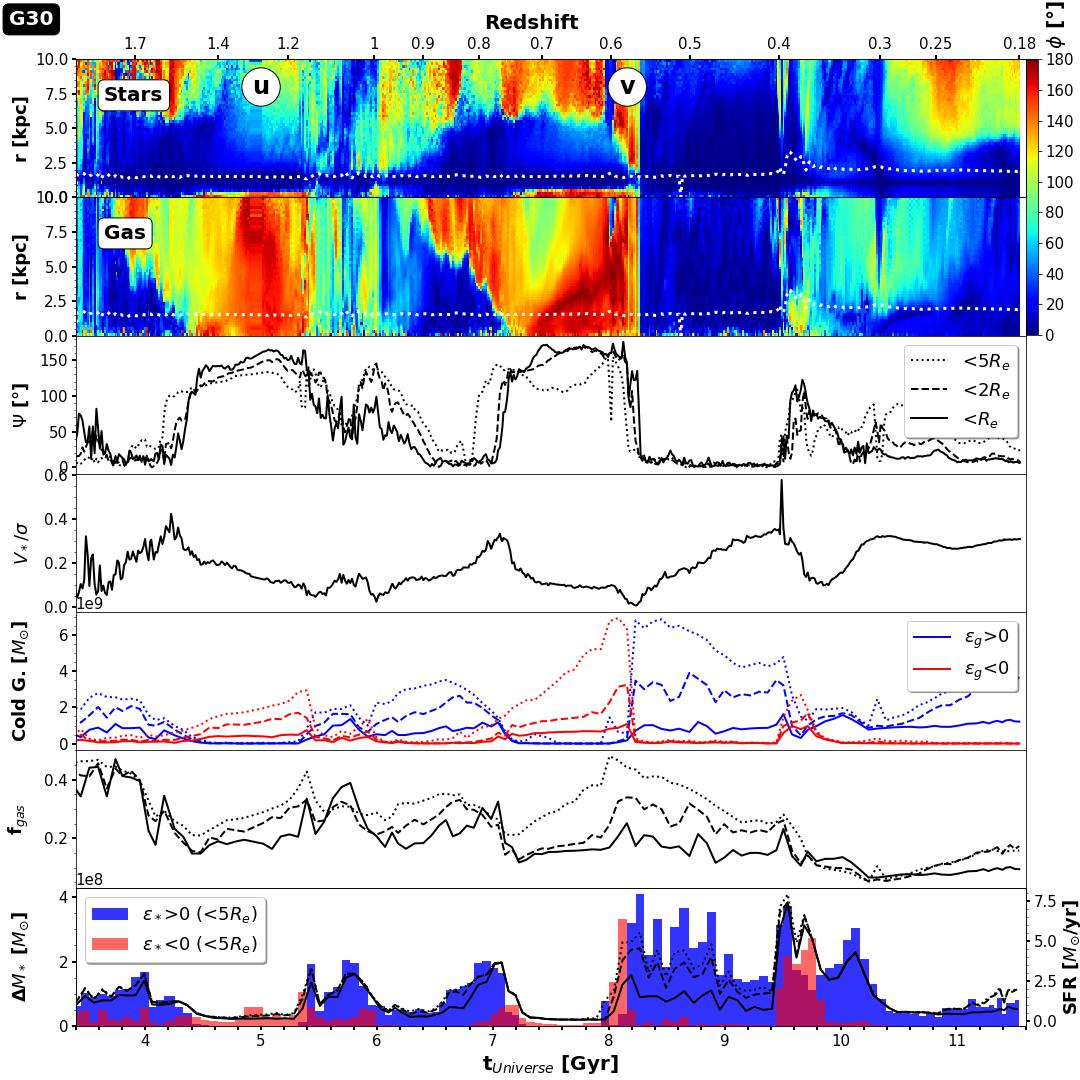}}
\rotatebox{0}{\includegraphics[width=9cm]{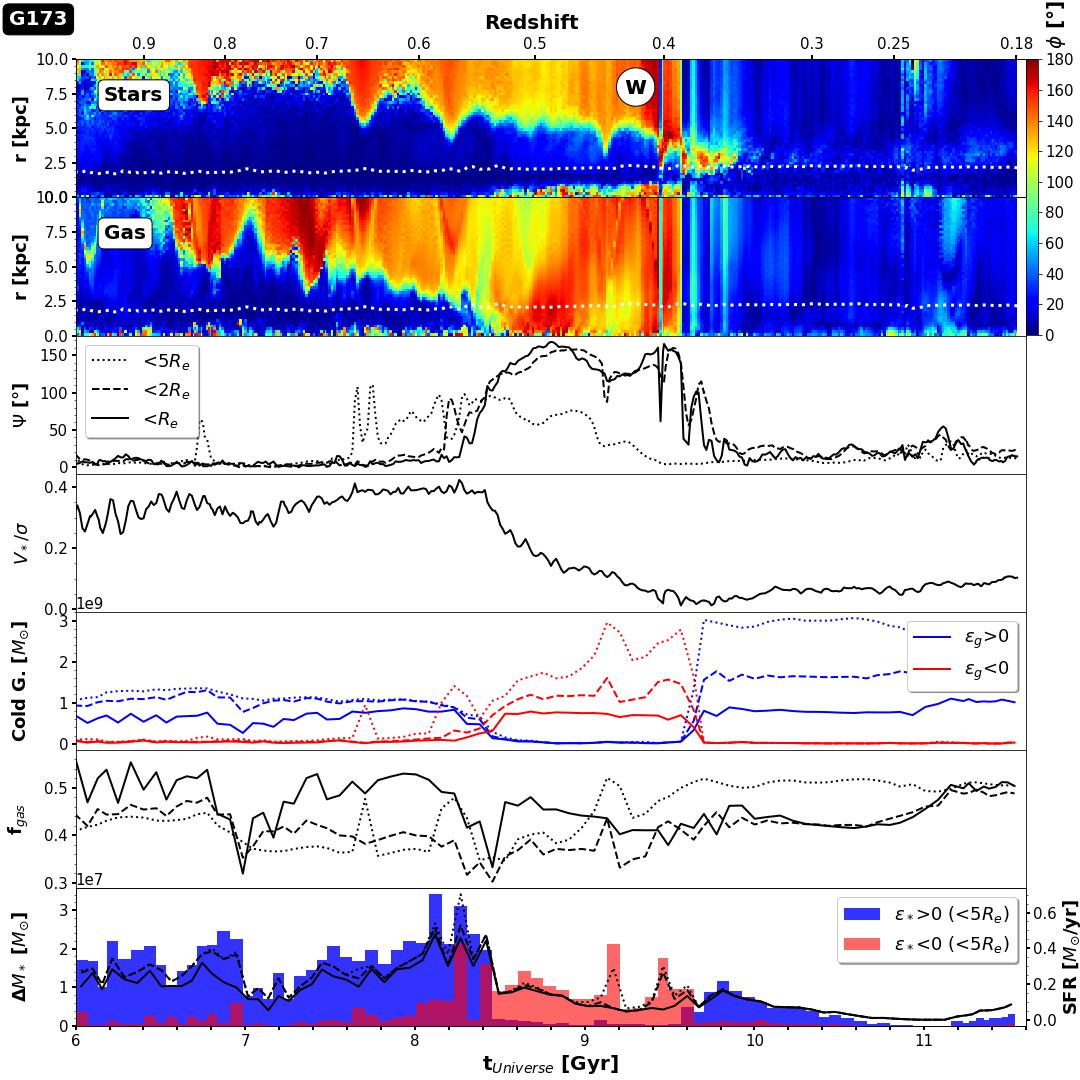}}
\caption{Same as Fig.~\ref{fig_g11_cosmic} but for G30, G173.
}
\label{fig_app2}
\end{center}
 \end{figure*}
%%%%%%%%%%%%%%%%%%%%%%%%%%%%%%%%%%%%%%%%%%%%%%%%%

%%%%%%%%%%%%%%%%%%%%%%%%%%%%%%%%%%%%%%%%%%%%%%%%%%%%%%%%%%%%%
%     FIG APPENDIX 3
%%%%%%%%%%%%%%%%%%%%%%%%%%%%%%%%%%%%%%%%%%%%%%%%%%%%%%%%%%%%%%
\begin{figure*}
\begin{center}
\rotatebox{0}{\includegraphics[width=9cm]{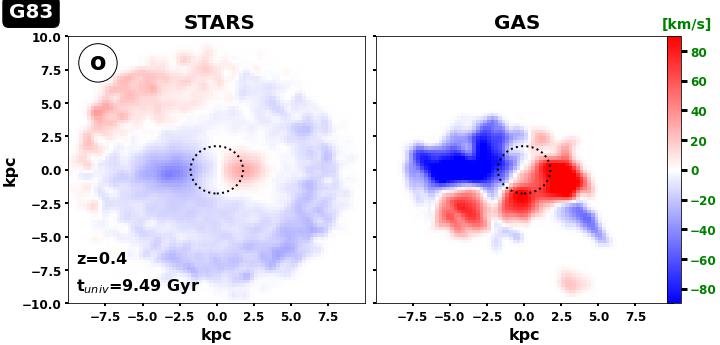}}
\rotatebox{0}{\includegraphics[width=9cm]{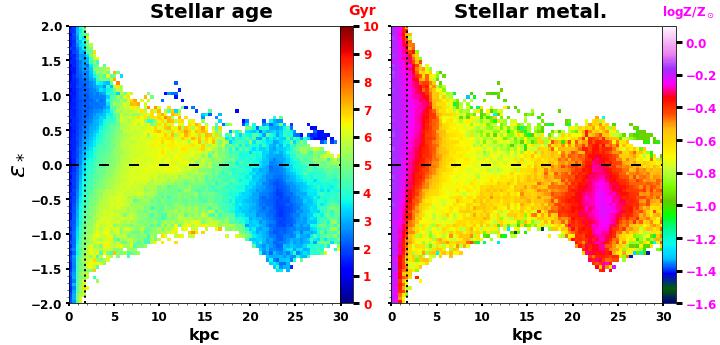}}
\rotatebox{0}{\includegraphics[width=9cm]{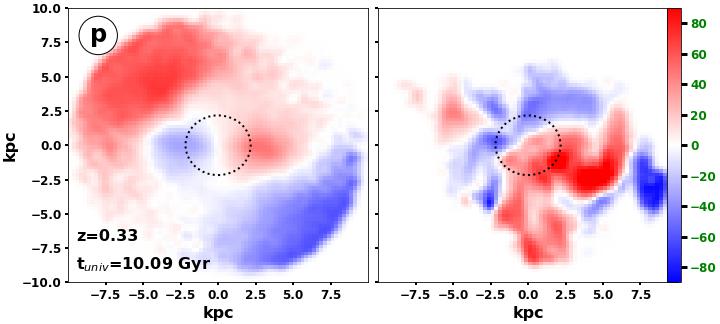}}
\rotatebox{0}{\includegraphics[width=9cm]{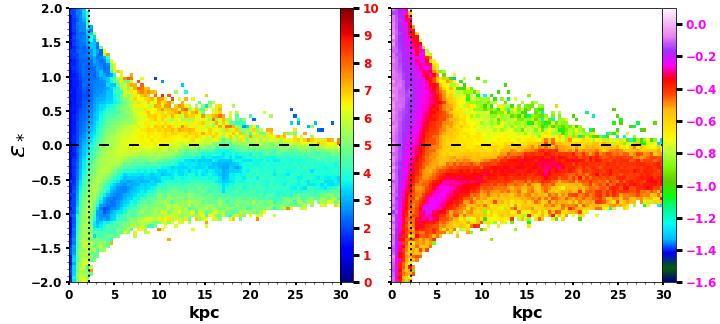}}
\rotatebox{0}{\includegraphics[width=9cm]{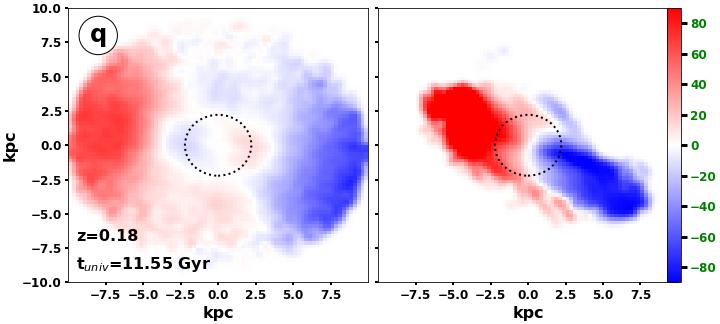}}
\rotatebox{0}{\includegraphics[width=9cm]{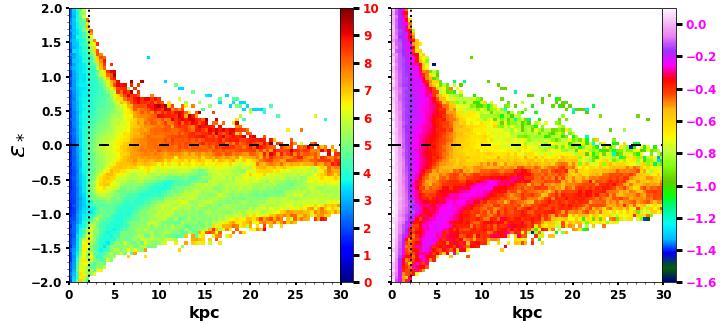}}
\caption{Same as Fig.~\ref{fig_g11_diag} but for G83.
We however do not show here  the time evolution
of the stellar mass and the mean circularity parameters for stars and gas.
}
\label{fig_app3}
\end{center}
 \end{figure*}
%%%%%%%%%%%%%%%%%%%%%%%%%%%%%%%%%%%%%%%%%%%%%%%%%

%%%%%%%%%%%%%%%%%%%%%%%%%%%%%%%%%%%%%%%%%%%%%%%%%%%%%%%%%%%%%
%     FIG APPENDIX 3
%%%%%%%%%%%%%%%%%%%%%%%%%%%%%%%%%%%%%%%%%%%%%%%%%%%%%%%%%%%%%%
\begin{figure*}
\begin{center}
\rotatebox{0}{\includegraphics[width=9cm]{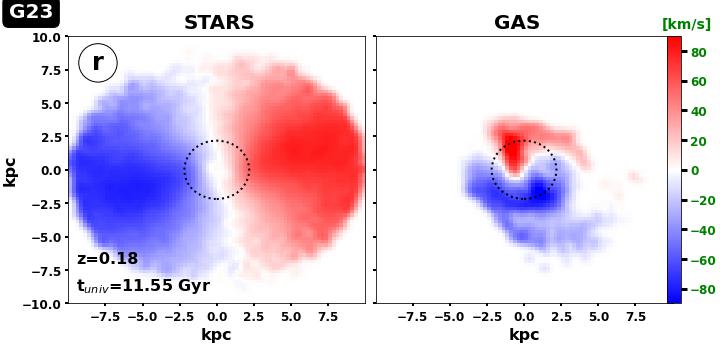}}
\rotatebox{0}{\includegraphics[width=9cm]{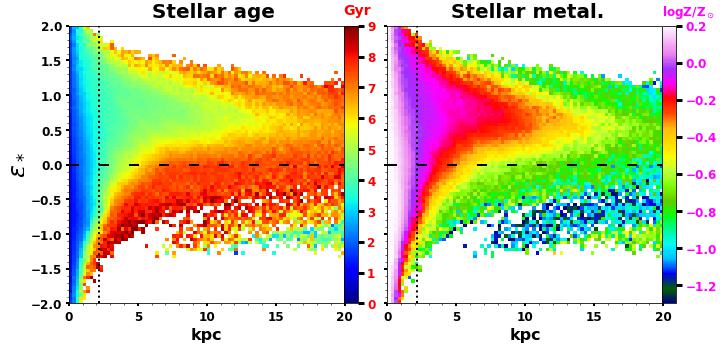}}
\rotatebox{0}{\includegraphics[width=9cm]{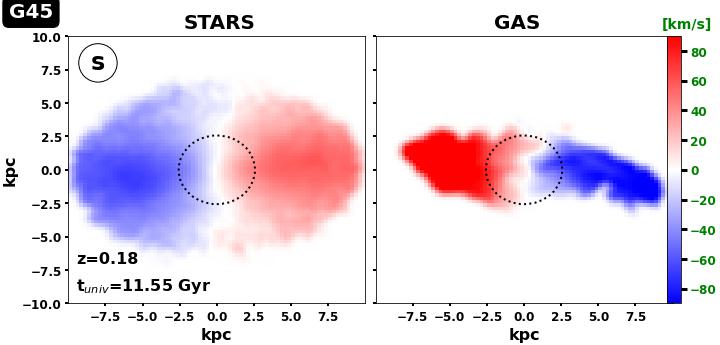}}
\rotatebox{0}{\includegraphics[width=9cm]{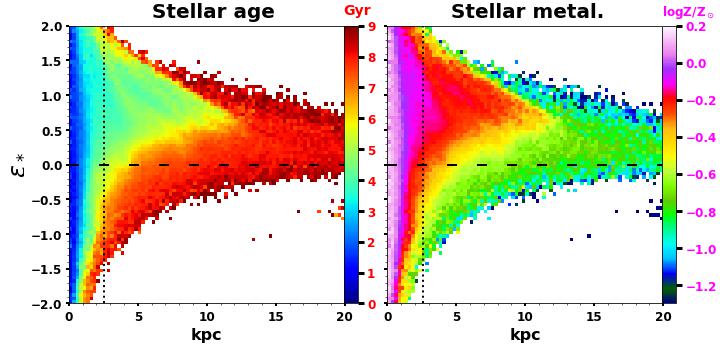}}
\rotatebox{0}{\includegraphics[width=9cm]{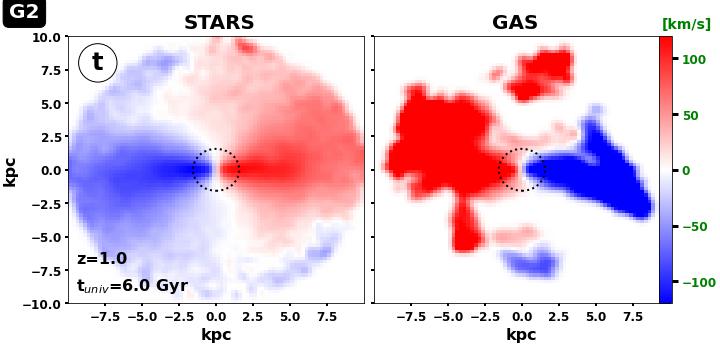}}
\rotatebox{0}{\includegraphics[width=9cm]{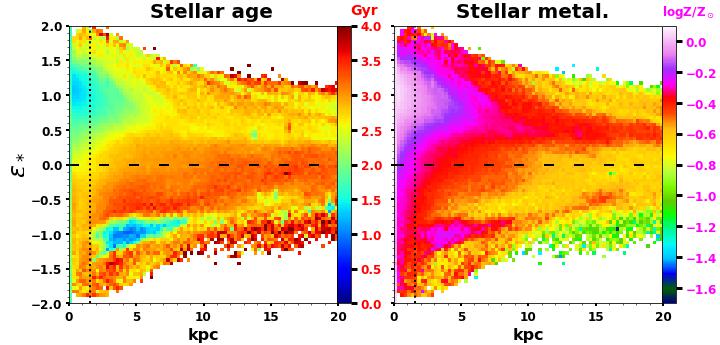}}
\rotatebox{0}{\includegraphics[width=9cm]{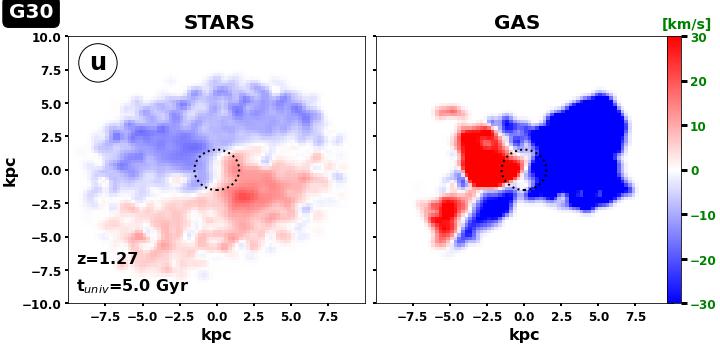}}
\rotatebox{0}{\includegraphics[width=9cm]{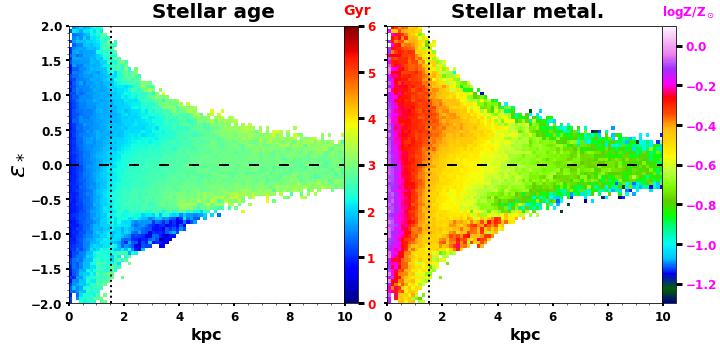}}
\rotatebox{0}{\includegraphics[width=9cm]{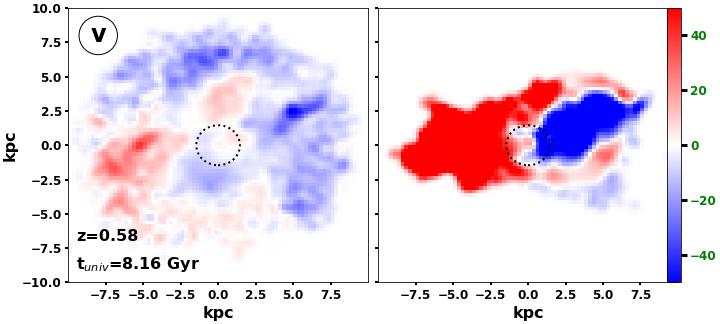}}
\rotatebox{0}{\includegraphics[width=9cm]{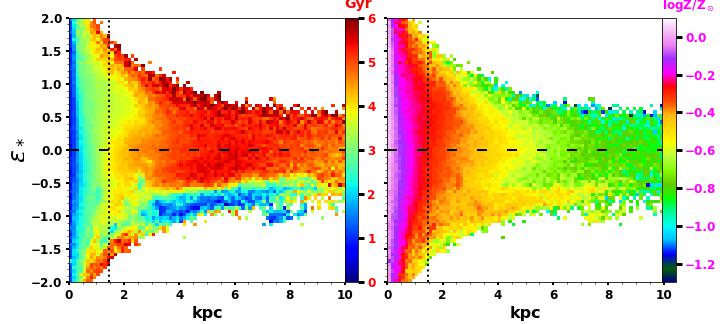}}
\caption{Same as Fig.~\ref{fig_g11_diag} but for G23,
G45, G2 and G30.
We, however, do not show here  the time evolution
of the stellar mass and the mean circularity parameters for stars and gas.
}
\label{fig_app3}
\end{center}
 \end{figure*}
%%%%%%%%%%%%%%%%%%%%%%%%%%%%%%%%%%%%%%%%%%%%%%%%%

%%%%%%%%%%%%%%%%%%%%%%%%%%%%%%%%%%%%%%%%%%%%%%%%%%%%%%%%%%%%%
%     FIG APPENDIX 4
%%%%%%%%%%%%%%%%%%%%%%%%%%%%%%%%%%%%%%%%%%%%%%%%%%%%%%%%%%%%%%
\begin{figure*}
\begin{center}
\rotatebox{0}{\includegraphics[width=9cm]{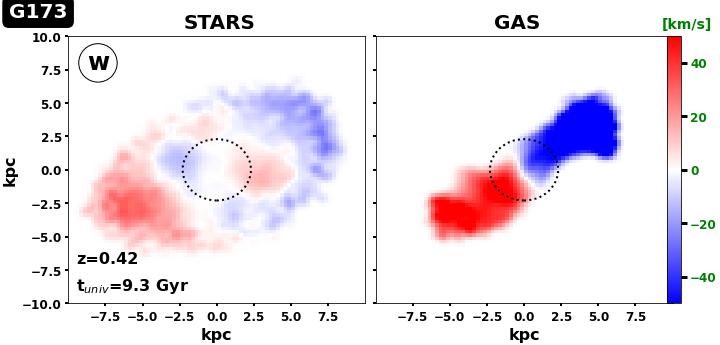}}
\rotatebox{0}{\includegraphics[width=9cm]{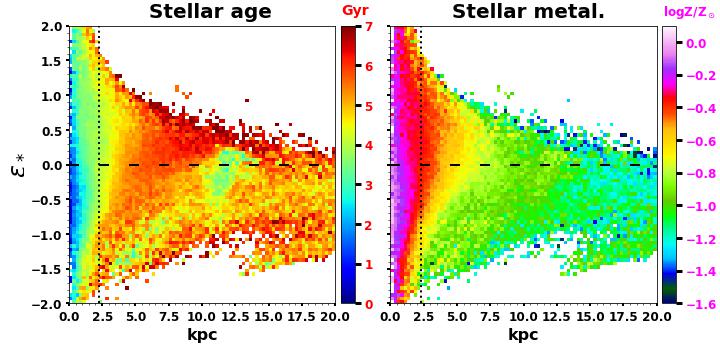}}

%\rotatebox{0}{\includegraphics[width=9cm]{FIG_G45_2.jpg}}
%\rotatebox{0}{\includegraphics[width=9cm]{FIG_G45_4.jpg}}
\caption{Same as Fig.~\ref{fig_g11_diag} but for G173.
We however do not show here  the time evolution
of the stellar mass and the mean circularity parameters for stars and gas.
}
\label{fig_app4}
\end{center}
 \end{figure*}
%%%%%%%%%%%%%%%%%%%%%%%%%%%%%%%%%%%%%%%%%%%%%%%%%

%%%%%%%%%%%%%%%%%%%%%%%%%%%%%%%%%%%%%%%%%%%%%%%%%%%%%%%%%%%%%%
%          FIG APPENDIX 5
%%%%%%%%%%%%%%%%%%%%%%%%%%%%%%%%%%%%%%%%%%%%%%%%%%%%%%%%%%%%%%
\begin{figure*}
\begin{center}
\rotatebox{0}{\includegraphics[width=6cm]{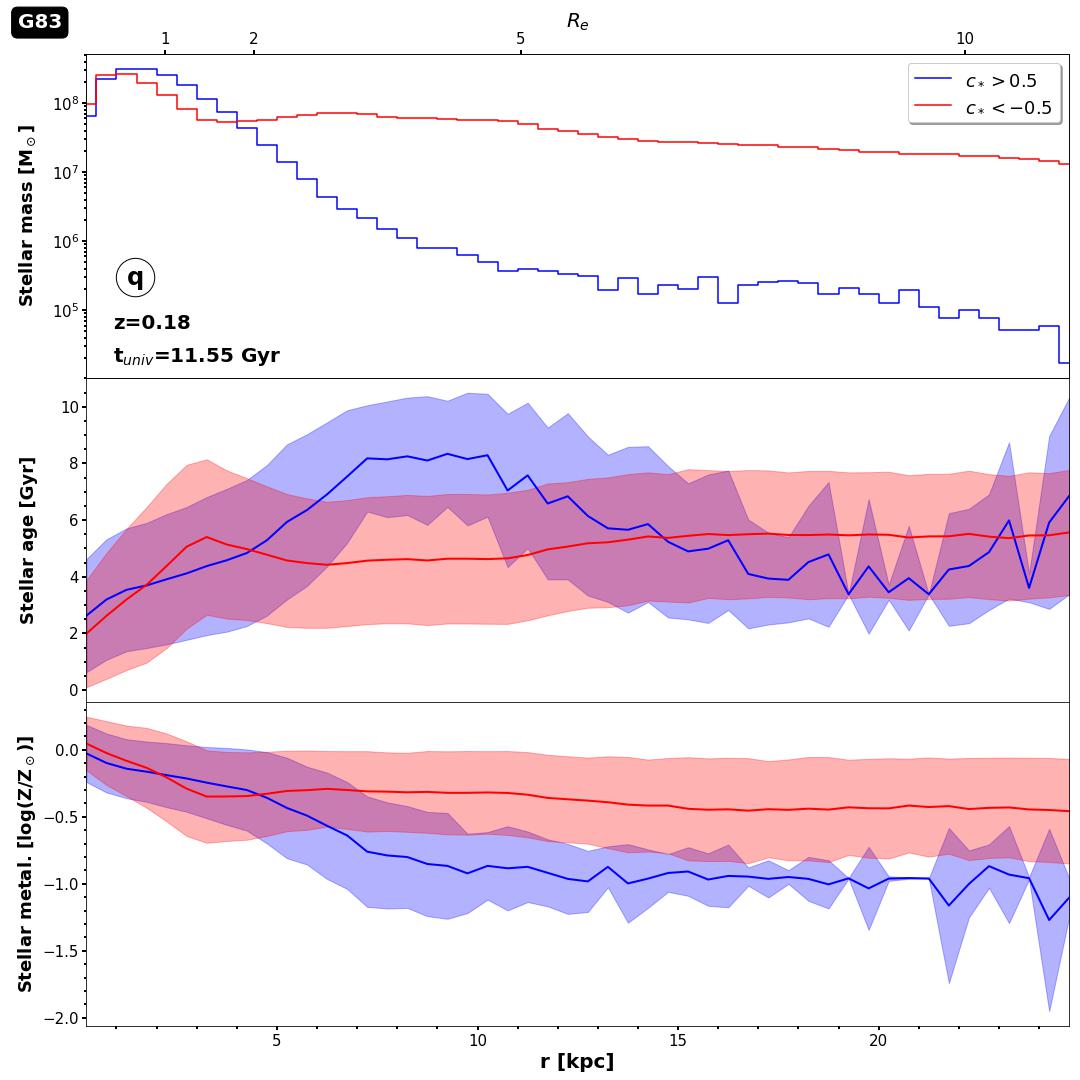}}
\rotatebox{0}{\includegraphics[width=6cm]{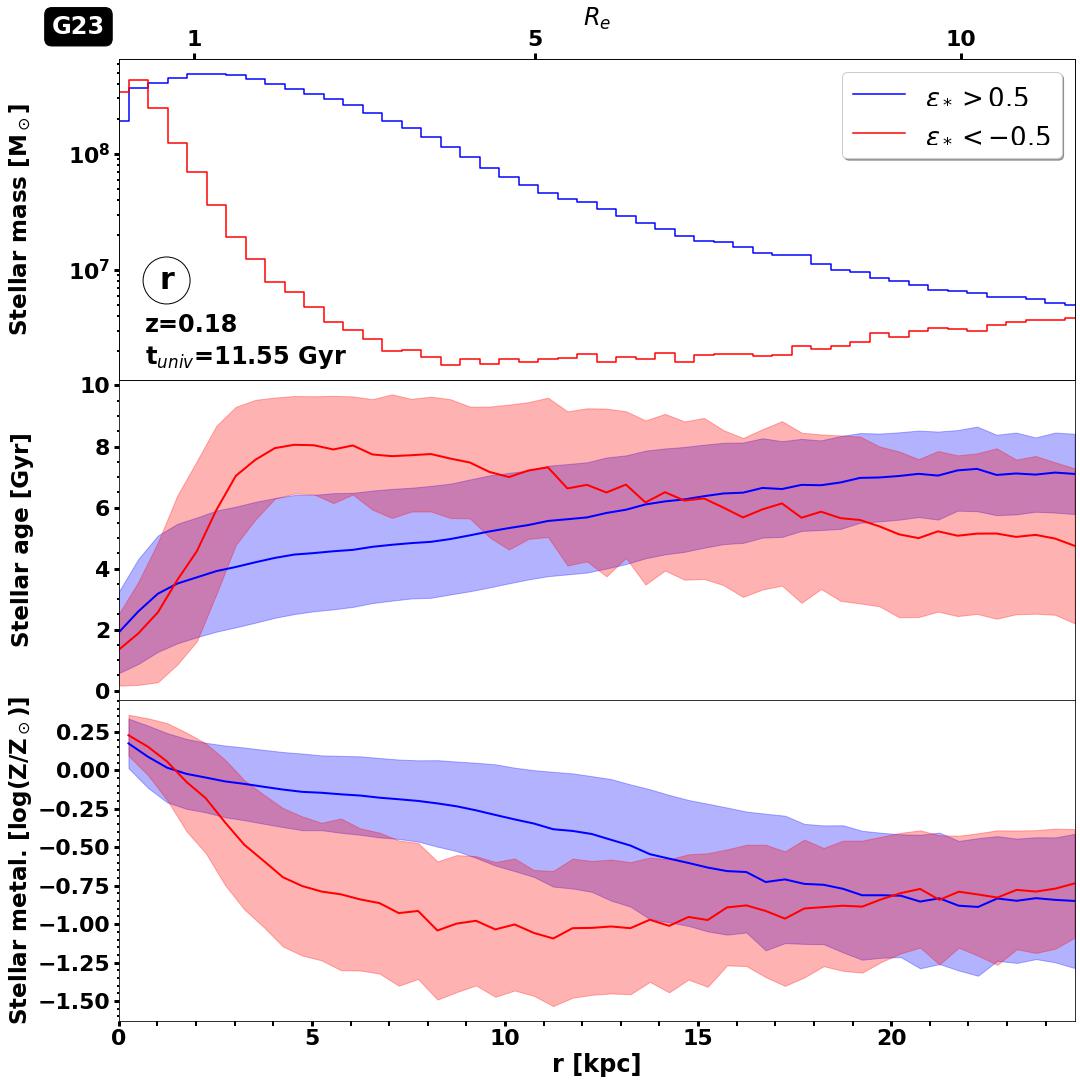}}
\rotatebox{0}{\includegraphics[width=6cm]{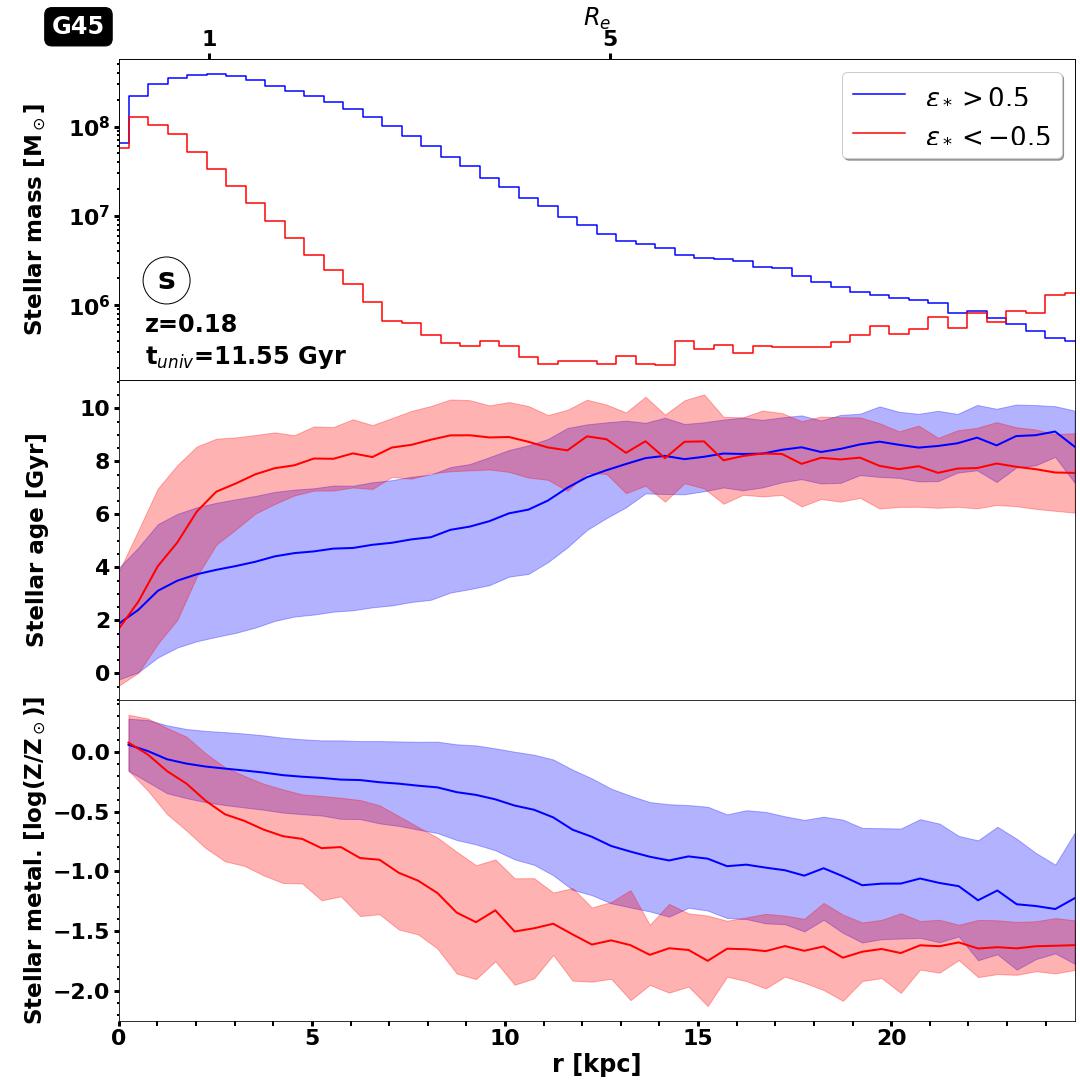}}
\rotatebox{0}{\includegraphics[width=6cm]{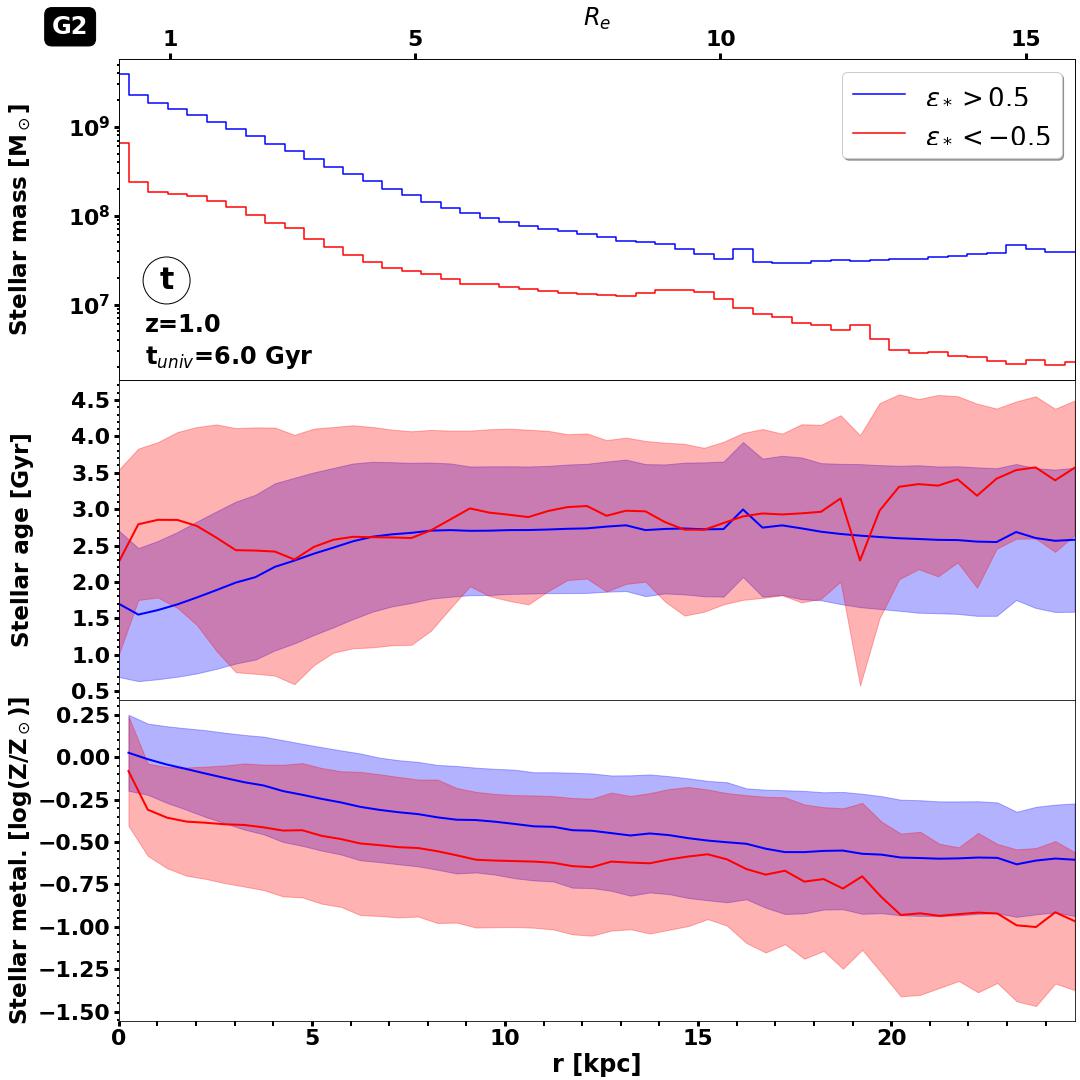}}
\rotatebox{0}{\includegraphics[width=6cm]{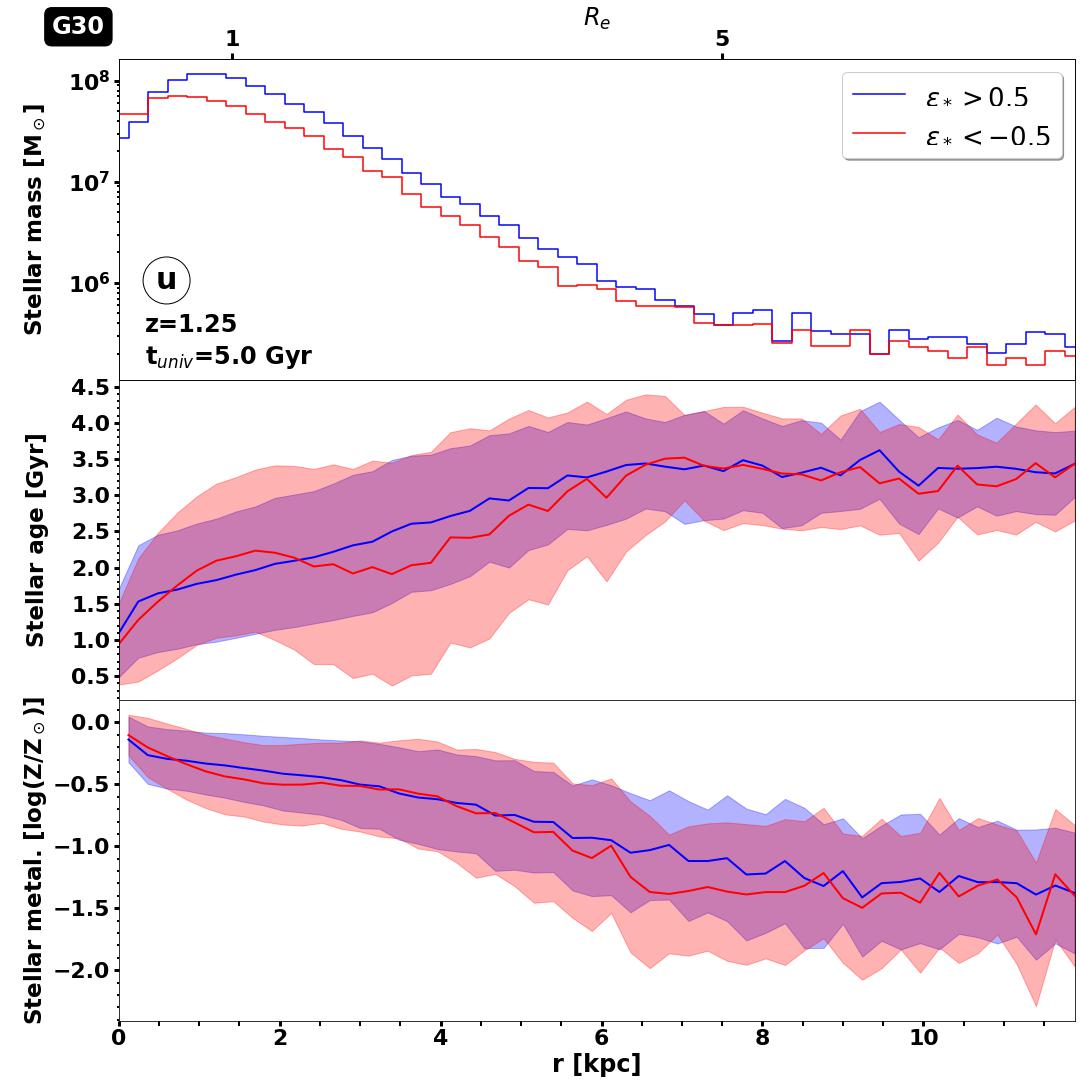}}
\rotatebox{0}{\includegraphics[width=6cm]{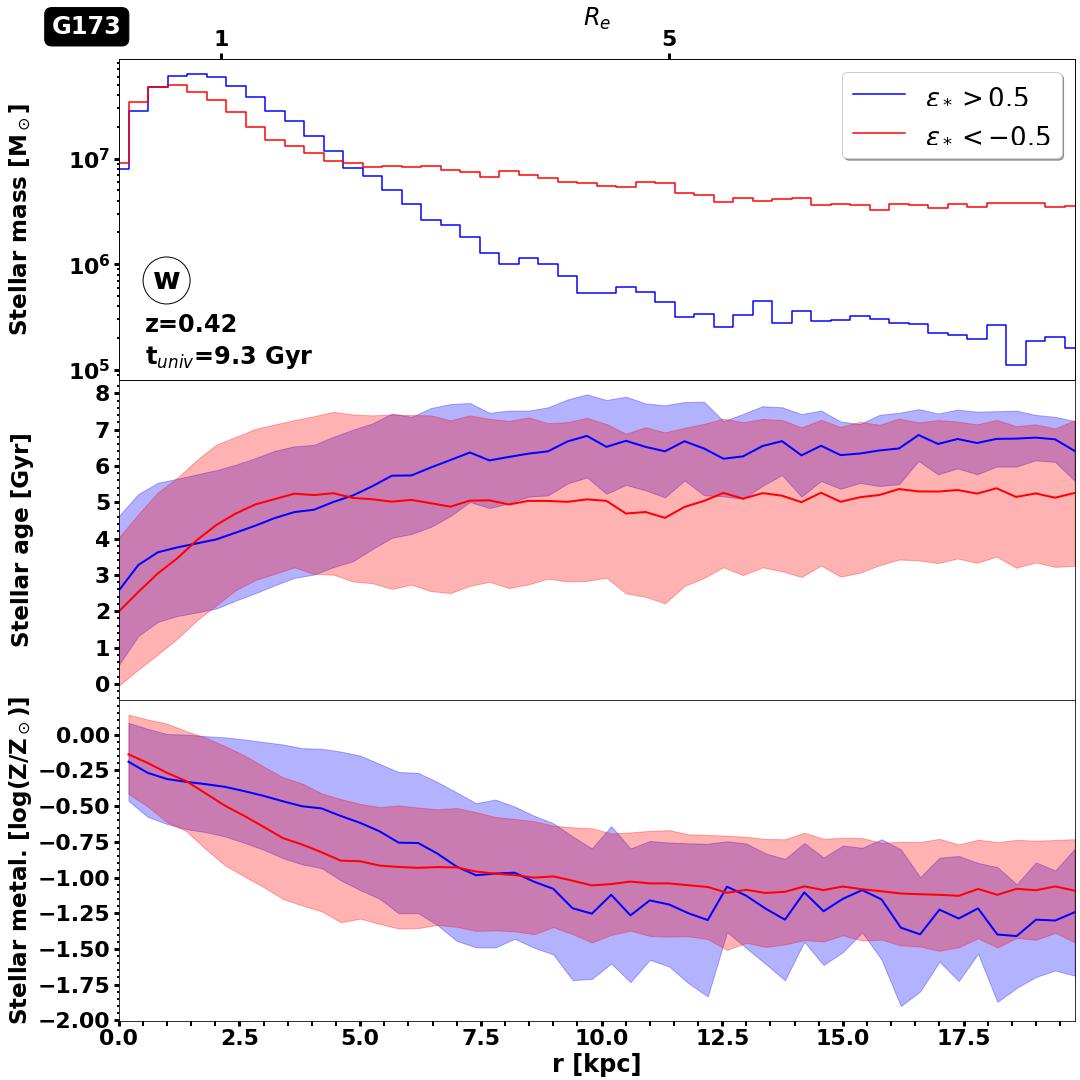}}
\caption{Same as Fig.~\ref{fig_g11_maz} but for G23, G45, G83,
G2, G30 and G173.
}
\label{fig_app5}
\end{center}
 \end{figure*}
%%%%%%%%%%%%%%%%%%%%%%%%%%%%%%%%%%%%%%%%%%%%%%%%%

%%%%%%%%%%%%%%%%%%%%%%%%%%%%%%%%%%%%%%%%%%%%%%%%%%%%%%%%%%%%%%
%          FIG 
%%%%%%%%%%%%%%%%%%%%%%%%%%%%%%%%%%%%%%%%%%%%%%%%%%%%%%%%%%%%%%
%\begin{figure*}
%\begin{center}
%\rotatebox{0}{\includegraphics[width=9cm]{FIG_Cmaz_BH132.jpg}}
%\rotatebox{0}{\includegraphics[width=9cm]{FIG_Cmaz_BH4015.jpg}}

%\caption{Same as Fig.~\ref{fig_g11_maz} but for G30 and G173.
%}
%\label{fig_app6}
%\end{center}
% \end{figure*}
%%%%%%%%%%%%%%%%%%%%%%%%%%%%%%%%%%%%%%%%%%%%%%%%%

\section{Evolution of BH796}
\label{appendix2}

Similarly to \cite{peirani+24}, we present here the cosmic evolution of the relevant properties of the central balck hole BH796, hosted by
G2.

%%%%%%%%%%%%%%%%%%%%%%%%%%%%%%%%%%%%%%%%%%%%%%%%%%%%%%%%%%%%%%
%          FIG 
%%%%%%%%%%%%%%%%%%%%%%%%%%%%%%%%%%%%%%%%%%%%%%%%%%%%%%%%%%%%%%
\begin{figure*}
\begin{center}
\rotatebox{0}{\includegraphics[width=12cm]{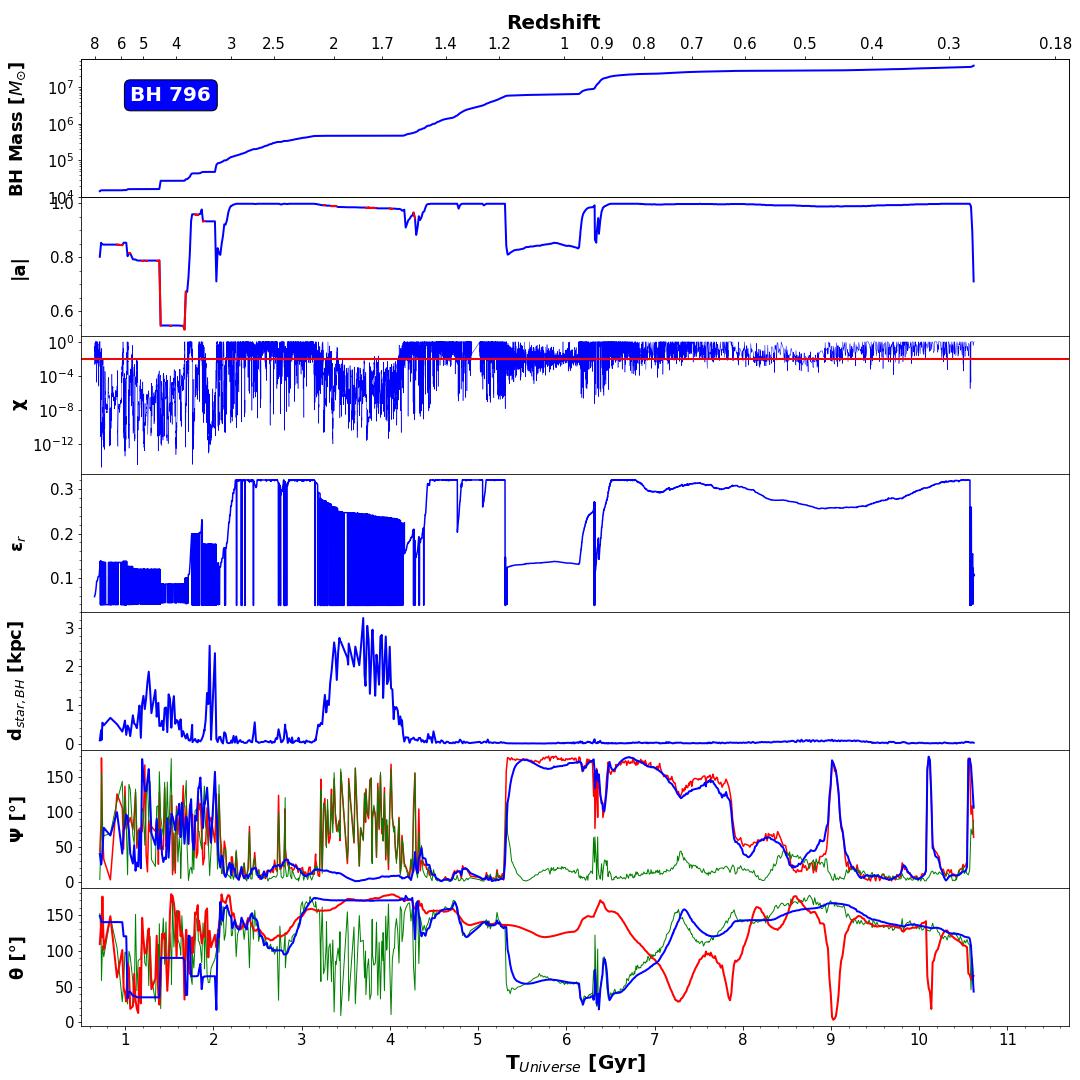}}
\caption{Evolution of relevant properties of the central black
  hole BH796.  \textit{From top to bottom:}
  {\bf 1)} BH mass evolution. The red and green dashed lines indicate
  epochs of major and minor mergers episodes respectively; {\bf 2)} BH
  spin evolution |a|. Red colors indicates when the gas accretion disk is in
  counter-rotation (i.e., a<0); {\bf 3} The Eddington ratio evolution
  ($\chi$). The red line delimits the quasar dominant mode
  ($\chi>$0.01) and the radio dominant mode ($\chi<$0.01); {\bf 4)}
  The evolution of the radiative efficiency of the gas accretion disk
  ($\epsilon_r$); {\bf 5)} The variation of the distance separation
  between the location of the BH and the center of its host galaxy
  (\dsbh); {\bf 6)} the evolution of the angle \PSIS between the BH
  spin and the angular momentum of the stellar component (blue)
  estimated within one effective radius.  We also plot the angle
  between the BH spin and the angular momentum of the gas accretion disk
  (green) as well as the angle between the stellar and the accreted
  gas component (red); {\bf 7)} The evolution of the polar angle
  $\theta$ describing the orientation of the BH spin (blue), the
  stellar (red) and gas accretion disk (green) angular momenta relative to a
  fixed reference frame of the simulation.
}
\label{fig_app7}
\end{center}
 \end{figure*}
%%%%%%%%%%%%%%%%%%%%%%%%%%%%%%%%%%%%%%%%%%%%%%%%%

\end{appendix}

\end{document}